\documentclass{WileyMSP-template}

\usepackage[binary-units]{siunitx}
\usepackage{chemformula}

\usepackage{ragged2e}

\usepackage{graphicx}
\usepackage{siunitx}
\graphicspath{{./Figures/}}
\DeclareGraphicsExtensions{.pdf,.jpg,.png,.eps}

\begin{document}

\pagestyle{fancy}
\rhead{\includegraphics[width=2.5cm]{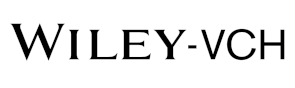}}
\setlength{\headheight}{25pt}

\title{Recent Advances and Future Perspectives of Single-Photon \\ Avalanche Diodes for Quantum Photonics Applications}

\maketitle

\author{Francesco Ceccarelli}
\author{Giulia Acconcia}
\author{Angelo Gulinatti}
\author{Massimo Ghioni}
\author{Ivan Rech}
\author{Roberto Osellame*}

\begin{affiliations}

Dr. F. Ceccarelli, Dr. R. Osellame\\
Istituto di Fotonica e Nanotecnologie – Consiglio Nazionale delle Ricerche (IFN-CNR) and Dipartimento di Fisica – Politecnico di Milano\\
Piazza Leonardo da Vinci 32, 20133 Milano, Italy\\
E-mail: roberto.osellame@polimi.it

Dr. G. Acconcia, Prof. A. Gulinatti, Prof. M. Ghioni, Prof. I. Rech\\
Dipartimento di Elettronica, Informazione e Bioingegneria – Politecnico di Milano\\
Piazza Leonardo da Vinci 32, 20133 Milano, Italy\\

\end{affiliations}


\justifying

\begin{abstract}

Photonic quantum technologies promise a revolution of the world of information processing, from simulation and computing to communication and sensing, thanks to the many advantages of exploiting single photons as quantum information carriers. In this scenario, single-photon detectors play a key role. On the one hand, superconducting nanowire single-photon detectors (SNSPDs) are able to provide remarkable performance on a broad spectral range, but their applicability is often limited by the need of cryogenic operating temperatures. On the other hand, single-photon avalanche diodes (SPADs) overcome the intrinsic limitations of SNSPDs by providing a valid alternative at room temperature or slightly below. In this paper, we review the fundamental principles of the SPAD operation and we provide a thorough discussion of the recent progress made in this field, comparing the performance of these devices with the requirements of the quantum photonics applications. In the end, we conclude with our vision of the future by summarizing prospects and unbeaten paths that can open new perspectives in the field of photonic quantum information processing.

\end{abstract}


\section{Introduction}
\label{sec:introduction}
Today quantum photonics represents one of the most fascinating and discussed field of research, with many efforts directed toward the development of novel technologies aimed at overcoming the limits of their classical counterparts. As an example, photonic quantum technologies promise the implementation of a global quantum network based on unconditionally secure communications \cite{Zhang2018qkd}, thanks to protocols like quantum key distribution (QKD) \cite{Lo2014} and components like quantum repeaters \cite{Briegel1998}. In this scenario, photons are the ideal candidates for carrying the quantum information, both for their flying nature and for the possibility to exploit the existing technological framework already developed for classical optical communications. Simulation and computing are other applications that photonic quantum technologies promise to overturn \cite{Aspuru-Guzik2012,OBrien2007}. The advantage of the quantum approach with respect to the classical one is now demonstrated on a specific task \cite{Arute2019}, while a general purpose quantum computer still looks far in the future. Nevertheless, many researchers, both in industry (see companies like Xanadu \cite{Xanadu2020} or PsiQuantum \cite{Psiquantum2020}) and academia \cite{Rudolph2017}, believe that the benefits provided by the photonic quantum technologies, like the operation in standard conditions (i.e. room temperature and pressure) and the weak interaction of photons with the external environment, will be the key for reaching such an astonishing achievement. Last, but not least, the exploitation of quantum systems to estimate unknown physical parameters with precision beyond the classical limits is the idea at the basis of quantum metrology and sensing \cite{Pirandola2018}. Also for this application, the photonic regime is often considered the most appropriate setting, exploited by mapping one or more physical parameters into the measurement of a photonic quantum state. Furthermore, this is also the most natural approach for the implementation of remote sensing protocols based on quantum technologies \cite{Lloyd2008,Maccone2020}.

A generic photonic quantum system consists of three different parts, corresponding to the generation, manipulation and detection of quantum states of light typically based on single photons. In this Progress Report we focus on single-photon detectors, also referred to as single-photon counters. Among the many technologies that are exploited for the implementation of single-photon detectors \cite{Hadfield2009}, an important class of devices is the one based on superconducting materials. In particular, superconducting nanowire single-photon detectors (SNSPDs) are now an established technology able to provide excellent performance on a broad set of wavelengths and to allow the realization of fruitful experiments in different quantum photonics applications \cite{Boaron2018,Wang2019}. SNSPDs can achieve photon detection efficiency (PDE) beyond 90\% \cite{Zadeh2017}, dark count rate (DCR) lower than \SI{0.01}{cps} \cite{Wollman2017}, timing jitter down to \SI{2.6}{\pico\second} full width at half maximum (FWHM) \cite{Korzh2020} and dead time compatible with photon count rates in the \si{\giga cps} range \cite{Vetter2016}. However, it is worth stressing that all these characteristics have never been demonstrated in the same device due to the technological trade-offs coming from both fabrication and operation of the detector \cite{Holzman2019}. This fact is evident if we consider SNSPDs with large active area (e.g. \SI{100}{\micro\meter\ diameter} \cite{Lv2017}), a feature that is particularly useful for simplifying the optical alignment in long-range free-space quantum communications \cite{Dequal2016}. Moreover, when we consider the high PDE they provide, it is worth also noting that this value is usually dependent on the polarization of light \cite{Marsili2013}, a characteristic that is not desirable when the photonic quantum information is encoded in degrees of freedom like path or time \cite{Flamini2019}. Another detector whose operation is based on superconducting materials is the transition-edge sensor (TES). This device, in addition to an exquisite sensitivity \cite{Fukuda2011}, provides also the ability of distinguishing the number of photons impinging on the detector at the same time, i.e. TESs are photon-number resolving (PNR) detectors. These interesting capabilities are highly appreciated in quantum photonics \cite{Giustina2013,Namekata2010}, but, on the other hand, TESs, being calorimetric devices, are also affected by a stringent trade-off between sensitivity and timing performance, that results in a timing jitter in the order of \si{\nano\second} \cite{Lamas-Linares2013} and a recovery time rarely lower than \SI{1}{\micro\second} \cite{Calkins2011}. Furthermore, both SNSPDs and TESs need a cryostat to operate at temperatures typically in the \si{1} - \SI{4}{\kelvin} range for SNSPDs and below \SI{1}{\kelvin} for TESs. This requirement dramatically increases the cost and the complexity of the experimental setup, preventing the exploitation of these detectors in components like satellite optical receivers \cite{Bedington2017}. In addition, the long-term vision in the field is the development of a fully integrated quantum system \cite{Wang2020}, where generation, manipulation and detection are all included in the same chip. Although single-photon detection has been already demonstrated on a waveguide-based photonic integrated circuit (PIC) for both SNSPDs \cite{Pernice2012} and TESs \cite{Gerrits2011}, yet the use of thermo-optic phase shifting (i.e. the most common approach for the implementation of programmable quantum PICs \cite{Harris2018}) has never been reported to date in a cryostat and, even assuming the compatibility between these two technologies, the typical cooling power of compact cryostats (i.e. in the order of hundreds of milliwatts \cite{Wang2014}) and the typical power consumption of a thermo-optic shifter (i.e. in the order of tens of milliwatts \cite{Harris2014}) would limit the number of reconfigurable integrated components to a few tens.

In this Progress Report, we review the recent evolution of what is currently considered the most important single-photon detector able to work at room temperature or slightly below: the single-photon avalanche diode (SPAD). Indeed, SPADs have currently superseded photomultiplier tubes (PMTs) for the detection of light at the level of individual photons thanks to their reliability, compactness, lower operation voltage, higher detection efficiency in the near infrared (NIR) region of the electromagnetic spectrum and possibility to be integrated in detector arrays \cite{Ghioni2007}. As of today, they are successfully employed in several applications like fluorescence microscopy \cite{Bruschini2019} and spectroscopy \cite{Michalet2014}, light detection and ranging (LiDAR) \cite{Buller2007} and even nuclear medicine \cite{Karami2017} in diagnostic tools such as positron emission tomography (PET) and single-photon emission computed tomography (SPECT). Speaking of quantum photonics, SPADs have been playing a prominent role in this field since the first years of the 2000s, being present in seminal works like the first demonstrations of deterministic single-photon sources \cite{Lounis2000,Yuan2002}, the first experiments in integrated quantum photonics \cite{Politi2008,Matthews2009} and, more recently, the first practical realizations of boson sampling experiments \cite{Broome2013,Spring2013}. Today, SPADs still play upfront in the field, as witnessed by the fact that they continue to be successfully employed for novel demonstrations in quantum photonics \cite{Liao2017satellite,Sparrow2018}, especially when a cryostat can not be employed in the experimental setup or when features like wide photoactive area or perfect insensitivity to the polarization of light are required. In addition, today SPADs are also the solution that guarantees the best scalability for the implementation of detector arrays (see as an example a comparison between \cite{Morimoto2020} and \cite{Wollman2019}).

The paper starts with a brief review of the SPAD fundamentals and of the relevant characteristics of its operation. After that, the discussion continues with the presentation of the main technological results achieved in the last decade, with a special focus on the application in quantum photonics. In the end, we conclude the paper by presenting our vision on SPADs integrated in waveguide-based quantum PICs.

\section{SPAD Fundamentals}
\label{sec:fundamentals}
In this section we report the fundamentals of SPADs: starting from the basic principle of operation, we discuss SPADs' main features, highlighting their dependence on the device structure and/or operating conditions.  

\subsection{Principle of Operation}
The investigation of pn junctions reverse-biased above the breakdown voltage started with some studies of their bistable behavior in the late 1950s \cite{Champlin1959}. A few years later, Webb et al. clearly refer to single-photon detection with such structures for the first time \cite{Webb1970spd} and nowadays this type of photodetector is widely known as SPAD, or sometimes Geiger-mode avalanche photodiode (Geiger-mode APD).

The typical I-V characteristic of a SPAD is sketched in Figure \ref{fig:fundamentals}a. As soon as a reverse bias $V_{rev} = V_{bd} + V_{ov}$ is applied (1), being $V_{bd}$ the breakdown voltage and being $V_{ov}$ usually referred to as overvoltage, no free carriers are present in the depletion region and therefore no current is flowing. In this condition, the absorption of a single photon can initiate the impact-ionization process giving rise to a macroscopic, self-sustaining avalanche current (2). At this point, the detector is completely blind, meaning that the absorption of another photon does not significantly change the amount of current flowing into the SPAD and therefore it can not be detected by the electronics; an external circuit able to quench the avalanche current by lowering the voltage below the breakdown level (3) is necessary. The quenching element can be a simple high-value resistor (in the order of \SI{100}{\kilo\ohm}) or a more complex circuit including active elements, usually referred to as active quenching circuit (AQC) \cite{Cova1996}. In the end, the SPAD must be reset to its initial bias condition (1) to make it ready to detect another photon.

In this scenario, single-photon sensitivity depends only on the detector since the macroscopic avalanche current that is generated by a SPAD can easily overcome the noise floor of the front end electronics. As a result, the output of the read out electronics is typically a digital pulse corresponding to the detection of a photon. When only the information about a photon impinging on the detector is retrieved, the approach is usually referred to as photon counting. 
Nevertheless, SPADs can also provide the information about the time of arrival of each photon with picosecond precision, a feature that is exploited in the so-called photon timing. Both the structure and the operating condition of a SPAD play a key role to determine the performance of this device either in photon counting and in photon timing applications. Therefore, in the following subsections the main performance of the SPAD are reviewed, with particular emphasis on their dependence on design and operating parameters.

\subsection{Photon Detection Efficiency}
The PDE of a SPAD is defined as the probability that a photon impinging on the active area of the detector succeeds in triggering an avalanche current that can be detected by the electronics. The study of the multiple factors involved in this mechanism was initiated in the 1970s \cite{Oldham1972triggering, Mcintyre1973avalanche} and extensive general models were recently reported \cite{Gulinatti2011,Pancheri2014}.

In order to have a macroscopic photogenerated current at the SPAD terminals, two conditions have to be met: the photon impinging on the detector must be absorbed into the active region and at least one of the two released carriers (an electron-hole pair) must reach the high-field region and trigger the avalanche current. The photon absorption, in turn, depends on two different aspects. First of all, the photon must not be reflected by the device. Unfortunately, this event is not rare due to the discontinuity of the refractive index between air and semiconductors. However, the exploitation of antireflective coatings (ARCs) has been proven effective to reduce the reflection coefficient provided that the coating material is properly chosen to smooth the refraction index discontinuity. Secondly, the photon must be absorbed within the device. For the sake of clarity, we can refer to the basic structure of a pn junction that is shown in Figure \ref{fig:fundamentals}b. The nature of the light-matter interaction results into an exponential dependence of the absorption probability density $\eta_{abs}$ on the absorption depth $x$ as described by the Lambert-Beer law:
\begin{equation}
    \eta_{abs}(x) = \alpha e^{-\alpha x},
    \label{eq:LB_law}
\end{equation}
where $\alpha$ is the absorption coefficient of the light in the device material, generally a function of the wavelength, with longer wavelengths typically requiring thicker detectors to be absorbed.

Now, photons that have been absorbed within the device still have only some probability to trigger an avalanche current due to two different factors, the collection probability $P_{coll}(x)$ and the triggering probability $P_{trig}(x)$. The first term takes into account that the photogenerated electron-hole pair may or may not be collected by the high electric field. Indeed, referring to Figure \ref{fig:fundamentals}b, three different regions can be identified, i.e. the two neutral regions and the depletion region. In the latter, the presence of an electric field makes $P_{coll}(x)$ unitary, while a carrier that is generated in the neutral regions moves only by diffusion and, thus, has a limited probability to reach the depletion layer. Since in the upper region the carrier that can trigger the avalanche is the hole, whereas in the lower one is the electron, we can state that:
\begin{equation}
    P_{coll}(x)=
    \begin{cases}
        P_{coll,h}, &\text{if 0 $< x < x_{n}$},\\
        1, &\text{if $x_{n} < x < x_{p}$},\\
        P_{coll,e}, &\text{if $x_{n} < x < x_{end}$},
    \end{cases}
\end{equation}
where $P_{coll,e}(x)$ and $P_{coll,h}(x)$ are the collection probabilities for the electron and the hole, respectively.

Finally, we must take into account that not every carrier reaching the depletion region triggers an avalanche. Indeed, there is only some probability that it succeeds in generating one or more daughter electron-hole pairs by impact ionization \cite{Oldham1972triggering}, then at least one of the daughter carriers have to cause another impact ionization event and so on until a self-sustaining mechanism is established. By using the results reported by Oldham et al. in \cite{Oldham1972triggering} and already used by Gulinatti et al. in \cite{Gulinatti2011} in their model, the overall triggering probability can be expressed as follows:
\begin{equation}
    P_{trig}(x) = P_{t,e}(x)+ P_{t,h}(x)- P_{t,e}(x) P_{t,h}(x),
    \label{eq:trig_prob}
\end{equation}
where $P_{t,e}(x)$ and $P_{t,h}(x)$ are the probabilities that an electron or a hole trigger an avalanche, respectively, while the product of the two terms is subtracted to avoid counting twice the events initiated by both an electron and a hole. The calculation of $P_{trig}(x)$ is usually not trivial, as the impact-ionization coefficients for both electrons and holes strongly depend on the applied electric field and on the history of the carrier itself \cite{Mcintyre1973avalanche, Okuto1974}. At the same time, the dependence of the triggering efficiency on the electric field, and therefore on the applied overvoltage \cite{Ghioni1996}, makes it possible to increase this contribution to the PDE by simply boosting the overvoltage. Unfortunately, this approach can easily cost an increment of DCR (see Section \ref{subsec:dcr}), afterpulsing (see Section \ref{subsec:ap}) and crosstalk in SPAD arrays (see Section \ref{subsec:crosstalk}).

\subsection{Timing Response}
\label{subsec:timing_response}
One of the best features of a SPAD is that it is not only able to detect a single photon but it can also provide the information about its arrival time with picosecond precision. The FWHM of the instrument response function (IRF) is a good indicator of the precision of the detector. In addition, some applications can also have tight requirements on the overall duration of the detector temporal response. In QKD experiments, for example, a long duration of the IRF can be a major source of bit errors at high rates \cite{Takesue2006}.

In the previous section, we have briefly recalled the different types of paths that a photogenerated carrier can travel before it succeeds in starting an avalanche. Here we will now analyze the impact of such different paths on the overall timing response of the detector. The IRF of a SPAD consists of two components, namely a peak and a tail \cite{Ripamonti1985,Lacaita1993,Gulinatti2011} as illustrated in Figure \ref{fig:fundamentals}c. Ideally, the delay $T_{det}$ between the absorption of a photon and the corresponding avalanche current reaching a given threshold should be constant. Actually, $T_{det}$ is the result of three main components, each one affected by statistical fluctuations. Referring again to Figure \ref{fig:fundamentals}b, we have already recalled that the avalanche starts in the depletion region. However, we can further split this region into two different areas: a high-field multiplication region (with a triangular shape in Figure \ref{fig:fundamentals}b), where the carrier multiplication actually takes place, and the remaining low-field drift region, where carriers are collected and accelerated toward the multiplication region. A photon absorbed close to the electric field peak in the multiplication region can promptly trigger an avalanche, while an electron photogenerated in the relatively low-field region needs to drift for a certain portion of the space charge region before it reaches the multiplication region where it can trigger an avalanche. As a result, this so-called transit time $T_{transit}$ in the drift region depends on the absorption depth $x$: since a photon can be absorbed anywhere in the drift region, the spread of transit times of the photogenerated electrons directly contributes to the jitter of a SPAD.

A second source of jitter is due to the randomness of the impact-ionization mechanism: once the photogenerated carrier is in the multiplication region, the avalanche current can increase in different ways, thus reaching a given threshold with a random delay $T_{build}$. This phase is usually called build-up \cite{Spinelli1997,Tan2007,Ingargiola2009}. It is worth noting that the impact-ionization mechanism is a very directional process, that develops parallel to the electric field vector. Therefore, we can distinguish this initial build-up phase, where the current growth is confined within a very narrow filament that contains the absorption point, from the following lateral propagation, in which the current starts to spread across the whole SPAD area. There are two main mechanisms leading to the lateral propagation: first, the multiplication-assisted diffusion \cite{Lacaita1990}, which is due to the strong carrier density gradient between the initial filament and the lateral regions. As a result, the carriers progressively diffuse, triggering the avalanche also in the area surrounding the filament. The multiplication-assisted diffusion contributes to statistical fluctuations for two reasons \cite{Assanelli2011}: on the one hand, the process itself is noisy because it relies on random phenomena (diffusion and impact ionization); on the other hand, the geometry of propagation depends on the position of photon absorption (e.g. in the center or at the edge of the detector), which is random as well. The second contribution to lateral propagation is due to hot carriers in the filament, which can cause the emission and subsequent absorption of secondary photons potentially triggering another avalanche also in another region of the same detector \cite{Lacaita1993photon}. In this case, the temporal evolution of the current is affected by the number and the absorption position of the secondary photons. It is clear that both these mechanisms contribute to a fluctuation of the third contribution $T_{prop}$. Overall, $T_{det}$ is given by the sum of these three contributions ($T_{transit} + T_{build} + T_{prop}$) and its statistical dispersion $ \sigma(T_{det})$, measured as FWHM, represents the SPAD timing jitter. In addition, carriers generated in the neutral regions slowly diffuse, possibly reaching the depletion region and starting an avalanche, but since they are not accelerated by the electric field, they can experience a long random delay. This phenomenon causes the diffusion tail of the IRF. 

It is worth saying that build-up and lateral propagation jitter contributions decrease if the overvoltage is increased thanks to the higher efficiency of the triggering and propagation phenomena. On the other hand, it is clear that a design trade-off exists between the temporal response of a SPAD and its PDE, especially at long wavelenghts where the photon absorption length in the detector material can exceed \SI{10}{\micro\meter}. Indeed, the use of a substrate clearly separated from the pn junction (i.e. n-type substrate below the p-side of the junction, see Figure \ref{fig:fundamentals}b as reference) can be paramount to limit both the transit time and the diffusion tail, paying the price of a reduced PDE. Finally, the diameter of the device can also play a key role when the lateral propagation is the dominant contribution. Nevertheless, a small SPAD can be a limiting factor for the optical alignment, especially in space communication, and a low timing jitter can be attained only by using a suitable low-threshold front end electronics \cite{Gulinatti2005} able to detect the avalanche when it is still confined in its initial filament. On the contrary, when the photoactive area exceeds the application requirements, focusing the photons in a small spot can be the right solution to effectively reduce the lateral propagation contribution.


\subsection{Dark Count Rate}
\label{subsec:dcr}
The absorption of a photon is not the only event that can trigger an avalanche current in a SPAD: indeed, spurious events are observed even when the detector is kept in perfect dark conditions. The average number of dark pulses per unit time is usually referred to as DCR. Having a low DCR means that a long time interval can elapse before a self-generated event is detected, and this is paramount in many quantum applications to minimize the error probability.

The main contribution to DCR at room temperature is usually given by the Shockley-Read-Hall (SRH) mechanism, which basically consists in trap-assisted generation of electron-hole pairs. For a reverse biased pn junction, the SRH generation rate per unit volume can be modeled as reported in \cite{Ghioni2007}: 
\begin{equation}
    G_{srh} = N_{trap} \frac{r_{e}r_{h}}{r_{e}+r_{h}},
    \label{eq:dcr_simple}
\end{equation}
where $N_{trap}$ is the volume density of deep levels, while $r_{e}$ and $r_{h}$ are the probabilities per unit time of emitting an electron or a hole, respectively. From Equation \ref{eq:dcr_simple}, it is already clear that DCR depends on the volume of the detector and, in turn, on its active area, which is usually set by the requirements of the optical setup. Moreover, the model reported above is valid only if the electric field is so low that direct and phonon-assisted tunneling from the trap level toward the bands can be neglected. This is not usually the case for SPADs; nevertheless, the SRH model can be extended by introducing an effective carrier emission rate:
\begin{equation}
    \left \{
        \begin{array}{rl}
            r_{e,eff}=\Gamma_{e}(F) r_{e},\\
            r_{h,eff}=\Gamma_{h}(F) r_{h},
        \end{array}
    \right.
\end{equation}
where the enhancement factor $\Gamma_{e}(F)$ depends on the electric field $F$. 
The latter is determined both by detector design and biasing condition. So a larger overvoltage may increase the PDE, but results in a higher DCR. Finally, both the initial quality of the material and the fabrication steps deeply affect $N_{trap}$: therefore, the optimization of the process is crucial to achieve a low $G_{srh}$. In particular, the most common source of deep levels is the metal contamination that may occurr during wafer handling, ion implantation or high-temperature heat treatment \cite{Ghioni2007}.

Besides SRH generation, for high values of the electric field, electrons can directly tunnel from the valence into the conduction band. As the corresponding electron-hole pair can trigger an avalanche, this phenomenon, known as band-to-band tunneling (BBT), contributes as well to the detector DCR. BBT is usually negligible at room temperature, where the DCR is dominated by field-enhanced SRH generation. However, as the temperature is lowered, the contribution of the SRH process drops quickly and the DCR may become dominated by BBT. This is a very undesired situation because the BBT generation rate exhibits only a weak dependence on the temperature and therefore it limits the effectiveness of cooling on DCR reduction. As BBT strongly depends on the electric field, this situation can be avoided with a proper design of the detector. This is shown in Figure \ref{fig:fundamentals}d, where the dependence of the DCR on the temperature for two SPADs, without and with engineering electric field, is reported.


\subsection{Afterpulsing and Maximum Count Rate}
\label{subsec:ap}
Besides DCR, another phenomenon known as afterpulsing can lead to the recording of spurious events with a SPAD detector. In this case, the macroscopic effect is quite different, as a train of pulses could be observed following a single-photon absorption, as already shown by Haitz et al. in 1965 \cite{Haitz1965}. The minimization of afterpulsing is a key factor in SPADs for quantum applications as it can be particularly detrimental due to its correlated nature. Indeed, the physical origin of this phenomenon relies in the trapping of carriers during an avalanche event \cite{Cova1991,Giudice2003}; such trapped carriers can be subsequently released either when the SPAD is reverse biased below or above the breakdown voltage, and in the second case a spurious event correlated to a former count is registered.

The afterpulsing has been widely modeled by following a statistical approach. Considering the statistical time $\tau$ between a trapping event and the subsequent release of the trapped carrier, the total afterpulsing probability can be computed as follows:
\begin{equation}
    P_{ap} = \int_{0}^{+\infty} \eta_{ap}(\tau) d\tau,
\end{equation}
where $\eta_{ap}$ is the afterpulsing probability density. Since SPADs typically present different types of deep levels, $\eta_{ap}$ can be expressed as a sum of exponential decays:
\begin{equation}
    \eta_{ap}(\tau)= \sum_{i=1}^{N} A_{e,i} e^{-e_{e,i} \tau},
\end{equation}
where $N$ is the number of different defects, $A_{e,i}$ is a suitable succession of prefactors that depend on multiple properties as the deep-level capture probability, the triggering efficiency and the avalanche charge, and $e_{e,i}$ is the emission rate of $i$-th trap. While this formula has been expressed for the case of an emitted electron, the same considerations can be made also for the holes.

Both the fabrication and the SPAD operating conditions can have a remarkable impact on the afterpulsing performance of these detectors. On the one hand, the minimization of deep levels is crucial to limit the amount of trapped carriers. As for SRH-DCR, this means that having a high-quality wafer and a limited number of carefully designed fabrication steps are the main keys to success. On the other hand, the amount of charge that is trapped and its release time depend on the working condition of the device. To limit the trapping, the amount of charge that crosses the device during an avalanche must be kept as low as possible. On the detector side, this can pose a trade-off with the PDE, since both the charge and the detection efficiency are directly proportional to the overvoltage. At the same time, given a fixed $V_{ov}$, a high parasitic capacitance at the SPAD terminals would require a high amount of charge flowing through the device at each quench/reset transition.

On the electronics side, the trapping can be limited with a prompt quench of the avalanche current. In this scenario, the fast passive quenching provided by a high-value resistor may seem the most appropriate solution. Nevertheless, such approach poses severe limitations to the maximum count rate of the system, due to the slow recharge (i.e. reset) introduced by the use of a high-value resistor: for this reason, an AQC is actually preferable to achieve a high count rate \cite{Cova1996}. Moreover, an AQC can keep the detector below the breakdown voltage during the so-called hold-off time. Such feature is paramount especially when a significant amount of carrier trapping cannot be avoided. In this case, by using a proper hold-off time, it is possible to ensure that the release of most trapped carriers occurs when the detector is kept below the breakdown voltage, thus avoiding the ignition of a spurious avalanche. The sum of sensing, quenching, hold-off and reset times is usually referred to as dead time, since it represents the minimum time between two consecutive detections. Both the hold-off and the total dead time are highlighted in Figure \ref{fig:fundamentals}e, where the anode voltage waveform of a SPAD has been recorded upon the photon arrival. It is worth noting that the dead time sets the maximum achievable count rate, provided that the latter is not limited by other factors like the power dissipation. This means that a trade-off exists between afterpulsing probability and maximum count rate. When the afterpulsing is not the most stringent limit for the application, but, on the other hand, a very high count rate is required, the hold-off time can be set to zero and the total dead time is limited only by the quenching/reset voltage transients and, thus, by the overvoltage that is employed. Finally, low-temperature operation can have adverse effects on afterpulsing. Indeed, the release-time of traps typically increases when lowering the temperature, thus leading to a higher probability of having a correlated event when the detector has been re-biased above the breakdown voltage. To mitigate this effect, a longer hold-off time can be used, although paying the price of a reduced maximum count rate. 


\subsection{Crosstalk}
\label{subsec:crosstalk}
\begin{figure}
    \includegraphics[width=\linewidth]{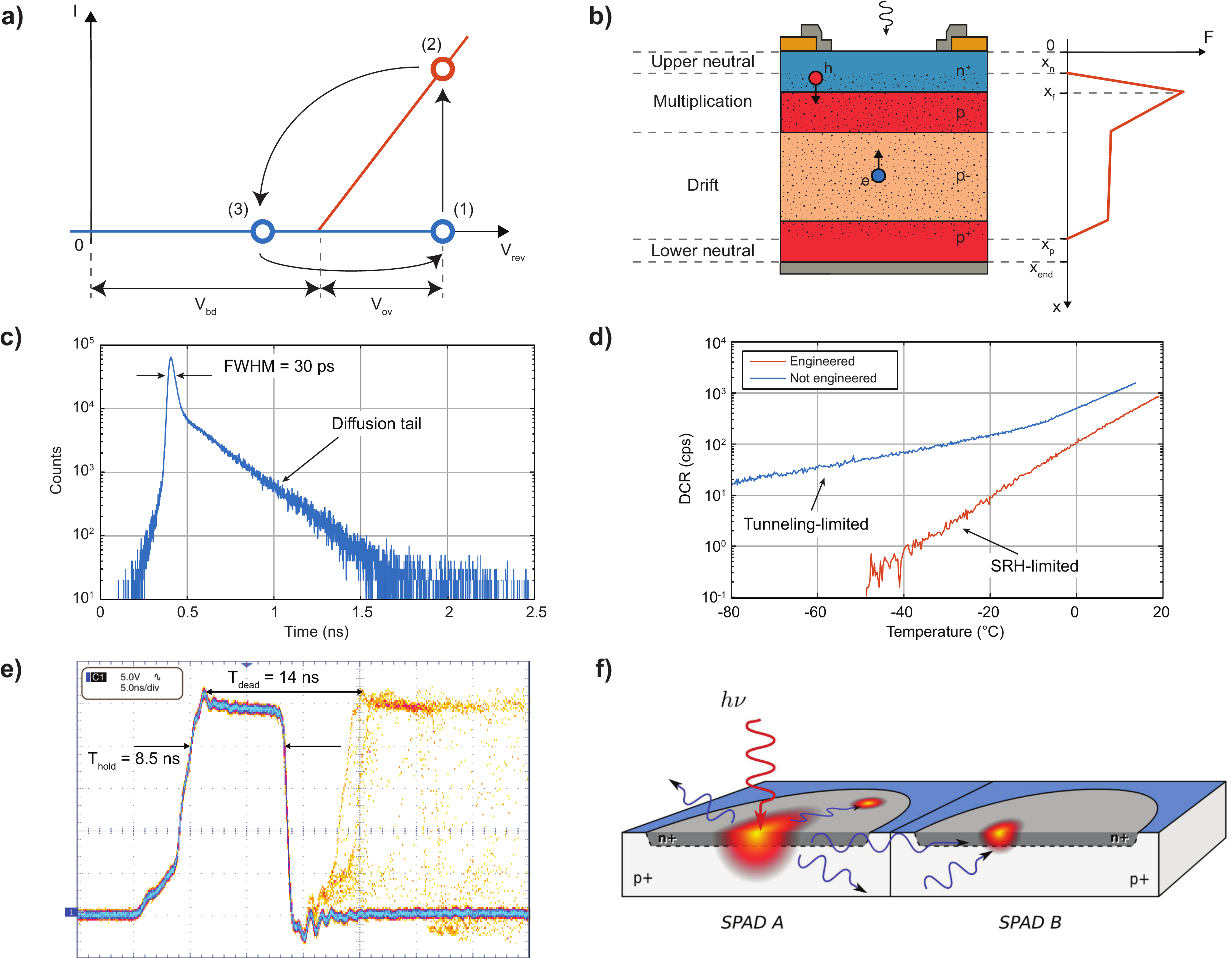}
    \caption{Fundamental principles about the operation of a SPAD detector. a) I-V characteristic of a SPAD; $V_{rev}$ is the reverse bias, $V_{bd}$ the breakdown voltage and $V_{ov}$ is the excess bias, also known as overvoltage. Different colors are used to highlight the bistability typical of this detector. b) Basic structure of a SPAD, with corresponding electric field $F$ on the rigth side. c) SPAD IRF: the logarithmic scale allows the reader to appreciate the presence of the diffusion tail. d) DCR dependence on the temperature for two SPADs: with and without engineering of the electric field. e) Voltage waveform recorded at the anode of the SPAD after the avalanche is sensed by the AQC; $T_{hold}$ is the hold-off time, while $T_{dead}$ is the dead time. f) Direct and indirect contributions to the optical crosstalk. Reproduced with permission \cite{Rech2008}. Copyright 2008, The Optical Society.}
    \label{fig:fundamentals}
\end{figure}

Finally, a key advantage of SPADs is the possibility of fabricating arrays. The development and exploitation of fully planar processes has been driving the fabrication of a wide variety of multi-pixel structures, potentially allowing faster and/or more complex analysis of light signals. In order to properly operate each detector independently in the array, crosstalk among pixels must be avoided. The main source of crosstalk in a SPAD array is the spurious triggering of one pixel caused by the operation of another one. To understand this phenomenon, first of all, photon emission from an avalanching junction must be considered. Indeed, when the avalanche current flows in a triggered pixel, photons at various wavelengths are emitted by intraband relaxation of hot-carriers crossing the junction \cite{Lacaita1993bremsstrahlung}. Such emitted photons can be reabsorbed by another pixel and cause a correlated ignition. This undesired event is known as optical crosstalk. In this scenario, both the distance between pixels and the substrate of the array can play a part. Considering the direct optical path between two adjacent pixels, it is clear that crosstalk increases with reducing the pixel-to-pixel distance, thus posing a limit to the maximum array density. To address this issue, several solutions have been successfully investigated to interrupt the internal light path, as deep trenches filled with metals \cite{Gulinatti2012}. However, Rech et al. in 2008 proved that also indirect optical paths can substantially contribute to the crosstalk \cite{Rech2008}. In this case, secondary photons are basically reflected off the bottom of the chip, potentially bypassing trenches. Indirect optical crosstalk is much less sensitive to pixel-to-pixel distance thus requiring some other technological solution. Both direct and indirect optical crosstalk is depicted in Figure \ref{fig:fundamentals}f.

While the spurious ignition of a SPAD represents the more evident effect of crosstalk, in photon timing applications it must be also taken into account that the timing performance of a SPAD can be threatened by the presence of other pixels on the same chip. This issue is particularly important in SPADs that require a low-threshold timing operation \cite{Ghioni2008}. In this case, charge sharing between pixels through the substrate capacitance must be avoided, as it could cause an unpredictable variation of the timing threshold potentially affecting the recorded photon-arrival time. Therefore, the substrate must be biased at a fixed potential, making the substrate capacitance a viable path where the avalanche current can get lost \cite{Acconcia2017}. This issue can make the design of the front end circuit particularly challenging to achieve low-threshold operation, especially if no solutions are used to minimize such parasitic capacitance \cite{Labanca2018}.

Finally, attention must be paid to the exploitation of a thick low-doped substrate directly connected to one of the SPADs terminals \cite{Mandai2012}: aimed at increasing the PDE at long wavelengths, this solution can not only lead to a long tail in the IRF, but it can also result in delayed crosstalk events \cite{Acerbi2019}.


\section{Fabrication Technologies}
\label{sec:fabrication}
In this section we review the main fabrication technologies employed for the realization of SPAD detectors. In particular, we analyze the most performing solutions proposed in literature in the last decade, arranging the discussion on the basis of the materials, whose optical absorption properties play a decisive part in choosing the detector that is most suited for a given application.

\subsection{SPADs for Visible/NIR Detection}
As of today, the silicon industry is considered one of the largest in the world, allowing micro and nanoelectronic devices to pervade all aspects of our life. Such a vast and thriving technological framework has provided the finest environment to silicon detectors to flourish and establish. Silicon SPADs are indeed a feather in the cap of single-photon detection, representing today one of the most performing solution for operation in the wavelength range between \si{400} and \SI{1000}{\nano\meter} \cite{Ghioni2007}. A large number of applications in quantum photonics exploits single photons in this range, from boson sampling \cite{Wang2017} to free-space QKD (based either on satellite-to-ground \cite{Liao2017satellite} or ground-to-ground \cite{Restelli2010} links), from quantum imaging \cite{Howland2013} to quantum memories \cite{Lvovsky2009}, that are considered today a key technology for the implementation of quantum repeaters or for the synchronization of probabilistic events in quantum computing. Such wavelength range is indeed very attractive for the applications; the reason behind this is not only related to the remarkable performance of silicon SPADs, but also to the availability of high-performance single-photon sources, both heralded \cite{Wang2016,Kaneda2015} and deterministic \cite{Senellart2017}, and low-loss PICs, based either on silica \cite{Politi2009,Meany2015} or silicon nitride \cite{Munoz2017} waveguides.

Silicon SPADs have been demonstrated in the literature following one of two orthogonal approaches \cite{Ghioni2007}. The first relies on a custom fabrication process, purposely optimized for the implementation of SPADs, while the second one is based on the exploitation of a standard technology, already developed for the fabrication of other devices (e.g. transistors). The best solutions reported in the literature to date are here selected and proposed for the employment in quantum photonics.

\subsubsection{Custom Technologies}
Since the very first works published by Webb and McIntyre \cite{McIntyre1985,Dautet1993}, SPADs have been implemented by employing a custom fabrication process, purposely designed for the optimization of the detector performance. Nowadays, custom SPADs can be divided into two main families: reach-through and planar SPADs, even though the latter can be further classified into thin, red-enhanced (RE-SPADs) and resonant-cavity-enhanced SPADs (RCE-SPADs).

Before going into details of the characteristics of these devices, it is worth saying that reach-through SPADs have been giving the most important contribution in quantum applications so far, thanks to the high PDE in the NIR, the low DCR, the large photoactive area and the availability of relatively low-cost commercial modules. However, not all the applications in this field have requirements that can be fulfilled by these detectors. As an example, the experimental characterization of single-photon sources \cite{Brida2012, Ates2012,Ates2013,Versteegh2014,Jons2017} often requires detectors with low timing jitter ($<$ \SI{100}{\pico\second}), while in other applications like QKD clocked at \si{\giga\hertz} rates it is not only important a low timing jitter, but also a maximum count rate higher than \SI{100}{\mega cps} \cite{Rogers2007}. For these reasons, a great research effort has been devoted to design a new generation of SPADs, i.e. planar SPADs, capable of combining high PDE and low DCR with low timing jitter and high count rate. Even more, the scalability of the detection system has been gaining a prominent role in quantum photonics. As an example, experiments of photonic boson sampling today exploit tens of optical modes \cite{Wang2019}, where each of them requires an independent detector. For these reasons, new planar SPADs offering not only high performance but also the possibility of fabricating arrays are envisioned to become more and more popular in quantum applications.

\begin{figure}
    \includegraphics[width=\linewidth]{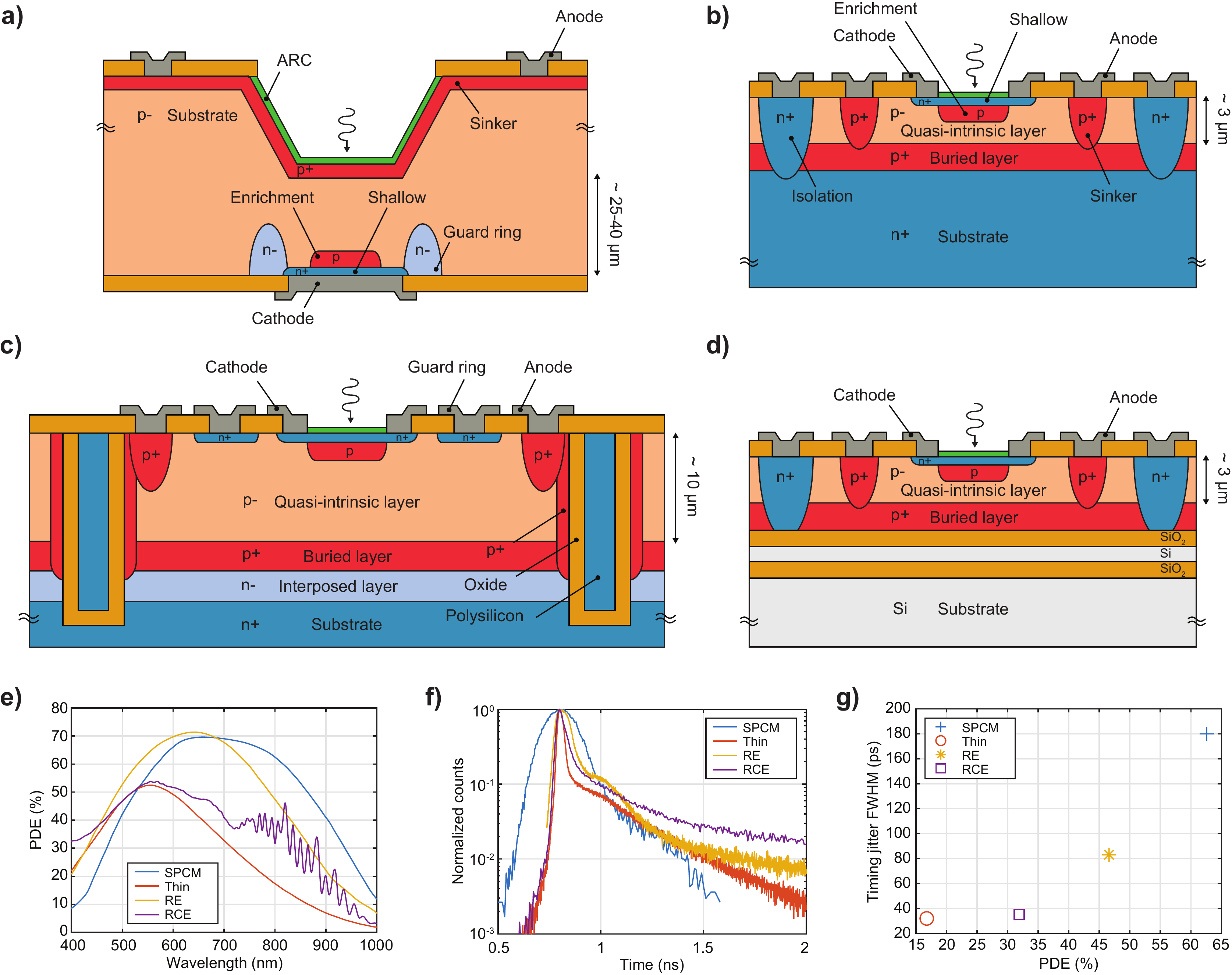}
    \caption{Device structure and performance of SPADs based on a customized fabrication process. a) Structure of the reach-through SPAD developed at former RCA Electro-Optics (now Excelitas Technologies) \cite{Dautet1993}. b) Structure of the planar thin SPAD developed at Politecnico di Milano \cite{Ghioni2007}. c) Structure of the planar RE-SPAD proposed by Gulinatti et al. \cite{Gulinatti2020}. d) Structure of the planar RCE-SPAD proposed by Ghioni et al. \cite{Ghioni2008rce}. e) PDE as a function of the wavelength reported for custom-technology SPADs: reach-through (SPCM) \cite{SPCM-AQRH2020}, thin \cite{Ghioni2007}, RE- \cite{Gulinatti2020} and RCE-SPAD \cite{Ghioni2008rce}. f) Timing response at $\lambda =$ \SI{820}{\nano\meter} reported for custom-technology SPADs: reach-through (SPCM, selected and modified by Rech et al.) \cite{Rech2006}, thin \cite{Ghioni2007}, RE- \cite{Ceccarelli2018timing} and RCE-SPAD \cite{Ghioni2008rce}. g) Final comparison on the performance of custom-technology SPADs at $\lambda=$ \SI{800}{\nano\meter} for the PDE and $\lambda =$ \SI{820}{\nano\meter} for the timing jitter.}
    \label{fig:custom}
\end{figure}

\paragraph{Reach-through SPADs}
Devised by McIntyre and Webb at former RCA Electro-Optics (now Excelitas Technologies), the basic structure of a reach-through SPAD (also referred to as thick SPAD) is reported in Figure \ref{fig:custom}a \cite{Dautet1993}. The cathode is defined by a n+ phosphorous diffusion on a quasi-intrinsic p- substrate, while a deeper p boron diffusion (i.e. enrichment) is exploited to suitably tailor the electric field in the substrate. Both the enrichment and the lateral lightly-doped n- phosphorous guard rings are needed to prevent edge breakdown. Finally, the wafer is flipped and the back side is etched down to a thickness usually in a range between \SIrange{25}{40}{\micro\meter} \cite{Zappa2007,Stipcevic2013}. The p+ boron diffusion on this side acts as a low-resistivity sinker to the charge carriers and, in addition, it guarantees a good ohmic contact with the anode metallization. The resulting electric field profile consists of the multiplication region, right beneath the front side of the detector, and the drift region, that extends through the entire quasi-intrinsic p- substrate. Since photons enter the SPAD from the back side, the detector is classified as back-illuminated. The final result is a p/n+ geometry that exploits the electrons as main carriers to initiate the avalanche, thus achieving a high triggering probability and, as a result, a high PDE. In addition, the detection efficiency can be further enhanced by exploiting the cathode metallization as a back-reflecting layer \cite{Stipcevic2013}.

Several commercial solutions based on reach-through SPADs have been made commercially available, as the SPCM modules by Excelitas Technologies \cite{SPCM2019}, the COUNT and SAP500 series distributed by Laser Components \cite{COUNT2019, SAP5002020} and the ID120 by IDQuantique \cite{ID1202019}. On the one hand, such photon counting modules can provide remarkable performance as a PDE exceeding 60\% up to a wavelength of \SI{800}{\nano\meter}, a DCR down to \SI{25}{cps} for a device diameter of \SI{180}{\micro\meter} (modules are equipped with a thermoelectric cooler) and an afterpulsing probability lower than 1\%. However, reach-through SPADs are also characterized by some important flaws. First of all, the timing jitter of such modules is usually in the order of some hundreds of \si{\pico\second} FWHM, although optimized versions do exist \cite{SPCM-TR2020} that can attain a \si{200} - \SI{250}{\pico\second} timing jitter, provided that the light is focused in a \SI{10}{\micro\meter} diameter spot. Even selecting best devices, focusing the light in a small region of the photoactive area and using a low-threshold front-end circuit, the timing jitter could not be reduced below \SI{150}{\pico\second} FWHM \cite{Rech2006,Stipcevic2013}. Indeed, such thick structure providing excellent performance in terms of PDE is also intrinsically affected by a remarkable dispersion of the transit time with detrimental effects on the timing response as explained in Section \ref{sec:fundamentals}. Secondly, the maximum speed of these SPADs is limited to a few tens of \si{\mega cps} \cite{SPCM-AQRH2020} due to the high power dissipated at each detection event caused by the high breakdown voltage (in the order of a few hundreds of volt). Finally, the high power dissipation, along with the complexity of such structure, has prevented the fabrication of reach-trough SPAD arrays so far \cite{Zappa2007}. 

\paragraph{Thin SPADs}
A possible structure aimed at solving the issues of reach-through SPADs has been devised at Politecnico di Milano. The first-generation planar device is usually referred to as thin SPAD, and it is based on the front-illuminated n+/p structure reported in Figure \ref{fig:custom}b \cite{Ghioni2007}. The avalanche region is fabricated again by means of a shallow n+ phosphorous diffusion for the cathode, while the p enrichment region in this case is defined by ion implantation. The most important difference with respect to reach-through SPADs is the substrate, that is realized by double epitaxy on a n+ wafer: a thin quasi-intrinsic p- region ($\simeq$ \SI{3}{\micro\meter}), that acts as an absorption layer, and a p+ buried layer that, along with the p+ sinker, minimizes the series resistance of the SPAD toward the anode terminal. The n+ substrate in this case plays a very important role not only for obtaining a sharp temporal response free of slow diffusion tails, but, together with the n+ diffusion surrounding the active area, it allows also the full electrical isolation of the SPAD and, in turn, the realization of detector arrays featuring fully independent pixels. Thin SPAD arrays have been demonstrated to date up to a format of $8\times8$, with a separation pitch of \SI{250}{\micro\meter} \cite{Ceccarelli2016}. Indeed, the number of thin SPADs that can be integrated is currently limited by the electrical connections with the external electronics, even though other research groups have already started the investigation of 3D integration techniques to increase the array format up to $256\times256$ and beyond \cite{Aull2018}. Other advantages of thin SPADs are the high count rate they can achieve (up to \SI{160}{\mega cps} with an afterpulsing probability lower than 5\% \cite{Ceccarelli2019}) and the low timing jitter (down to \SI{32}{\pico\second} FWHM even for a detector diameter as large as \SI{200}{\micro\meter} \cite{Gulinatti2005}) that, thanks to the employment of a suitable front end circuit \cite{Acconcia2017}, can be obtained also in thin SPADs having a relatively low avalanche electric field. This is indeed very important to obtain at the same time a sharp timing response and a low DCR of few \si{cps} (demonstrated for circular SPADs having \SI{50}{\micro\meter} diameter and operating at \SI{-20}{\celsius} \cite{Ceccarelli2018}). On the other hand, due to the limited extension of the drift region, thin SPADs can not provide the excellent PDE that is guaranteed by thicker structures and, indeed, this quantity is usually limited to 50\% at \SI{550}{\nano\meter} and 15\% at \SI{800}{\nano\meter}. Today, thin SPADs are available on market thanks to the single-photon detection modules distributed by Micro Photon Devices (MPD) \cite{PDM2019}.

\paragraph{RE-SPADs}
In 2012, Gulinatti et al. proposed a new SPAD structure \cite{Gulinatti2012red} usually referred to as RE-SPAD, aimed at combining high PDE in the NIR range with low timing jitter. As depicted in Figure \ref{fig:custom}c, the RE-SPAD is based on the thin SPAD structure, yet with some important modifications \cite{Gulinatti2020}. First of all, the epitaxial quasi-intrinsic region is expanded to \SI{10}{\micro\meter} in order to improve the PDE up to 70\% at \SI{650}{\nano\meter} (47\% at \SI{800}{\nano\meter}), with no side effect on the DCR, that is fully determined by the electric field of the avalanche region. However, this modification results in a higher breakdown voltage, that increases from about \SI{35}{\volt} for a thin SPAD up to \SI{70}{\volt} for a RE-SPAD, and in a larger overvoltage, that increases from a few volt up to \SI{20}{\volt}. In order to fully accommodate such higher voltages it is necessary, on the one hand, to avoid the edge breakdown by introducing guard rings around the cathode diffusion and, on the other hand, to increase the breakdown voltage of the substrate junction by interposing a lightly-doped n- layer between the n+ substrate and the p+ buried layer. Moreover, this additional layer reduces also the capacitive parasitics of the junction \cite{Labanca2018}, with beneficial effects in terms of both timing jitter and afterpulsing. Unfortunately, the increased thickness of the quasi-intrinsic layer also prevents the n+ isolation from reaching the n+ substrate, resulting in a SPAD that is no more electrically isolated from the rest of the substrate and, thus, that can not be employed in detector arrays of fully independent pixels. The solution to this issue is the replacement of junction isolation with a dielectric approach based on deep trenches whose sidewalls are covered with silicon dioxide (in order to isolate the p well of the RE-SPAD) and whose volume is filled with n+ polysilicon (in order to prevent direct optical crosstalk). Deep isolation trenches can be also exploited to minimize the volume of the p well in which the RE-SPAD is fabricated, with a further beneficial effect on the substrate junction capacitance. RE-SPAD arrays based on this approach have been demonstrated up to a format of $32\times1$, with a separation pitch again of \SI{250}{\micro\meter} \cite{Ceccarelli2018redarray}. Moreover, the same problem described for the n+ isolation diffusion affects also the p+ sinker, that does not reach the p+ buried layer. This problem results in a higher series resistance of the SPAD that can slow down the avalanche current growth and, in turn, impair the temporal jitter. Deep trenches are the right solution also to this problem: indeed, by implanting boron on the trench sidewalls it is possible to recover the low-resistivity path between the sinker and the buried layer. Thanks to the reduction of both the capacitive and resistive parasitics of the RE-SPAD, a timing jitter as low as \SI{83}{\pico\second} FWHM has been recently reported \cite{Ceccarelli2018timing}. Finally, RE-SPADs can be also employed in high-count-rate applications up to \SI{100}{\mega cps}, even though the afterpulsing probability exceeds 10\% for count rates beyond \SI{80}{\mega cps} \cite{Ceccarelli2019}.

\paragraph{RCE-SPADs}
SPAD structures reported in previous paragraphs show that the depth of the depletion region can easily set a trade-off between PDE and timing jitter/count rate. To overcome such limitation, Ghioni et al. in 2008 \cite{Ghioni2008rce} investigated the exploitation of a buried Bragg reflector fabricated in a double silicon-on-insulator (SOI) substrate (Figure \ref{fig:custom}d). This evolution of the thin SPAD is usually referred to as RCE-SPAD. Compared to traditional thin SPADs, the p well in which the detector is fabricated behaves as a resonant cavity able to confine the photons and to make them transit multiple times through the depletion layer. The final result is an enhancement of the PDE in the range between \SIrange{750}{950}{\nano\meter} up to a factor of 3, while preserving the excellent timing jitter, the low operating voltage and the compatibility to be fabricated in detector arrays that are intrinsic of the thin structure. A comparison between the PDE and the timing response provided by reach-through, thin, RE- and RCE-SPADs is reported in Figure \ref{fig:custom}e and Figure \ref{fig:custom}f, respectively. As a matter of fact, a disadvantage of the RCE-SPAD is the high sensitivity of the PDE on the wavelength, that makes a fine tuning of the cavity necessary in order to maximize the enhancement factor in terms of absorption. Secondly, the higher defectivity of the SOI substrates can be a limiting factor for the DCR. Nevertheless, dark counts lower than \SI{30}{cps} at \SI{-20}{\celsius} have been demonstrated \cite{Ghioni2008rce} for a RCE-SPADs having a diameter of \SI{20}{\micro\meter} and even better results have been reported shortly afterwards by modifying the position of the gettering region that surrounds the active area of the detector \cite{Ghioni2009}. A final summary about PDE and timing jitter is reported in Figure \ref{fig:custom}g.

\subsubsection{Standard Technologies}
Since the first years of 2000s, many researchers have been looking at the complementary metal-oxide-semiconductor (CMOS) industry to investigate the exploitation of such mature and reliable technology, yet available at an affordable cost, to design complex SPAD-based systems on chip thanks to the possibility of integrating the detector along with the quenching and pre-processing electronics. Being a SPAD essentially a pn junction, working single-pixel devices have been demonstrated in several CMOS technology nodes so far \cite{Rochas2003, Finkelstein2006, Niclass2007, Xiao2007, Faramarzpour2008, Gersbach2009, Field2010, Karami2010, Pancheri2011, Gersbach2012, Villa2014, Dutton2016,RonchiniXimenes2019}. Starting from these results, systems featuring up to a million of pixels have been reported in literature \cite{Morimoto2020, Ximenes2018, Gyongy2017, Ulku2018, Perenzoni2015, Villa2014bis, Field2014, Braga2013,  Maruyama2013, Niclass2012, Veerappan2011}. However, the potential drawbacks of this CMOS approach were clear from the beginning \cite{Ghioni2007}. As a matter of fact, the evolution of CMOS circuits is driven by mass applications with demands inherently different from the ones imposed by the optimization of the SPAD performance. Indeed, the relentless scaling of CMOS devices toward shallow depth of wells and tubs, high doping concentrations and low supply voltages is not directly compatible with the implementation of the large absorption layers and the high breakdown voltages that are needed by SPAD detectors in order to provide high PDE, especially in the NIR, and low noise.

While large and densely-integrated CMOS SPAD arrays are successfully exploited in LiDAR \cite{Morimoto2020, Zhang2018, Hutchings2019} and fluorescence microscopy/spectroscopy \cite{Stoppa2009,Bruschini2019, Schwartz2008, Ulku2018}, the use of single-pixel CMOS SPADs in quantum photonics is still limited and, despite the remarkable progresses made in recent years, custom-technology solutions are often preferred thanks to the higher sensitivity to red and NIR photons and to their wide active area. CMOS SPADs are considered a suitable option only when the application requires a detector array, as an example to realize quantum imaging systems \cite{Lubin2019}, to perform protected \cite{Piacentini2017} and weak value quantum measurements \cite{Piacentini2016,Piacentini2016bis}, or to increase the throughput of quantum random number generators (QRNGs) \cite{Burri2013, Tisa2015}. In principle, CMOS SPADs could also be a suitable option to achieve a high count rate as the typically low overvoltage along with the minimization of electrical parasitics that can be achieved by integrating the detector along with the electronics on the same chip makes it easier to perform a fast quenching/reset transition \cite{Eisele2011,Niclass2010}. Nevertheless, the actual exploitation of a very short dead time with CMOS SPADs is typically prevented by the rapid increase of the afterpulsing \cite{Steindl2017}.

Concerning the structure, till the beginning of the 2010s, typical CMOS SPADs relied on one-sided p+/n structures based on a p+ diffusion (i.e. the anode) over a standard/deep n well (i.e. the cathode) and resulting in shallow junctions at few hundreds of \si{\nano\meter} from the surface. As a result, the peak of the PDE was biased toward the blue/green region of the spectrum, while the PDE for NIR photons ($\lambda >$ \SI{800}{\nano\meter}) was limited to no more than 10\%. In recent years, a significant research effort has been devoted to the design of structures aimed at overcoming this limitation.

\begin{figure}
    \includegraphics[width=\linewidth]{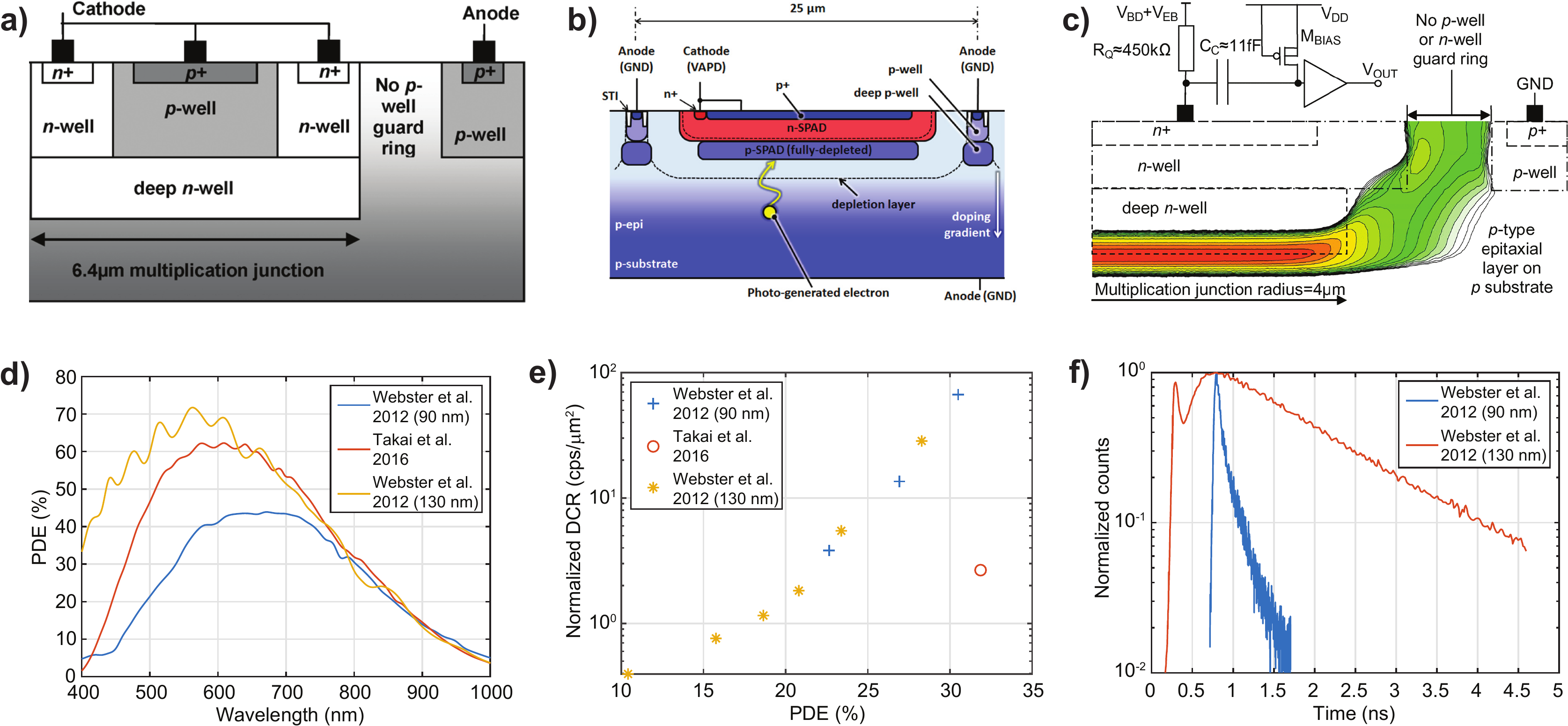}
    \caption{Device structure and performance of SPADs based on a standard fabrication process, with no substrate isolation. a) Device structure of the SPAD developed by Webster et al. in a \SI{90}{\nano\meter} CMOS imaging technology. Adapted with permission \cite{Webster2012}. Copyright 2012, IEEE. b) Device structure of the SPAD developed by Takai et al. in a \SI{180}{\nano\meter} CMOS technology, with additional custom layers. Adapted under terms of the CC-BY license \cite{Takai2016}. Copyright 2016, The Authors, published by MDPI. c) Device structure of the SPAD developed by Webster et al. in a \SI{130}{\nano\meter} CMOS imaging technology. Reproduced with permission \cite{Webster2012bis}. Copyright 2012, IEEE. d) PDE as a function of the wavelength reported for \cite{Webster2012,Takai2016,Webster2012bis}. e) Comparison on the PDE and DCR performance at $\lambda =$ \SI{800}{\nano\meter} for \cite{Webster2012,Takai2016,Webster2012bis}. f) Timing response in the blue region of the spectrum ($\lambda =$ \SI{470}{\nano\meter} \cite{Webster2012} and \SI{443}{\nano\meter} \cite{Webster2012bis}, respectively).}
    \label{fig:standard_noIso}
\end{figure}

\paragraph{SPADs with no substrate isolation}
The first successful attempt to increase the PDE at the longer wavelengths was a reverse approach based on an n/p geometry. As an example, Webster et al. \cite{Webster2012} exploited a \SI{90}{\nano\meter} CMOS imaging technology to demonstrate a PDE having a maximum value of 44\% at $\lambda =$ \SI{690}{\nano\meter} ($>$ 30\% at $\lambda =$ \SI{800}{\nano\meter}). As depicted in Figure \ref{fig:standard_noIso}a, this result was achieved by burying the depletion layer of the detector at the interface between a deep n well and the underneath p epitaxial layer, with no isolation from the p substrate, while a shallow p well was also introduced to prevent the dark carriers generated at the surface from reaching the multiplication region and, in turn, from contributing to the DCR. In addition, such a region collects also the slow carriers mostly generated from photons at the shorter wavelengths, providing in this spectral region a sharp timing response free from slow diffusion phenomena. The same approach was also adopted by Takai et al. \cite{Takai2016}, achieving a similar PDE in the NIR, yet with a fundamental difference: indeed, they exploited two custom layers (Figure \ref{fig:standard_noIso}b) in a less scaled CMOS technology (i.e. \SI{180}{\nano\meter}) in order to fully optimize the doping profile and, in turn, the electric field in the multiplication region. It is worth noting that this approach partially reduces the advantages of having a standard and thus reliable and low-cost process. Nevertheless, the importance of the optimization of the electric field is evident when looking at the DCR. On the other hand, the PDE of both solutions is strongly limited at the shorter wavelengths by the presence of the shallow p diffusion, which is responsible for losing a not-negligible part of the photogenerated carriers. A few wide spectral range solutions were proposed in the literature with no p diffusion \cite{Mandai2012,Webster2012bis}. In particular, it is worth mentioning the result obtained again by Webster et al. \cite{Webster2012bis} (Figure \ref{fig:standard_noIso}c) who demonstrated a PDE having a maximum value of 72\% at $\lambda =$ \SI{560}{\nano\meter} (Figure \ref{fig:standard_noIso}d), with values for the NIR region ($\simeq$ 30\% at $\lambda =$ \SI{800}{\nano\meter}) that are basically unchanged with respect to \cite{Webster2012} (Figure \ref{fig:standard_noIso}e). Unfortunately, the timing performance at \SI{443}{\nano\meter} is completely spoiled by the presence of slow diffusing carriers (Figure \ref{fig:standard_noIso}f) and the practical use of this SPAD in timing applications can be considered only for longer wavelengths: indeed, at \SI{654}{\nano\meter} the FWHM of the timing response is lower than \SI{60}{\pico\second} and the detrimental effect of the slow carriers is visible only if we look at the diffusion tail, which has a time constant in the order of \si{\nano\second} (see the discussion on substrate-isolated SPADs). It is worth observing that the slow diffusion tail at the longer wavelengths is a problem intrinsically related to the choice of having a non-isolated substrate, which also results in the possibility of optically-induced afterpulsing events and, in the case of SPAD arrays, delayed optical crosstalk \cite{Acerbi2019}.

\begin{figure}
    \includegraphics[width=\linewidth]{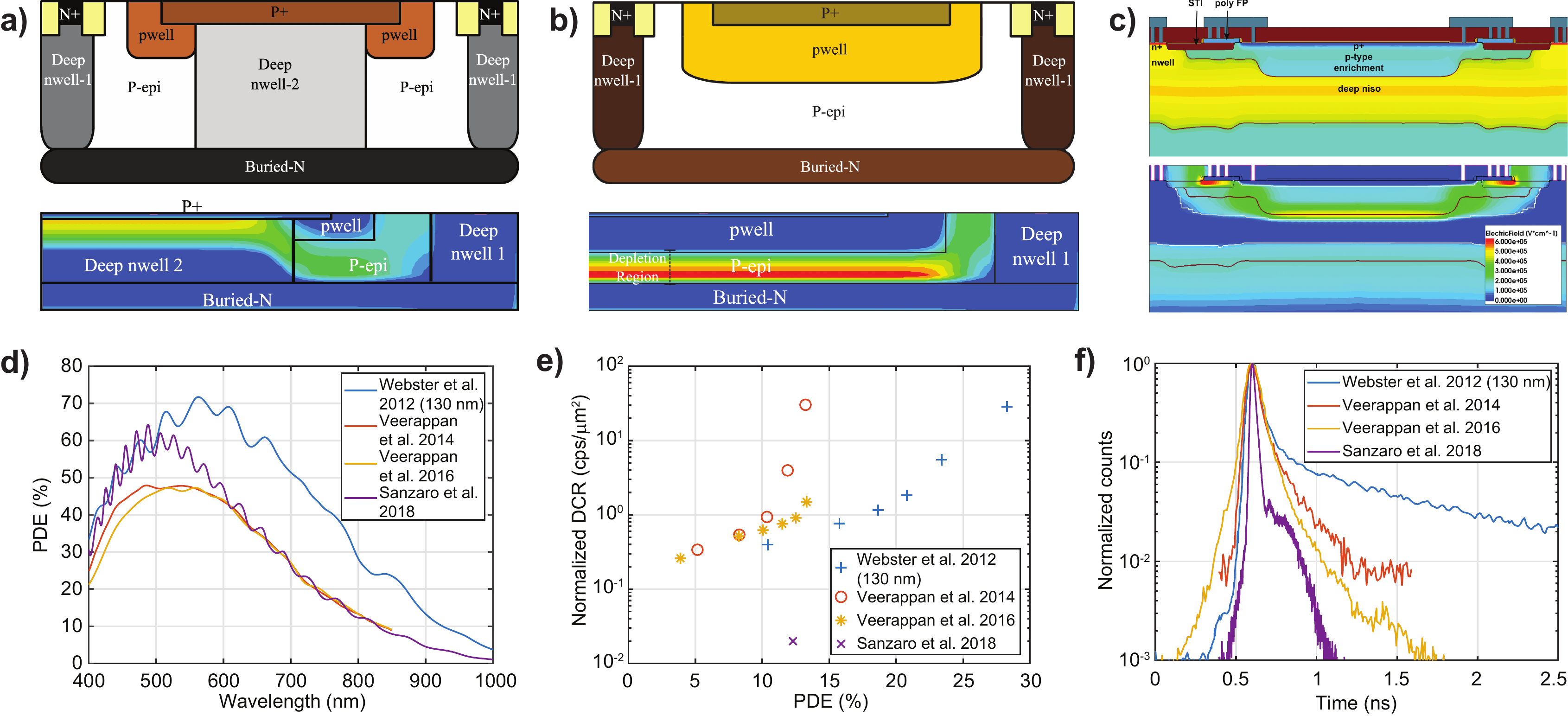}
    \caption{Device structure and performance of SPADs based on a standard fabrication process, with substrate isolation. a) Device structure and electric field profile of the SPAD developed by Veerappan et al. in a \SI{180}{\nano\meter} CMOS technology. Adapted with permission \cite{Veerappan2014}. Copyright 2014, IEEE. b) Device structure and electric field profile of the p-i-n SPAD developed by Veerappan et al. in a \SI{180}{\nano\meter} CMOS technology. Adapted with permission \cite{Veerappan2016}. Copyright 2015, IEEE. c) Device structure and electric field profile of the SPAD developed by Sanzaro et al. in a \SI{0.16}{\micro\meter} BCD technology. Reproduced with permission \cite{Sanzaro2018}. Copyright 2017, IEEE. d) PDE as a function of the wavelength reported for \cite{Veerappan2014,Veerappan2016,Sanzaro2018}. Webster et al. \cite{Webster2012bis} is reported for comparison with a SPAD having the substrate not isolated. e) Comparison on the PDE and DCR performance at $\lambda =$ \SI{800}{\nano\meter} for \cite{Veerappan2014,Veerappan2016,Sanzaro2018}. Webster et al. \cite{Webster2012bis} is reported for comparison with a SPAD having the substrate not isolated. f) Timing response in the red region of the spectrum ($\lambda =$ \SI{637}{\nano\meter} \cite{Veerappan2014,Veerappan2016} and \SI{820}{\nano\meter} \cite{Sanzaro2018}, respectively). Webster et al. ($\lambda =$ \SI{654}{\nano\meter}) \cite{Webster2012bis} is reported for comparison with a SPAD having the substrate not isolated.}
    \label{fig:standard_iso}
\end{figure}

\paragraph{Substrate-isolated SPADs}
Better results in terms of diffusion tail have been obtained by Veerappan et al. in 2014 \cite{Veerappan2014} by resorting to a \SI{180}{\nano\meter} CMOS technology, with a p+/n SPAD isolated from the p substrate by means of a buried n layer (Figure \ref{fig:standard_iso}a). The adoption of a deep n well to host the depletion region, along with the high excess bias enabled by the guard ring optimization, allowed the demonstration of a PDE of more than 40\% from \SIrange{460}{620}{\nano\meter} and about 13\% at \SI{800}{\nano\meter}, which is consistent with the fact that now the slow carriers generated in the substrate are not collected. Further improvement was then achieved by the same authors \cite{Veerappan2016} and with the same technology by resorting to a completely different structure, i.e. a p-i-n diode (Figure \ref{fig:standard_iso}b). Such SPAD relies on the avalanche multiplication that originates at the interface between a p- epitaxial layer and a n+ buried layer used also to isolate the SPAD from the rest of the p substrate. Thanks again to the high excess bias, they achieve the same PDE over the whole spectral range, but, on the other hand, the lower electric field developed over the p- epitaxial layer allowed they to demonstrate a low DCR ranging from \SIrange{60}{300}{cps} for SPADs having the same diameter of \SI{12}{\micro\meter}. Finally, it is worth noting that, despite the lower electric field, the timing jitter of the p-i-n SPAD is still remarkably low (about \SI{100}{\pico\second} FWHM for both \si{405} and \SI{637}{\nano\meter} light), while a low afterpulsing (i.e. $<$ 1\%) is yet to be demonstrated.

In most CMOS processes, the lack of steps specifically conceived to reduce the concentration of the defects in the SPAD active area, combined with high electric fields, has been limiting the linear dimension of SPADs’ active area to less than \SI{15}{\micro\meter}. Nevertheless, a large area is often required in quantum photonics, especially when only a single-pixel SPAD is needed or when the pixels of an array are employed as single independent detectors. For this reason, some research effort has been devoted also to the design of CMOS SPADs featuring a large area. In 2014 Villa et al. \cite{Villa2014} demonstrated devices with a diameter of the active area up to \SI{500}{\micro\meter} by exploiting a \SI{0.35}{\micro\meter} technology. Such remarkable geometry was possible thanks to the exceptional improvement of the dark counts with respect to the previous generation \cite{Guerrieri2010}, from \SI{4}{cps \per\micro\meter\squared} down to only \SI{0.05}{cps \per\micro\meter\squared}, which makes the DCR of these SPADs comparable to that demonstrated for detectors fabricated with custom technologies. Such a result is achieved thanks to the ultralow concentration of defects and contaminants and it is even more remarkable if we think that it is reported along with a low afterpulsing probability ($<$1\%), a fair timing jitter ($<$ \SI{90}{\pico\second} FWHM for SPADs having diameter up to \SI{50}{\micro\meter}) and a sensitivity that peaks at \SI{450}{\nano\meter} with a PDE of 53\%. Unfortunately, the PDE in the NIR is limited by the p+/n structure with substrate isolation (only 5\% at \SI{800}{\nano\meter}).

To fill this gap, Sanzaro et al. \cite{Sanzaro2018} in 2018 devised a novel SPAD design, based on the \SI{0.16}{\micro\meter} Bipolar-CMOS-DMOS (BCD) technology provided by STMicroelectronics (BCD8sP). The recent increased availability of high-voltage CMOS (HV-CMOS) technologies, pushed by automotive and control applications, provide ultralow concentrations of defects and deep low-doping diffusion regions that are more suitable for SPADs design. In particular, Sanzaro et al. reported the design of a circular SPAD with diameter up to \SI{80}{\micro\meter}. The structure (Figure \ref{fig:standard_iso}c) is fully enclosed in a deep n well and the doping profile is designed by acting on a retrograde custom n implant, used to provide a low resistivity path to the avalanche current, and on a subsequent high-energy p implant, harnessed to properly tailor the electric field in the multiplication region. The final device is a p/n+ junction, with the multiplication region underneath the drift region: such an electric field profile is adopted to fully exploit the high triggering efficiency of the electrons for detecting red/NIR photons \cite{Gulinatti2011}: the result is a PDE that reaches a maximum value of about 60\% at \SI{500}{\nano\meter} and still a fair 13\% at \SI{800}{\nano\meter} (Figure \ref{fig:standard_iso}d). Also in this case two custom implantation steps have been added to the standard process in order to optimize the electric field and to reduce both the DCR (Figure \ref{fig:standard_iso}e) and the series resistance. As a result, this SPAD features a remarkable timing response (less than \SI{30}{\pico\second} FWHM along with less than \SI{50}{\pico\second} diffusion tail time constant, as reported in Figure \ref{fig:standard_iso}f) provided that a dedicated front end circuit able to sense the avalanche at the initial stage of its growth is used. The role played by the substrate isolation in achieving such a sharp timing response is clear from the comparison with Webster et al. \cite{Webster2012bis} (see again Figure \ref{fig:standard_iso}f).

\subsection{SPADs for Infrared Detection}
Silicon offers multiple advantages for the development of SPAD detectors, guaranteeing the best performance demonstrated to date. Unfortunately, this material can not be exploited for single-photon detection at a wavelength over \SI{1000}{\nano\meter} due to its large electronic bandgap. At the same time, a non-negligible set of quantum photonics applications relies on single photons at these wavelengths: first and foremost, all the applications based on silicon photonics \cite{Silverstone2016}, such as large-scale quantum computing \cite{Qiang2018}, and, secondly, also long-range QKD based on single-mode optical fibers \cite{Korzh2015} or daylight free-space propagation \cite{Liao2017}, in which the use of infrared photons allows one to reduce the bright background originating from the sunlight. Experiments and demonstrations in these fields are most commonly carried out around \SI{1310}{\nano\meter} and \SI{1550}{\nano\meter}, often relying on weak coherent light \cite{Yuan2016} or heralded single-photon sources \cite{Kaneda2019}, even though also deterministic sources are currently under investigation \cite{Portalupi2019}, with notable results especially at \SI{1310}{\nano\meter} \cite{Kim2016}.

To overcome this limitation, a first approach based on sum-frequency generation (SFG) in periodically poled lithium niobate (PPLN) waveguides has been explored \cite{Tanzilli2005,Rakher2010}. In this case, infrared photons are converted to the visible range thus enabling the exploitation of silicon SPADs in such applications. This solution is already employed in low-noise free-space QKD \cite{Liao2017}. On the other hand, specific SPADs have been developed for direct infrared detection by exploiting different materials. While the performance of currently-available SPADs in this field are far from the ones demonstrated with superconducting detectors, remarkable steps have been made in the last few years. Thanks also to the possibility of operating these devices close to room temperature, infrared SPADs can be considered a viable alternative today. On the basis of current literature and future perspective, the materials that we consider the most interesting for quantum photonics are indium gallium arsenide (InGaAs) and germanium (Ge).

\subsubsection{InGaAs/InP}
The most common solution for detecting photons at a wavelength beyond \SI{1000}{\nano\meter} is based on the employment of InGaAs instead of silicon, in order to exploit the narrow electronic bandgap of this material ($E_g =$ \SI{0.75}{\electronvolt} at room temperature, corresponding to a cut-off wavelength of about \SI{1653}{\nano\meter}). The investigation of InGaAs-based single-photon detection started in the mid 1990s, when commercially available APDs, developed for communication and ranging applications, started being considered for Geiger-mode operation \cite{Lacaita1996,Hiskett2000}. Nevertheless, SPAD operation above the breakdown voltage inherently requires significantly different design criteria \cite{Acerbi2013,Ma2016}. For this reason, starting from the second half of the 2000s, InGaAs-based detectors optimized for operation in Geiger mode started to appear in the literature \cite{Pellegrini2006,Jiang2007}. Besides InGaAs, a wide-bandgap material is needed for the avalanche multiplication. This way, the high electric field in InGaAs can be avoided, thus limiting the BBT effect. While some alternatives like silicon \cite{Kang2004} or indium aluminium arsenide (InAlAs) \cite{Meng2016} have been considered, best results to date have been achieved with indium phosphide (InP) resulting into the widely diffused InGaAs/InP structure. It is worth noting that the design of these detectors is strictly connected to the concurrent development of dedicated quenching solutions, due to the high DCR and afterpulsing that can easily affect these devices. Initially, InGaAs/InP SPADs were only used under the control of a gate command, i.e. enabling the Geiger-mode operation only for a narrow time window \cite{Cova1996}. Unfortunately, this solution can be only employed for synchronous single-photon detection. To overcome this limitation, some solutions suitable for free-running operation have been recently reported in the literature. Finally, the integration of InGaAs/InP SPADs in large and densely-integrated detector arrays has been a hot topic in the last years.

Since the beginning of their development, InGaAs-based SPADs have found their most natural application in quantum communications \cite{Zhang2015} and, more specifically, in fiber-based QKD, whose progress in performance has been always closely linked to the progress made on single-photon detectors. In particular, SPADs operated in gated mode have played a very important role in the development of the QKD techniques since the late 1990s \cite{Ribordy1998, Bourennane1999} and, lately, milestones like the first demonstration of decoy-state QKD with one-way communication \cite{Peng2007} has been achieved thanks to these detectors. In addition, fast gating operation of these SPADs has given a dramatic contribution to increase the secure key rate of the communication \cite{Walenta2012,Frohlich2017} and the development of low-noise free-running devices \cite{Korzh2014} has culminated with the first distribution of a quantum cryptographic key over an optical fiber \SI{307}{\kilo\meter} long \cite{Korzh2015}. While SNSPDs are currently employed for longer distances \cite{Boaron2018}, InGaAs/InP SPADs are still considered a more viable solution in terms of both cost and complexity for practical exploitation in a quantum network. Besides QKD, such SPADs have given important contributions also to the study of other quantum communication protocols \cite{Liu2012,Williams2019}, they have been employed for the experimental characterization of single-photon sources \cite{Davanco2012} and even for the implementation of QRNGs \cite{Dynes2008}. In the end, it is worth mentioning that these devices can be harnessed also in quantum applications requiring PNR capabilities not only thanks to the possibility of spatially multiplexing the photons on a detector array \cite{Jiang2015}, but also by means of fast gating techniques able to exploit the dependence of the avalanche build-up on the number of carriers initiating the avalanche itself \cite{Kardyna2008}.

\begin{figure}
    \includegraphics[width=\linewidth]{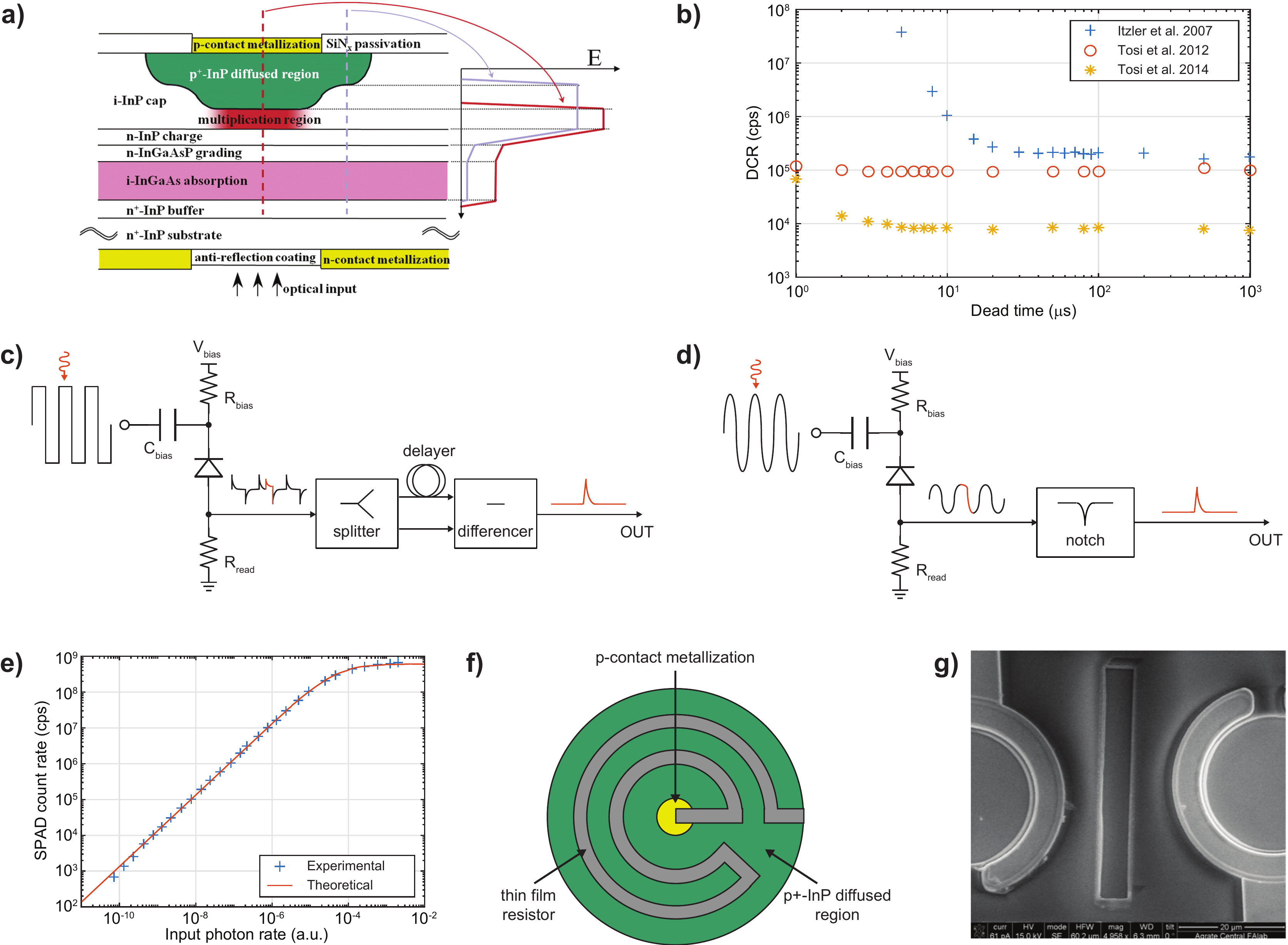}
    \caption{Device structure, front end electronics and performance of InGaAs/InP SPADs. a) Device structure of the back-illuminated InGaAs/InP SPAD proposed by Itzler et al. \cite{Itzler2007}. Reproduced with permission \cite{Itzler2011}. Copyright 2011, Taylor \& Francis. b) DCR as a function of the dead time (i.e. the time interval between two gate pulses) for the three best solutions selected among InGaAs/InP gated at low frequency (gate time $T_{on} =$ \SI{20}{\nano\second}, repetition rate $f_g <$ \SI{1}{\mega\hertz}). Temperature is $T =$ \SI{220}{\kelvin} for Itzler et al. \cite{Itzler2007} and $T =$ \SI{225}{\kelvin} for the others \cite{Tosi2012,Tosi2014}. c) Schematic illustration of the front end electronics necessary to implement the SD read out mechanism. d) Schematic illustration of the front end electronics necessary to operate the SPAD with a sinusoidal gate. e) Sinusoidally-gated SPAD count rate as a function of the impinging photon flux as reported by Scarcella et al.\cite{Scarcella2015}. The count rate saturates at \SI{650}{\mega cps}. f) Schematic illustration (top view) of the InGaAs/InP NFAD developed by Itzler et al. \cite{Itzler2009}. g) Scanning electron microscopy image of an isolation trench between two adjacent InGaAs/InP SPAD pixels as reported by Calandri et al. The image has been taken before the metallization. Adapted with permission \cite{Calandri2016}. Copyright 2016, IEEE.}
    \label{fig:ingaas}
\end{figure}

\paragraph{Gated SPADs}
The most common structure of InGaAs/InP SPAD reported to date is based on a separate absorption, charge and multiplication (SACM) structure, as reported by Itzler et al. \cite{Itzler2007} (see Figure \ref{fig:ingaas}a). An \ch{In_{0.53}Ga_{0.47}As} absorption layer is lattice-matched to an InP multiplication layer to limit the generation of carriers through band-to-band tunneling. The two layers are separated by a thin grading layer of InGaAsP, with the aim of reducing hole trapping effects due to the valence band offset between the two materials, and by a charge n layer, used to provide an additional degree of freedom for a flexible tailoring of the electric field profile and to avoid the penetration of the multiplication field in the InGaAs layer. The fabrication of the InGaAs/InP stack is carried on by metalorganic chemical vapor deposition (MOCVD). The main pn junction is fabricated by diffusing zinc in the intrinsic InP cap region, while premature breakdown is avoided by subsequent zinc diffusions aimed at suitably tailoring the shape of the p-doped well. 

The back-illuminated InGaAs/InP SPAD proposed by Itzler et al. \cite{Itzler2007}, along with other front-illuminated works \cite{Tosi2009,Tosi2012,Tosi2014}, achieved an unprecedented level of performance at \SI{1550}{\nano\meter}, with a PDE higher than 30\%, a timing jitter lower than \SI{70}{\pico\second} FWHM and a DCR in the order of some \si{\kilo cps} for a circular SPAD having a diameter of \SI{25}{\micro\meter}. The noise performance of these devices are indeed reported in Figure \ref{fig:ingaas}b. Compared to silicon SPADs, these detectors are operated at a lower temperature (i.e. \SI{225}{\kelvin} or even less), with a gate time window usually in the order of tens of \si{\nano\second} and at a repetition rate usually in the order of tens of \si{\kilo\hertz} by using a dedicated quenching electronics developed on purpose. While lowering the temperature is needed to reduce the DCR, the low-frequency gate is exploited to allow full carrier detrapping, thus mitigating the high afterpulsing originating within the high-field InP multiplication region \cite{Tosi2009} and responsible for the increasing DCR at repetition rates approaching \SI{1}{\mega\hertz} (i.e. corresponding to a dead time of \SI{1}{\micro\second}). However, the trade-off between afterpulsing probability and maximum gate frequency of these SPADs is particularly limiting for the applications operating at high photon count rate. Given the many challenges in reducing the density of traps in the materials, in order to relax this trade-off the research of the last years has focused its attention on limiting the charge flowing during each avalanche, thus reducing the potential number of carriers that can be trapped \cite{Itzler2011,Zhang2015}. The most natural way to pursue this goal is by reducing the gate window down to, or even below, the time needed to trigger the avalanche, usually in the order of hundreds of \si{\pico\second}. However, by operating the SPAD in this regime (usually referred to as fast gating) forces one to use exquisite electronics solutions able to detect a faint avalanche among the large spurious pulses originating from the fast transients that couple to the output line through the parasitic capacitance of the detector itself. A first solution is called self-differencing (SD) \cite{Yuan2007} (Figure \ref{fig:ingaas}c) and consists of subtracting from the SPAD output response the same signal delayed by exactly one gate period. In this way, the capacitive transient is subtracted, while the avalanche waveform can be correctly discriminated. Such a technique has been employed to demonstrate fast gating at a repetition rate of \SI{2}{\giga\hertz} \cite{Yuan2010}, yet with an afterpulsing probability that is lower than 5\%. A second solution is instead based on using a sinusoidal gate command \cite{Namekata2006} (Figure \ref{fig:ingaas}d). Since the avalanche waveform is usually characterized by a broader frequency spectrum, the resulting sinusoidal capacitive transient can be easily removed with the aid of a notch filter. Repetition rates in the \si{\giga\hertz} range have been demonstrated also for this technique, even though with afterpulsing probabilities that are slightly higher \cite{Namekata2010,Zhang2010}. A similar performance has been achieved also with techniques relying on a mixed approach \cite{Zhang2009} or by coherent addition of discrete higher-order harmonics to the fundamental sinusoidal gate \cite{Restelli2013}. However, it is worth remarking that all these solutions enable the exploitation of the SPAD in \si{\giga\hertz}-clocked applications, yet this does not mean that a \si{\giga cps} count rate can be achieved. Speaking of maximum count rate, remarkable results have been reported by Comandar et al. \cite{Comandar2015}, that exploited a SD front end circuit together with a rectangular gate command (repetition rate of \SI{1}{\giga\hertz}) to demonstrate a count rate as high as \SI{500}{\mega cps} (the maximum count rate achievable with such a SD system), along with an afterpulsing probability of 7\% and a PDE as high as 50\% at \SI{1550}{\nano\meter}. Nevertheless, best in class results to date have been demonstrated by Scarcella et al. \cite{Scarcella2015} by employing a sinusoidal gate at \SI{1.3}{\giga\hertz} frequency, with a resulting maximum count rate as high as \SI{650}{\mega cps}, an afterpulsing probability as low as 1.5\% and a PDE higher than 30\% at \SI{1550}{\nano\meter}. The linearity curve of this device is reported in Figure \ref{fig:ingaas}e.

\paragraph{Free-running SPADs}
In order to reduce the speed limitations of gated SPADs, researchers have recently achieved many interesting results in the design of InGaAs/InP SPADs that can be operated in free-running mode, by employing passive \cite{Warburton2009}, active \cite{Thew2007} and even mixed \cite{Liu2008} quench/reset architectures. Commercial modules are now available in both gated and free-running mode \cite{IDQube2020,PDM-IR2019}. In this case, the objective is mainly pursued by integrating a quenching resistor within the SPAD die in order to minimize the capacitive load \cite{Itzler2009} (Figure \ref{fig:ingaas}f). This approach is beneficial for the count rate since it allows both the reduction of the avalanche charge flowing during the detection and a faster reset of the quiescent condition after the avalanche quenching. These SPADs are usually referred to as negative feedback avalanche diodes (NFADs) to highlight the role played by the quenching resistor in contrasting the positive feedback at the origin of the avalanche current. Itzler et al. \cite{Itzler2009} firstly proposed this solution by exploiting a thin film meandering NiCr resistor (see again Figure \ref{fig:ingaas}f) and demonstrating single-photon operation with circular NFADs having diameter ranging from \SIrange{22}{82}{\micro\meter} \cite{Itzler2010}. However, such detectors are subject to the typical constraints of passively-quenched SPADs, so that the value of the integrated resistor has to be high enough to allow proper quenching at the overvoltage necessary to achieve high PDE and, at the same time, low enough not to limit the reset time and thus the maximum achievable count rate. In addition, purely passive quenching does not provide a given hold-off time in which trapped carriers can be released without triggering an avalanche. For all these reasons, the best performance with NFADs have been reached by operating them with an external AQC \cite{Lunghi2012}, namely by using an hybrid quenching approach that, for a maximum count rate of \SI{100}{\kilo cps}, limits the afterpulsing probability at no more than 20\%. The same research group has been also the first to demonstrate that the negative feedback does not prevent the extraction of the timing information with a jitter lower than \SI{100}{\pico\second}. Indeed, Amri et al. \cite{Amri2016} reported on a NFADs coming from the same family, operating at \SI{223}{\kelvin} and able to detect \SI{1550}{\nano\meter} photons with a PDE of 30\% and a timing jitter between \si{52} and \SI{67}{\pico\second} FWHM. In this configuration the DCR is about \SI{7}{\kilo cps}, yet it can be decreased down to almost \SI{10}{cps} by operating the NFAD at \SI{163}{\kelvin} and by tolerating a slightly lower PDE of 27.7\%, a timing jitter of \SI{129}{\pico\second} FWHM and an afterpulsing probability that is kept at 20\% for a maximum count rate of \SI{50}{\kilo cps} \cite{Korzh2014}. Finally, it is worth mentioning that NFADs have been also realized by fabricating the integrated resistor with the same zinc diffusion used to tailor the shape of the p-doped well \cite{Sanzaro2016} or by suitably engineering the InGaAs/InP heterostructure \cite{Cheng2011}.

\paragraph{SPAD arrays}
Pushed by the research activity on eye-safe LiDAR optical receivers and space laser communications, a significant effort has been devoted to the integration of InGaAs/InP SPADs in large and densely-integrated detector arrays in the past few years. To date, 3D-integrated InGaAs/InP SPAD arrays have been demonstrated up to a format of $128\times32$ \cite{Jiang2015}, with a separation pitch of \SI{100}{\micro\meter}. However, both the integration density and the number of SPADs fabricated in the same die are today limited by the large optical crosstalk probability, that can even exceed 80\% \cite{Calandri2016}. Metal-filled trenches (Figure \ref{fig:ingaas}g) have been proposed as an optical isolation solution, with a resulting crosstalk probability of 37\% when the pixels are separated by only \SI{60}{\micro\meter} \cite{Calandri2016}. Finally, the development of NFADs arrays \cite{Jiang2015} is the second approach investigated for the development of large InGaAs/InP SPAD arrays, with the perspective of a simplification of the read out circuitry, yet, as already explained, at the cost of a reduced PDE. Such devices have been demonstrated to date up to a format of $8\times8$.

\subsubsection{Germanium}
A second material that is investigated today for the detection of single photons in the infrared range is germanium ($E_g =$ \SI{0.66}{\electronvolt} at room temperature, corresponding to a cut-off wavelength of about \SI{1879}{\nano\meter}). Even though preliminary studies on germanium SPADs date back to 1971 \cite{Haecker1971}, the practical application of these detectors starts only in the 1990s \cite{Lacaita1994}, when a commercial APD designed for optical communications is operated in gated Geiger mode in order to perform time-resolved photoluminescence measurements \cite{Buller1996}. Gated mode is indeed the most common way also to operate germanium-based SPADs.

\begin{figure}
    \includegraphics[width=\linewidth]{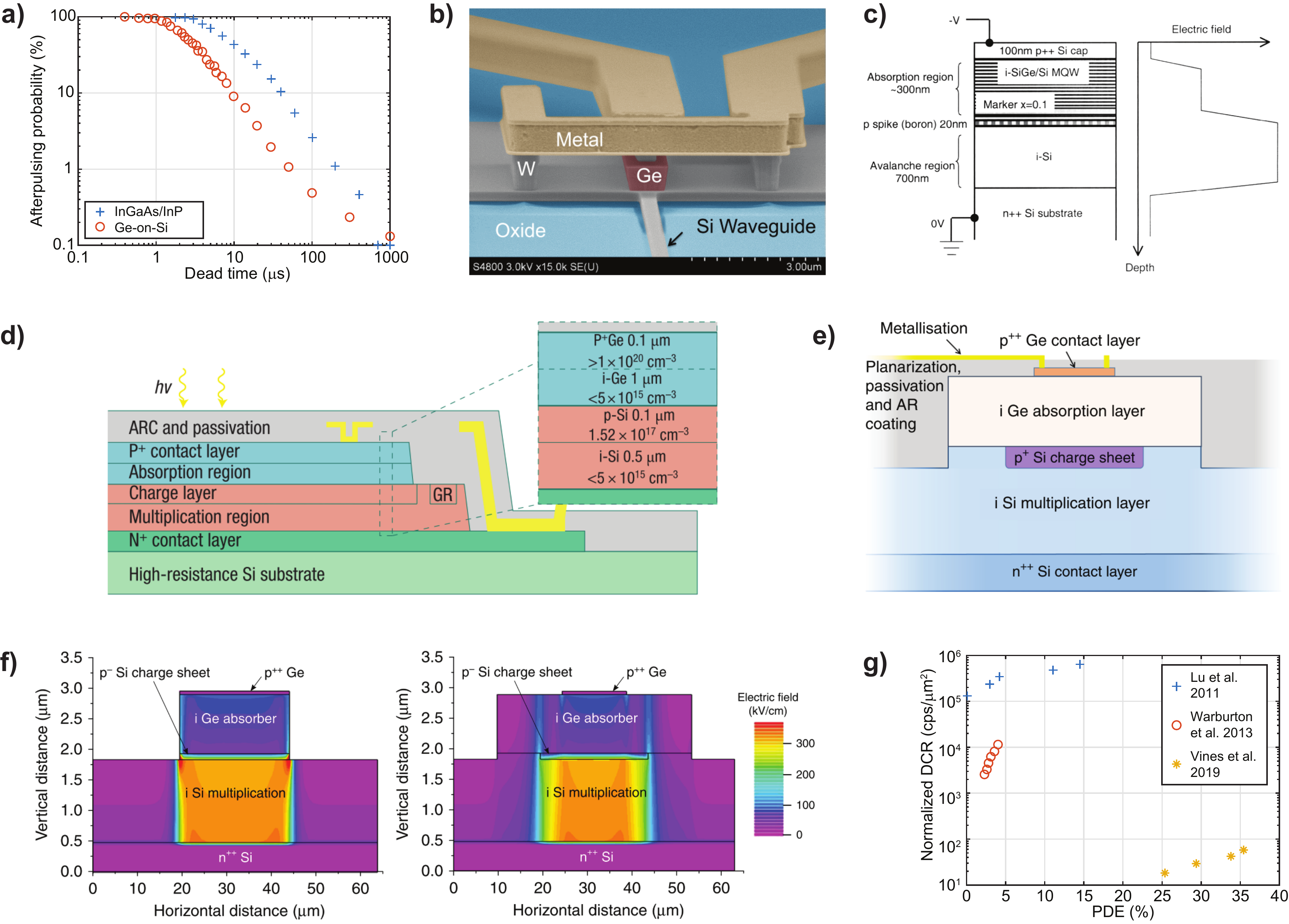}
    \caption{Device structure and performance of germanium SPADs. a) Afterpulsing probability as a function of the dead time (i.e. the time interval between two gate pulses) for a commercially available InGaAs/InP SPAD and the planar Ge-on-Si SPAD reported by Vines et al. \cite{Vines2019}. b) Structure of the mesa Ge-on-Si APD operated in Geiger mode by Martinez et al. A silicon waveguide is butt coupled to the detector. Reproduced with permission \cite{Martinez2017}. Copyright 2017, The Optical Society. c) Structure and electric field profile of the SPAD based on a mesa Si/\ch{Si_{0.7}Ge_{0.3}} MQW reported by Loudon et al. Reproduced with permission \cite{Loudon2002}. Copyright 2002, The Optical Society. d) Structure of the mesa Ge-on-Si APD operated in Geiger mode by Lu et al. \cite{Lu2011}. Reproduced with permission \cite{Kang2009}. Copyright 2008, Springer Nature Limited. e) Structure of the planar Ge-on-Si SPAD proposed by Vines et al. Reproduced under terms of the CC-BY license \cite{Vines2019}. Copyright 2019, The Authors, published by Springer Nature Limited. f) Comparison of the electric field profile between a mesa (left) and a planar (right) Ge-on-Si SPAD as reported by Vines et al. Adapted under terms of the CC-BY license \cite{Vines2019}. Copyright 2019, The Authors, published by Springer Nature Limited. g) Comparison among the performance (DCR and PDE at \SI{1310}{\nano\meter}) of the three best Ge-on-Si SPADs reported in literature \cite{Lu2011,Warburton2013,Vines2019}. Temperature is $T$ = \SI{200}{\kelvin} for \cite{Lu2011}, while the others are reported for $T$ = \SI{100}{\kelvin}.}
    \label{fig:ge-on-si}
\end{figure}

From the following discussion it will be clear that, despite the important progress made in the last decade, germanium SPADs still look a step behind InGaAs/InP ones. This is particularly evident if we consider that germanium SPADs achieve DCRs comparable to those of their InGaAs/InP counterparts only if cooled at much lower temperatures (e.g. \SI{100}{\kelvin}), not attainable with typical thermoelectric coolers. Moreover, at these temperatures, their PDE is lower than in InGaAs/InP SPADs, especially at \SI{1550}{\nano\meter}. Nevertheless, today the research on this material for SPADs can open new perspectives in many applications, especially if we look at quantum photonics. Firstly, high-speed QKD requires infrared detectors able to operate at high count rates \cite{Rogers2007} and this quantity is usually limited in InGaAs/InP SPADs by the high afterpulsing probability as recalled in the previous section. On the contrary, some solutions have already been proposed to break the trade-off between maximum count rate and afterpulsing with germanium-based SPADs. Vines et al. \cite{Vines2019} have indeed compared their germanium-on-silicon (Ge-on-Si) SPAD structure with commercially available InGaAs/InP ones operating in identical conditions of temperature and PDE. As showed in Figure \ref{fig:ge-on-si}a, the Ge-on-Si SPAD compare favorably with the InGaAs/InP detector, despite the smaller active area of the latter (diameter: \SI{25}{\micro\meter} vs \SI{100}{\micro\meter}). This result suggests that, under the same conditions, germanium is a better material in terms of afterpulsing compared to InP. So, if the Ge-on-Si SPADs will ever be operated at the same temperature of InGaAs SPADs, they will be able to provide an advantage in terms of afterpulsing and maximum count rate. Secondly, Ge-on-Si detectors open the way to the integration of SPADs in a standard silicon photonics technology, that is a hot topic in this field as will be discussed in Section \ref{sec:waveguide_SPADs}. While the performance of the first Ge-on-Si SPAD butt coupled to a lateral silicon waveguide \cite{Martinez2017} (Figure \ref{fig:ge-on-si}b) is still far from best-in-class results, with an overall PDE that is only 5.27\% at \SI{1310}{\nano\meter}, such work represents the first experimental demonstration of a SPAD integrated with an optical waveguide.

\paragraph{Mesa SPADs}
Since the very first studies, researchers have abandoned the idea of germanium homojunctions to fabricate SPADs since the narrow bandgap of this material would easily lead to a high BBT noise.. To overcome this problem, a SACM structure based on silicon for the avalanche multiplication and on germanium for the absorption of photons has been preferred, similarly to what happens for InGaAs/InP SPADs. Such an approach is also interesting for the potential integration with CMOS circuitry \cite{Sammak2016}. In particular, Loudon et al. \cite{Loudon2002} were the first to report on a SPAD structure based on these principles, with a circular mesa detector featuring a SiGe absorption layer (Figure \ref{fig:ge-on-si}c). The DCR showed a promising improvement, with \SI{100}{\kilo cps} at a temperature as high as \SI{200}{\kelvin} and for a large diameter of \SI{120}{\micro\meter}, yet with a very limited PDE (i.e. lower than 0.01\% at \SI{1210}{\nano\meter}, with no data for the most relevant wavelengths \SI{1310}{\nano\meter} and \SI{1550}{\nano\meter}). The main reason of such a low value is that the thickness of the absorption layer in Si/Ge heterostructures is limited by the large lattice mismatch between these two materials, that results in a trade-off between concentration of defects at the interface and absorption probability. Loudon et al. exploited a Si/\ch{Si_{0.7}Ge_{0.3}} multiple quantum well (MQW) structure in order to attain a larger thickness, yet with a limited concentration of defects. Nevertheless, the low fraction of germanium, along with a thickness still restricted to only \SI{300}{\nano\meter}, resulted in a poor performance in terms of PDE.

The most notable breakthrough for germanium-based SPADs has been the possibility of growing a pure crystalline layer of germanium directly on silicon with thickness larger than \SI{1}{\micro\meter} and concentration of defects down to \SI{5E6}{\centi\meter^{-2}} \cite{Kang2009,Tan2012}. This approach is usually referred to as Ge-on-Si. In particular, Lu et al. \cite{Lu2011} were the first to demonstrate single-photon operation with a circular Ge-on-Si APD \cite{Kang2009} (\SI{30}{\micro\meter} diameter) designed for linear mode operation, but operated in gated Geiger mode. The detector was based on a mesa geometry grown by performing subsequent chemical vapor depositions (CVDs) of Si/Ge layers in order to obtain the SACM structure reported in Figure \ref{fig:ge-on-si}d. However, the noise performance of the detector was quite poor, with a DCR higher than \SI{100}{\mega cps} at \SI{200}{\kelvin} that, most probably, was even impairing a correct operation of the device. Lately, Warburton et al. \cite{Warburton2013} exploited a similar approach, but designing the mesa stack (both doping and thickness of each layer) from scratch in order to optimize the operation of the detector in Geiger mode at a lower temperature. After the fabrication, they reported on a selected SPAD (\SI{25}{\micro\meter} diameter) operated at \SI{100}{\kelvin} and characterized by a DCR in the \si{\mega cps} range, along with a PDE of 4\% and a timing jitter of \SI{300}{\pico\second} FWHM for detection of photons at \SI{1310}{\nano\meter}. The worse PDE and timing performance with respect to both germanium homojunction and InGaAs/InP SPADs \cite{Tosi2007} was ascribed to the low excess bias at which the detector is operated, that can not be raised to keep a relatively low DCR. Single-photon counting at \SI{1550}{\nano\meter} was demonstrated for the first time with a mesa structure, again with poor results, since the PDE was only 0.15\%. This time, the bad performance has to be ascribed not only to the low excess bias, but also to the low operating temperature of the SPAD (i.e. \SI{125}{\kelvin}): indeed, for this temperature the bandgap of germanium is \SI{0.84}{\electronvolt} wide, comparable to the corresponding energy of a photon at \SI{1550}{\nano\meter} (i.e. \SI{0.80}{\electronvolt}). Nevertheless, Warburton et al. could not raise the temperature without increasing the DCR beyond a reasonable level.

\paragraph{Planar SPADs}
The work reported on mesa SPADs highlights how this geometry is not suitable for obtaining low noise performance. This is related to the fact that with this structure the sidewalls of the SPAD are depleted and the corresponding electric field enhances the generation of dark carriers, collecting and accelerating them toward the multiplication region. In order to solve this problem, high quality passivation of the sidewalls is required or, more effectively, the SPAD can be designed by concentrating the depleted region far away from the lateral surfaces, similarly to planar silicon SPADs. The latter is the approach followed by Vines et al. \cite{Vines2019}, that have recently reported on what is now the state of the art for Ge-on-Si SPADs. In particular, they exploited a selective boron implantation (Figure \ref{fig:ge-on-si}e) in order to focus the electric field far from the sidewalls: a comparison in terms of electric field profile between a mesa and a planar SPAD has been carried out by the same authors and is here reported in Figure \ref{fig:ge-on-si}f. The improvement in terms of DCR is impressive: indeed, a circular SPAD having a diameter as large as \SI{100}{\micro\meter} is operated at \SI{100}{\kelvin} achieving a DCR that is less than \SI{500}{\kilo cps}, along with a PDE of 35\% at \SI{1310}{\nano\meter}, that compares favorably with both mesa SPADs reported in \cite{Lu2011} and \cite{Warburton2013} (Figure \ref{fig:ge-on-si}g). In addition, the improvement in terms of DCR allowed Vines et al. to operate the SPADs at a temperature up to \SI{175}{\kelvin}, showing encouraging results for detection at \SI{1550}{\nano\meter}. On the other hand, the timing jitter is still relatively high, with \SI{310}{\pico\second} FWHM, but it is worth also noting that this value is certainly due to the large photoactive area, since the jitter drops to \SI{175}{\pico\second} for SPADs having a diameter of \SI{26}{\micro\meter}. Finally, the work reported by Vines et al. represents also a big step toward the demonstration of Ge-on-Si SPAD arrays thanks to the relatively high yield (i.e. around 90\%) achieved by the fabrication process that they use.

\section{Waveguide SPADs: applications and outlook}
\label{sec:waveguide_SPADs}
As already mentioned, the integration of single-photon detectors within the quantum PICs is subject to a significant interest from the scientific community today and the achievement of such a milestone would be of paramount importance for multiple reasons. First of all, to fully exploit all the advantages of the integration, like the unrivaled miniaturization and scalability \cite{Wang2020}. Secondly, to avoid the coupling with external optical fibers and, thus, to guarantee minimal photon losses, especially in the case of optical circuits based on high confinement platforms like silicon photonics \cite{Silverstone2016}. Furthermore, the on-chip single-photon detection would also allow a prompt sensing of a given photonic quantum state, a feature that is highly desirable for the implementation of feed-forward operations \cite{Prevedel2007} in measurement-based linear optical quantum computing \cite{Walther2005}.

A few important steps toward this goal have already been accomplished with superconducting detectors \cite{Pernice2012,Gerrits2011}, paving the way for the realization of fully integrated quantum optical circuits, with on-chip generation, manipulation and detection of quantum states of light. However, the peculiarities of the SPADs, discussed in the previous sections, make them especially appealing for such an integration. Indeed, SPADs can be operated at much higher temperatures, thus dispensing for the cryogenic cooling of the entire PIC and guaranteeing full compatibility with reconfigurable optical circuits featuring a large number of thermo-optic phase shifters. To reach this target, we envision the development of waveguide SPADs, i.e. detectors guiding the photon through its depletion region until the particle is absorbed and the avalanche is triggered. In this device, light propagation and electric field are typically orthogonal, breaking the trade-off between detection efficiency and transit time \cite{Bowers1986}. This concept has been already exploited in the past for the realization of waveguide photodiodes operating with classical light \cite{Bowers1986,Koester2006,Zhu2009} and, even though Martinez et al. do not mention it, the work reported in \cite{Martinez2017} can be considered as a first, yet rudimentary, example of waveguide SPAD monolithically integrated in an elementary waveguide circuit.

\begin{figure}
    \includegraphics[width=\linewidth]{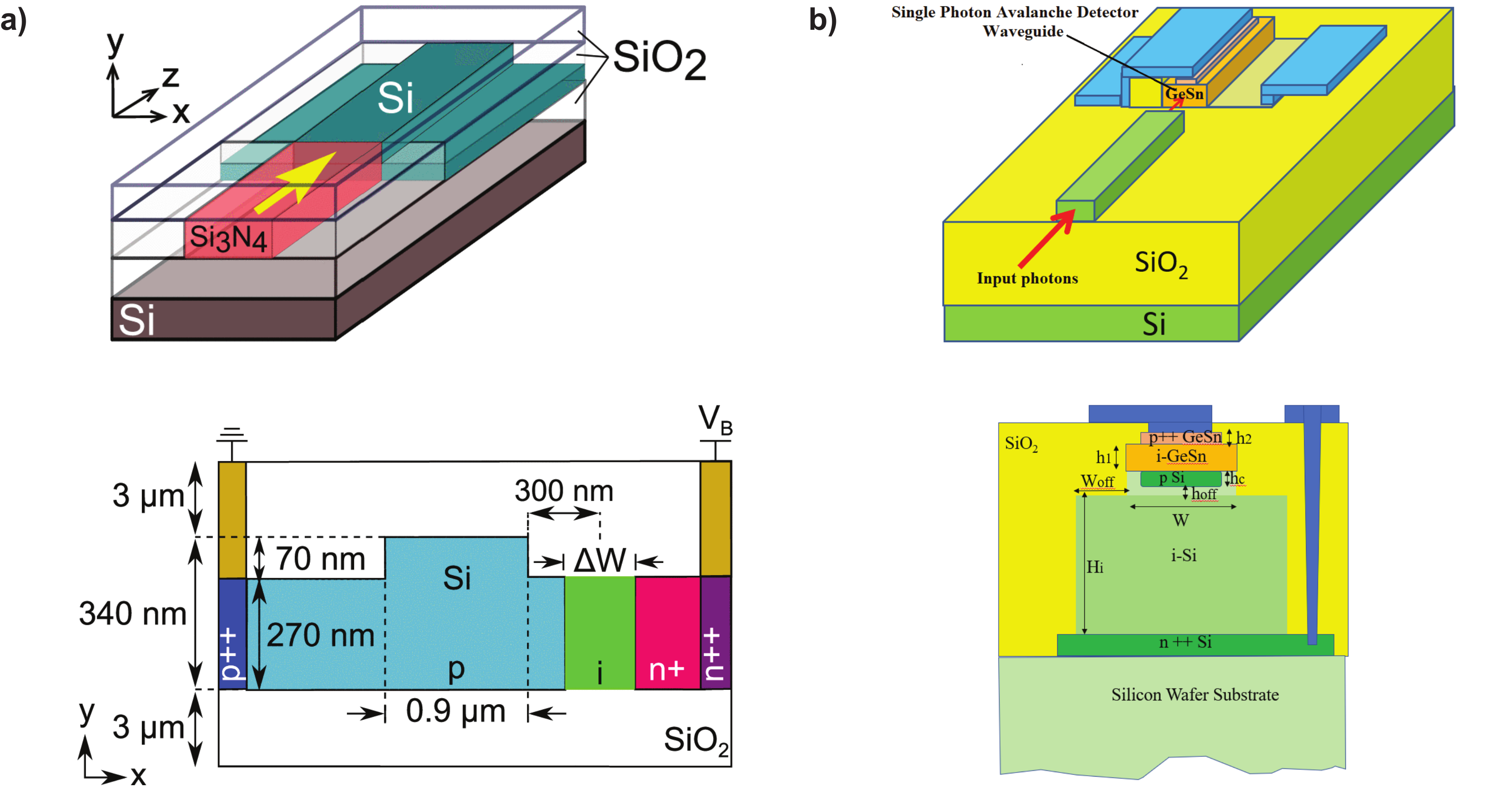}
    \caption{Theoretical works on waveguide SPADs recently proposed in the literature. a) Silicon waveguide SPAD butt-coupled to a silicon nitride waveguide for operation in the visible/NIR range: 3D structure (top) and cross-section (bottom). Adapted under terms of the CC-BY license \cite{Yanikgonul2020}. Copyright 2019, The Authors, published by IEEE. b) GeSn waveguide SPAD butt coupled to a silicon waveguide for operation in the infrared range: 3D structure (top) and cross-section (bottom). Adapted with permission \cite{Soref2019}. Copyright 2019, American Chemical Society.}
    \label{fig:waveguide}
\end{figure}

Given a certain operating wavelength, a PIC with integrated waveguide SPADs requires three main materials: a transparent material for the fabrication of the core region of the circuit, a material able to efficiently absorb the same photons within the SPAD core and a third transparent material with low refractive index that can act as cladding for both circuit and SPAD. For the visible/NIR range (e.g. $ \lambda =$ \SI{800}{\nano\meter}), silicon nitride \cite{Munoz2017} represents a solution enabling the fabrication of optical circuits with low propagation losses ($<$ \SI{0.1}{\deci\bel\per\centi\meter}), small bending radii ($<$ \SI{100}{\micro\meter}), compatible with the CMOS integration and in which it is possible to produce single photons from integrated sources \cite{Lu2019}. After the discussion carried out in Section \ref{sec:fabrication}, silicon is instead the most natural choice for the SPAD. Lastly, the cladding can be realized with silica. Even though all these materials are technologically compatible and can be integrated in a standard SOI platform \cite{Sacher2015}, different questions are still waiting for an answer. The most important one is how to efficiently couple light from a transparent silicon nitride waveguide, that must be designed for single-mode operation, to an absorbing silicon one, whose dimension is subject to the requirements of the SPAD operation. The simplest approach consists in butt coupling the two waveguides, as proposed by the theoretical works of Yanikgonul et al. \cite{Yanikgonul2018,Yanikgonul2020} (Figure \ref{fig:waveguide}a). Another approach relies on the evanescent coupling, that can be realized by placing the silicon nitride waveguide on top of the silicon one or, alternatively, by placing them side by side. All these solutions require a non-trivial design and the strong difference of refractive index between silicon nitride and silicon will play an important role in both the choice of the coupling approach and the sizing process of the geometrical parameters.

The situation for the infrared range (e.g. $ \lambda =$ \SI{1550}{\nano\meter}) is instead quite different. Silicon \cite{Silverstone2016}, with silica as cladding, is the material of choice for the fabrication of waveguide optical circuits in this wavelength range thanks to the superior miniaturization of the integrated photonic components, the feasibility of scaling to mass production and, also in this case, the possibility of integrating single-photon sources \cite{Qiang2018}. However, the choice of the material for the SPAD is not as trivial as for the circuit. Germanium is technologically compatible with both silicon and silica and today Ge-on-SOI is a well-established platform for the fabrication of PICs operating up to the mid infrared range \cite{Koester2006,Malik2014,Younis2016}. Nevertheless, as thoroughly discussed in Section \ref{sec:fabrication}, a lot of work is still necessary to improve the performance of germanium SPADs, especially at \SI{1550}{\nano\meter}. To this aim, Soref et al. \cite{Soref2019} presents a theoretical work on a waveguide SPAD whose structure is conceptually similar to the others (Figure \ref{fig:waveguide}b), but with an absorption layer fabricated in a germanium tin alloy (GeSn) in order to improve the sensitivity at this wavelength. On the other hand, a completely orthogonal approach is based on the hybrid integration \cite{Elshaari2020,Kim2020} instead of the monolithic one, in order to have an additional degree of freedom to improve the performance of germanium-based SPADs \cite{Ke2017} (e.g. by using a different substrate) or to take into consideration other materials like InGaAs/InP \cite{Roelkens2005}. In both cases, it is worth saying that waveguide-detector coupling of infrared photons could be easier than visible/NIR photons thanks to the similar refractive index of silicon (used as transparent material in this case) and germanium or InGaAs.

\section{Conclusion}

\label{sec:conclusion}
\begin{table}
\resizebox{\textwidth}{!}{%
\begin{tabular}{|c|c|l|c|c|c|c|c|}
\hline
\textbf{Technology} & \begin{tabular}[c]{@{}c@{}}\textbf{Diameter}\\ \textbf{(\si{\micro\meter})}\end{tabular} & \multicolumn{1}{c|}{\begin{tabular}[c]{@{}c@{}}\textbf{Measurement} \\ \textbf{conditions}\end{tabular}} & \begin{tabular}[c]{@{}c@{}}\textbf{PDE} \\ \textbf{(NIR or IR)}\end{tabular} & \begin{tabular}[c]{@{}c@{}}\textbf{DCR\textsuperscript{1}}\\ \textbf{(\si{cps})}\end{tabular} & \begin{tabular}[c]{@{}c@{}}\textbf{Max count rate\textsuperscript{2}}\\ \textbf{(\si{\mega cps})}\end{tabular} & \begin{tabular}[c]{@{}c@{}}\textbf{Afterpulsing}\\ \textbf{probability}\end{tabular} & \begin{tabular}[c]{@{}c@{}}\textbf{Timing jitter}\\ \textbf{(\si{\pico\second} FWHM)}\end{tabular} \\ \hline\hline
\begin{tabular}[c]{@{}c@{}}custom Si\\ reach-through\\ \cite{SPCM-AQRH2020,Rech2006}\end{tabular} & 180 & \begin{tabular}[c]{@{}l@{}}$V_{ov}$ n.r.\\ thermoelectrically cooled\\ free running\end{tabular} & \begin{tabular}[c]{@{}c@{}}@\SI{800}{\nano\meter}\\ 62\%  \end{tabular} & 25 & 37 & 0.5\% & \begin{tabular}[c]{@{}c@{}}180\textsuperscript{3} \end{tabular} \\ \hline
\begin{tabular}[c]{@{}c@{}}custom Si\\ thin SPAD\\ \cite{Ghioni2007,Ceccarelli2019,Acconcia2017} \end{tabular} & 50 & \begin{tabular}[c]{@{}l@{}}$V_{ov} =$ \SI{5}{\volt}\\ room temperature\\ free running\end{tabular} & \begin{tabular}[c]{@{}c@{}}@\SI{800}{\nano\meter}\\ 17\%\end{tabular} & 900(9) & 125(160) & 1.4\% & 32 \\ \hline
\begin{tabular}[c]{@{}c@{}}custom Si \\ RE-SPAD\\ \cite{Gulinatti2020,Ceccarelli2019,Ceccarelli2018timing}\end{tabular} & 50 & \begin{tabular}[c]{@{}l@{}}$V_{ov} =$ \SI{20}{\volt}\\ room temperature\\ free running\end{tabular} & \begin{tabular}[c]{@{}c@{}}@\SI{800}{\nano\meter}\\ 47\%\end{tabular} & \SI{3}{\kilo\relax}(125) & \begin{tabular}[c]{@{}c@{}}71(100)\end{tabular} & 5\% & 83 \\ \hline
\begin{tabular}[c]{@{}c@{}}custom Si\\ RCE-SPAD\\ \cite{Ghioni2008rce,Ghioni2009}\end{tabular} & 20 & \begin{tabular}[c]{@{}l@{}}$V_{ov} =$ \SI{5}{\volt}\\ room temperature\\ free running\end{tabular} & \begin{tabular}[c]{@{}c@{}}@\SI{800}{\nano\meter}\\ 32\%\end{tabular} & 700(25) & \begin{tabular}[c]{@{}c@{}}12.5 \end{tabular} & 2.2\% & 35 \\ \hline
\begin{tabular}[c]{@{}c@{}}CMOS (\SI{130}{\nano\meter})\\ w/o sub isolation\\ \cite{Webster2012bis}\end{tabular} & 8 & \begin{tabular}[c]{@{}l@{}}$V_{ov} =$ \SI{12}{\volt}\\ room temperature\\ free running\end{tabular} & \begin{tabular}[c]{@{}c@{}}@\SI{800}{\nano\meter}\\ 28\%\end{tabular} & \SI{1.4}{\kilo\relax} & 30 & 3.5\% & 52 \\ \hline
\begin{tabular}[c]{@{}c@{}}BCD (\SI{160}{\nano\meter})\\ with sub isolation\\ \cite{Sanzaro2018}\end{tabular} & 30 & \begin{tabular}[c]{@{}l@{}}$V_{ov} =$ \SI{5}{\volt}\\ room temperature\\ free running\end{tabular} & \begin{tabular}[c]{@{}c@{}}@\SI{800}{\nano\meter}\\ 13\%\end{tabular} & \begin{tabular}[c]{@{}c@{}}100(0.3) \end{tabular} & 20 & 0.8\% & 31 \\ \hline\hline
\begin{tabular}[c]{@{}c@{}}custom InGaAs/InP\\ \cite{Scarcella2015}\end{tabular} & 25 & \begin{tabular}[c]{@{}l@{}}$V_{ov} =$ \SI{7}{\volt}\\ $T =$ \SI{-33}{\celsius}\\ sinusoidal gating\end{tabular} & \begin{tabular}[c]{@{}c@{}}@\SI{1550}{\nano\meter}\\ 33\%\end{tabular} & \SI{26}{\kilo\relax} & 650 & 1.3\% & 65 \\ \hline
\begin{tabular}[c]{@{}c@{}}custom InGaAs/InP\\ \cite{Korzh2014}\end{tabular} & 25 & \begin{tabular}[c]{@{}l@{}}$V_{ov}$ n.r.\\ $T =$ \SI{-110}{\celsius}\\ free running\end{tabular} & \begin{tabular}[c]{@{}c@{}}@\SI{1550}{\nano\meter}\\ 28\%\end{tabular} & 10 & 0.05 & 20\% & 129 \\ \hline
\begin{tabular}[c]{@{}c@{}}custom Ge-on-Si\\ mesa \\ \cite{Warburton2013}\end{tabular} & 25 & \begin{tabular}[c]{@{}l@{}}$V_{ov} = 0.1V_{bd}$\\ $T =$ \SI{-173}{\celsius}\\ gated mode\end{tabular} & \begin{tabular}[c]{@{}c@{}}@\SI{1310}{\nano\meter}\\ 4\%\end{tabular} & \SI{5.4}{\mega\relax} & 0.01(1)\textsuperscript{4} & n.r.\textsuperscript{4,5} & 300 \\ \hline
\begin{tabular}[c]{@{}c@{}}custom Ge-on-Si\\ planar\\ \cite{Vines2019}\end{tabular} & 100 & \begin{tabular}[c]{@{}l@{}}$V_{ov} = 0.054V_{bd}$\\ $T =$ \SI{-173}{\celsius}\\ gated mode\end{tabular} & \begin{tabular}[c]{@{}c@{}}@\SI{1310}{\nano\meter}\\ 35\%\end{tabular} & \SI{450}{\kilo\relax} & 0.1(2.5)\textsuperscript{6} & 10\%\textsuperscript{6} & 310\textsuperscript{7} \\ \hline
\end{tabular}%
}
\footnotesize{(1) Data for silicon SPADs are reported at room temperature, possibly completed with the best results at $T =$ \SI{-20}{\celsius} in brackets. Data for other SPADs are reported at the temperature specified in the measurement conditions column.} \\ 
\footnotesize{(2) Count rate corresponding to the afterpulsing probability reported on the right. The best result ever reported is in brackets.} \\
\footnotesize{(3) Light focused on a \SI{30}{\micro\meter} spot.} \\
\footnotesize{(4) $T =$ \SI{-123}{\celsius}.} \\
\footnotesize{(5) The afterpulsing rate is negligible with respect to the DCR.} \\
\footnotesize{(6) $T =$ \SI{-148}{\celsius}.} \\
\footnotesize{(7) $T =$ \SI{-195}{\celsius}.} \\

\caption{Summary of the performance for the SPADs discussed throughout the paper.}
\label{tab:summary_table}
\end{table}

The ability to generate, detect and manipulate quanta of light is the essential ground of quantum applications. In this scenario, a great research effort has been devoted to the investigation of materials and solutions to design single-photon detectors able to provide a combination of high PDE, low noise, high timing precision and low crosstalk when integrated in arrays. In this Progress Report, we focused our attention on SPADs. The performance of most suitable SPADs for quantum photonics applications reported to date are summarized in Table \ref{tab:summary_table}.

Concerning visible/NIR detection, silicon SPADs have been playing a key role, especially with custom technologies that proved to be effective in combining high detection efficiency with low noise and afterpulsing, maximum count rate in the tens of \si{\mega cps} range and a timing jitter of few tens of \si{\pico\second}. While reach-through structures currently provide the highest PDE in this region of the spectrum, the planar process used to fabricate thin, RE- and RCE-SPADs has opened the way to the fabrication of large arrays with high performance. Beyond \SI{1000}{\nano\meter}, a noteworthy trend has been observed with SPADs based either on InGaAs or germanium. While InGaAs/InP SPADs have already been demonstrated both for synchronous detection and free-running operation, and also in arrays up to a thousand of pixels, germanium SPADs have been gaining attention from the research community, especially thanks to the possibility of growing a high-quality crystalline layer of germanium over silicon, that has opened the way to the design of Ge-on-Si SPADs with relatively low DCR.

Finally, the development of waveguide integrated SPADs poses exceptional challenges in terms of materials, fabrication technology, optical and electrical design. However, at the same time, it would allow the manufacturing of quantum PICs operated at room temperature and with a large number of thermo-optic phase shifters. Therefore, we think this is the path to take to fully exploit the tremendous advantages of integrated quantum photonics.

\medskip
\textbf{Acknowledgements} \par
F. C. acknowledges Dr. Simone Atzeni (Dipartimento di Fisica - Politecnico di Milano) for the enlightening discussions on the application, especially to the field of integrated quantum photonics. R. O. acknowledges funding from the European Research Council (ERC) under the European Union’s Horizon 2020 research and innovation programme (project CAPABLE - Grant agreement No. 742745). All the authors thank the researchers who have contributed to this manuscript by providing figures and data previously published by them.

\medskip
\textbf{Conflict of Interest} \par 
M. G. discloses equity in Micro Photon Devices S.r.l. (MPD). No resources or personnel from MPD were involved in this work.

\medskip
\bibliographystyle{MSP}
\bibliography{references}

\begin{thebibliography}{100}
\providecommand{\url}[1]{\texttt{#1}}
\providecommand{\urlprefix}{URL }

\bibitem{Zhang2018qkd}
Q.~Zhang, F.~Xu, Y.-A. Chen, C.-Z. Peng, J.-W. Pan,
\newblock \emph{Optics Express} \textbf{2018}, \emph{26}, 18 24260.

\bibitem{Lo2014}
H.-K. Lo, M.~Curty, K.~Tamaki,
\newblock \emph{Nature Photonics} \textbf{2014}, \emph{8}, 8 595.

\bibitem{Briegel1998}
H.-J. Briegel, W.~D{\"{u}}r, J.~I. Cirac, P.~Zoller,
\newblock \emph{Physical Review Letters} \textbf{1998}, \emph{81}, 26 5932.

\bibitem{Aspuru-Guzik2012}
A.~Aspuru-Guzik, P.~Walther,
\newblock \emph{Nature Physics} \textbf{2012}, \emph{8}, 4 285.

\bibitem{OBrien2007}
J.~L. O'Brien,
\newblock \emph{Science} \textbf{2007}, \emph{318}, 5856 1567.

\bibitem{Arute2019}
F.~Arute, K.~Arya, R.~Babbush, D.~Bacon, J.~C. Bardin, R.~Barends, R.~Biswas,
  S.~Boixo, F.~G. S.~L. Brandao, D.~A. Buell, B.~Burkett, Y.~Chen, Z.~Chen,
  B.~Chiaro, R.~Collins, W.~Courtney, A.~Dunsworth, E.~Farhi, B.~Foxen,
  A.~Fowler, C.~Gidney, M.~Giustina, R.~Graff, K.~Guerin, S.~Habegger, M.~P.
  Harrigan, M.~J. Hartmann, A.~Ho, M.~Hoffmann, T.~Huang, T.~S. Humble, S.~V.
  Isakov, E.~Jeffrey, Z.~Jiang, D.~Kafri, K.~Kechedzhi, J.~Kelly, P.~V. Klimov,
  S.~Knysh, A.~Korotkov, F.~Kostritsa, D.~Landhuis, M.~Lindmark, E.~Lucero,
  D.~Lyakh, S.~Mandr{\`{a}}, J.~R. McClean, M.~McEwen, A.~Megrant, X.~Mi,
  K.~Michielsen, M.~Mohseni, J.~Mutus, O.~Naaman, M.~Neeley, C.~Neill, M.~Y.
  Niu, E.~Ostby, A.~Petukhov, J.~C. Platt, C.~Quintana, E.~G. Rieffel,
  P.~Roushan, N.~C. Rubin, D.~Sank, K.~J. Satzinger, V.~Smelyanskiy, K.~J.
  Sung, M.~D. Trevithick, A.~Vainsencher, B.~Villalonga, T.~White, Z.~J. Yao,
  P.~Yeh, A.~Zalcman, H.~Neven, J.~M. Martinis,
\newblock \emph{Nature} \textbf{2019}, \emph{574}, 7779 505.

\bibitem{Xanadu2020}
{Xanadu Quantum Technologies Inc.},
\newblock {Xanadu Website}, \textbf{2020},
\newblock \urlprefix\url{https://xanadu.ai/}.

\bibitem{Psiquantum2020}
{PsiQuantum Ltd.},
\newblock {PsiQuantum Website}, \textbf{2020},
\newblock \urlprefix\url{https://psiquantum.com/}.

\bibitem{Rudolph2017}
T.~Rudolph,
\newblock \emph{APL Photonics} \textbf{2017}, \emph{2}, 3 030901.

\bibitem{Pirandola2018}
S.~Pirandola, B.~R. Bardhan, T.~Gehring, C.~Weedbrook, S.~Lloyd,
\newblock \emph{Nature Photonics} \textbf{2018}, \emph{12}, 12 724.

\bibitem{Lloyd2008}
S.~Lloyd,
\newblock \emph{Science} \textbf{2008}, \emph{321}, 5895 1463.

\bibitem{Maccone2020}
L.~Maccone, C.~Ren,
\newblock \emph{Physical Review Letters} \textbf{2020}, \emph{124}, 20 200503.

\bibitem{Hadfield2009}
R.~H. Hadfield,
\newblock \emph{Nature Photonics} \textbf{2009}, \emph{3}, 12 696.

\bibitem{Boaron2018}
A.~Boaron, G.~Boso, D.~Rusca, C.~Vulliez, C.~Autebert, M.~Caloz, M.~Perrenoud,
  G.~Gras, F.~Bussi{\`{e}}res, M.-J. Li, D.~Nolan, A.~Martin, H.~Zbinden,
\newblock \emph{Physical Review Letters} \textbf{2018}, \emph{121}, 19 190502.

\bibitem{Wang2019}
H.~Wang, J.~Qin, X.~Ding, M.-C. Chen, S.~Chen, X.~You, Y.-M. He, X.~Jiang,
  L.~You, Z.~Wang, C.~Schneider, J.~J. Renema, S.~H{\"{o}}fling, C.-Y. Lu,
  J.-W. Pan,
\newblock \emph{Physical Review Letters} \textbf{2019}, \emph{123}, 25 250503.

\bibitem{Zadeh2017}
I.~E. Zadeh, J.~W.~N. Los, R.~B.~M. Gourgues, V.~Steinmetz, G.~Bulgarini, S.~M.
  Dobrovolskiy, V.~Zwiller, S.~N. Dorenbos,
\newblock \emph{APL Photonics} \textbf{2017}, \emph{2}, 11 111301.

\bibitem{Wollman2017}
E.~E. Wollman, V.~B. Verma, A.~D. Beyer, R.~M. Briggs, B.~Korzh, J.~P.
  Allmaras, F.~Marsili, A.~E. Lita, R.~P. Mirin, S.~W. Nam, M.~D. Shaw,
\newblock \emph{Optics Express} \textbf{2017}, \emph{25}, 22 26792.

\bibitem{Korzh2020}
B.~Korzh, Q.~Y. Zhao, J.~P. Allmaras, S.~Frasca, T.~M. Autry, E.~A. Bersin,
  A.~D. Beyer, R.~M. Briggs, B.~Bumble, M.~Colangelo, G.~M. Crouch, A.~E. Dane,
  T.~Gerrits, A.~E. Lita, F.~Marsili, G.~Moody, C.~Pe{\~{n}}a, E.~Ramirez,
  J.~D. Rezac, N.~Sinclair, M.~J. Stevens, A.~E. Velasco, V.~B. Verma, E.~E.
  Wollman, S.~Xie, D.~Zhu, P.~D. Hale, M.~Spiropulu, K.~L. Silverman, R.~P.
  Mirin, S.~W. Nam, A.~G. Kozorezov, M.~D. Shaw, K.~K. Berggren,
\newblock \emph{Nature Photonics} \textbf{2020}, \emph{14}, 4 250.

\bibitem{Vetter2016}
A.~Vetter, S.~Ferrari, P.~Rath, R.~Alaee, O.~Kahl, V.~Kovalyuk, S.~Diewald,
  G.~N. Goltsman, A.~Korneev, C.~Rockstuhl, W.~H.~P. Pernice,
\newblock \emph{Nano Letters} \textbf{2016}, \emph{16}, 11 7085.

\bibitem{Holzman2019}
I.~Holzman, Y.~Ivry,
\newblock \emph{Advanced Quantum Technologies} \textbf{2019}, \emph{2}, 3-4
  1800058.

\bibitem{Lv2017}
C.~L. Lv, H.~Zhou, H.~Li, L.~X. You, X.~Y. Liu, Y.~Wang, W.~J. Zhang, S.~J.
  Chen, Z.~Wang, X.~M. Xie,
\newblock \emph{Superconductor Science and Technology} \textbf{2017},
  \emph{30}, 11 115018.

\bibitem{Dequal2016}
D.~Dequal, G.~Vallone, D.~Bacco, S.~Gaiarin, V.~Luceri, G.~Bianco,
  P.~Villoresi,
\newblock \emph{Physical Review A} \textbf{2016}, \emph{93}, 1 010301.

\bibitem{Marsili2013}
F.~Marsili, V.~B. Verma, J.~A. Stern, S.~Harrington, A.~E. Lita, T.~Gerrits,
  I.~Vayshenker, B.~Baek, M.~D. Shaw, R.~P. Mirin, S.~W. Nam,
\newblock \emph{Nature Photonics} \textbf{2013}, \emph{7}, 3 210.

\bibitem{Flamini2019}
F.~Flamini, N.~Spagnolo, F.~Sciarrino,
\newblock \emph{Reports on Progress in Physics} \textbf{2019}, \emph{82}, 1
  016001.

\bibitem{Fukuda2011}
D.~Fukuda, G.~Fujii, T.~Numata, K.~Amemiya, A.~Yoshizawa, H.~Tsuchida,
  H.~Fujino, H.~Ishii, T.~Itatani, S.~Inoue, T.~Zama,
\newblock \emph{IEEE Transactions on Applied Superconductivity} \textbf{2011},
  \emph{21}, 3 241.

\bibitem{Giustina2013}
M.~Giustina, A.~Mech, S.~Ramelow, B.~Wittmann, J.~Kofler, J.~Beyer, A.~Lita,
  B.~Calkins, T.~Gerrits, S.~W. Nam, R.~Ursin, A.~Zeilinger,
\newblock \emph{Nature} \textbf{2013}, \emph{497}, 7448 227.

\bibitem{Namekata2010}
N.~Namekata, S.~Adachi, S.~Inoue,
\newblock \emph{IEEE Photonics Technology Letters} \textbf{2010}, \emph{22}, 8
  529.

\bibitem{Lamas-Linares2013}
A.~Lamas-Linares, B.~Calkins, N.~A. Tomlin, T.~Gerrits, A.~E. Lita, J.~Beyer,
  R.~P. Mirin, S.~W. Nam,
\newblock \emph{Applied Physics Letters} \textbf{2013}, \emph{102}, 23 231117.

\bibitem{Calkins2011}
B.~Calkins, A.~E. Lita, A.~E. Fox, S.~W. Nam,
\newblock \emph{Applied Physics Letters} \textbf{2011}, \emph{99}, 24 241114.

\bibitem{Bedington2017}
R.~Bedington, J.~M. Arrazola, A.~Ling,
\newblock \emph{npj Quantum Information} \textbf{2017}, \emph{3}, 1 30.

\bibitem{Wang2020}
J.~Wang, F.~Sciarrino, A.~Laing, M.~G. Thompson,
\newblock \emph{Nature Photonics} \textbf{2020}, \emph{14}, 5 273.

\bibitem{Pernice2012}
W.~H.~P. Pernice, C.~Schuck, O.~Minaeva, M.~Li, G.~N. Goltsman, A.~V.
  Sergienko, H.~X. Tang,
\newblock \emph{Nature Communications} \textbf{2012}, \emph{3}, 1 1325.

\bibitem{Gerrits2011}
T.~Gerrits, N.~Thomas-Peter, J.~C. Gates, A.~E. Lita, B.~J. Metcalf,
  B.~Calkins, N.~A. Tomlin, A.~E. Fox, A.~L. Linares, J.~B. Spring, N.~K.
  Langford, R.~P. Mirin, P.~G.~R. Smith, I.~A. Walmsley, S.~W. Nam,
\newblock \emph{Physical Review A} \textbf{2011}, \emph{84}, 6 060301.

\bibitem{Harris2018}
N.~C. Harris, J.~Carolan, D.~Bunandar, M.~Prabhu, M.~Hochberg, T.~Baehr-Jones,
  M.~L. Fanto, A.~M. Smith, C.~C. Tison, P.~M. Alsing, D.~Englund,
\newblock \emph{Optica} \textbf{2018}, \emph{5}, 12 1623.

\bibitem{Wang2014}
C.~Wang, B.~Lichtenwalter, A.~Friebel, H.~X. Tang,
\newblock \emph{Cryogenics} \textbf{2014}, \emph{64} 5.

\bibitem{Harris2014}
N.~C. Harris, Y.~Ma, J.~Mower, T.~Baehr-Jones, D.~Englund, M.~Hochberg,
  C.~Galland,
\newblock \emph{Optics Express} \textbf{2014}, \emph{22}, 9 10487.

\bibitem{Ghioni2007}
M.~Ghioni, A.~Gulinatti, I.~Rech, F.~Zappa, S.~Cova,
\newblock \emph{IEEE Journal of Selected Topics in Quantum Electronics}
  \textbf{2007}, \emph{13}, 4 852.

\bibitem{Bruschini2019}
C.~Bruschini, H.~Homulle, I.~M. Antolovic, S.~Burri, E.~Charbon,
\newblock \emph{Light: Science {\&} Applications} \textbf{2019}, \emph{8}, 1
  87.

\bibitem{Michalet2014}
X.~Michalet, A.~Ingargiola, R.~A. Colyer, G.~Scalia, S.~Weiss, P.~Maccagnani,
  A.~Gulinatti, I.~Rech, M.~Ghioni,
\newblock \emph{IEEE Journal of Selected Topics in Quantum Electronics}
  \textbf{2014}, \emph{20}, 6 248.

\bibitem{Buller2007}
G.~Buller, A.~Wallace,
\newblock \emph{IEEE Journal of Selected Topics in Quantum Electronics}
  \textbf{2007}, \emph{13}, 4 1006.

\bibitem{Karami2017}
M.~A. Karami, M.~Ansarian,
\newblock \emph{Basic and Clinical Neuroscience Journal} \textbf{2017},
  \emph{8}, 1 19.

\bibitem{Lounis2000}
B.~Lounis, W.~E. Moerner,
\newblock \emph{Nature} \textbf{2000}, \emph{407}, 6803 491.

\bibitem{Yuan2002}
Z.~Yuan, B.~E. Kardynal, R.~M. Stevenson, A.~J. Shields, J.~L. Charlene,
  K.~Cooper, N.~S. Beattie, D.~A. Ritchie, M.~Pepper,
\newblock \emph{Science} \textbf{2002}, \emph{295}, 5552 102.

\bibitem{Politi2008}
A.~Politi, M.~J. Cryan, J.~G. Rarity, S.~Yu, J.~L. O'Brien,
\newblock \emph{Science} \textbf{2008}, \emph{320}, 5876 646.

\bibitem{Matthews2009}
J.~C.~F. Matthews, A.~Politi, A.~Stefanov, J.~L. O'Brien,
\newblock \emph{Nature Photonics} \textbf{2009}, \emph{3}, 6 346.

\bibitem{Broome2013}
M.~A. Broome, A.~Fedrizzi, S.~Rahimi-Keshari, J.~Dove, S.~Aaronson, T.~C.
  Ralph, A.~G. White,
\newblock \emph{Science} \textbf{2013}, \emph{339}, 6121 794.

\bibitem{Spring2013}
J.~B. Spring, B.~J. Metcalf, P.~C. Humphreys, W.~S. Kolthammer, X.-M. Jin,
  M.~Barbieri, A.~Datta, N.~Thomas-Peter, N.~K. Langford, D.~Kundys, J.~C.
  Gates, B.~J. Smith, P.~G.~R. Smith, I.~A. Walmsley,
\newblock \emph{Science} \textbf{2013}, \emph{339}, 6121 798.

\bibitem{Liao2017satellite}
S.-K. Liao, W.-Q. Cai, W.-Y. Liu, L.~Zhang, Y.~Li, J.-G. Ren, J.~Yin, Q.~Shen,
  Y.~Cao, Z.-P. Li, F.-Z. Li, X.-W. Chen, L.-H. Sun, J.-J. Jia, J.-C. Wu, X.-J.
  Jiang, J.-F. Wang, Y.-M. Huang, Q.~Wang, Y.-L. Zhou, L.~Deng, T.~Xi, L.~Ma,
  T.~Hu, Q.~Zhang, Y.-A. Chen, N.-L. Liu, X.-B. Wang, Z.-C. Zhu, C.-Y. Lu,
  R.~Shu, C.-Z. Peng, J.-Y. Wang, J.-W. Pan,
\newblock \emph{Nature} \textbf{2017}, \emph{549}, 7670 43.

\bibitem{Sparrow2018}
C.~Sparrow, E.~Mart{\'{i}}n-L{\'{o}}pez, N.~Maraviglia, A.~Neville, C.~Harrold,
  J.~Carolan, Y.~N. Joglekar, T.~Hashimoto, N.~Matsuda, J.~L. O'Brien, D.~P.
  Tew, A.~Laing,
\newblock \emph{Nature} \textbf{2018}, \emph{557}, 7707 660.

\bibitem{Morimoto2020}
K.~Morimoto, A.~Ardelean, M.-L. Wu, A.~C. Ulku, I.~M. Antolovic, C.~Bruschini,
  E.~Charbon,
\newblock \emph{Optica} \textbf{2020}, \emph{7}, 4 346.

\bibitem{Wollman2019}
E.-E. Wollman, V.~B. Verma, A.~E. Lita, W.~H. Farr, M.~D. Shaw, R.~P. Mirin,
  S.~W. Nam,
\newblock \emph{Optics Express} \textbf{2019}, \emph{27}, 24 35279.

\bibitem{Champlin1959}
K.~S. Champlin,
\newblock \emph{Journal of Applied Physics} \textbf{1959}, \emph{30}, 7 1039.

\bibitem{Webb1970spd}
P.~Webb, R.~McIntyre,
\newblock \emph{Bulletin of the American Physical Society} \textbf{1970},
  \emph{15}, 6 813.

\bibitem{Cova1996}
S.~Cova, M.~Ghioni, A.~Lacaita, C.~Samori, F.~Zappa,
\newblock \emph{Applied Optics} \textbf{1996}, \emph{35}, 12 1956.

\bibitem{Oldham1972triggering}
W.~G. Oldham, R.~R. Samuelson, P.~Antognetti,
\newblock \emph{IEEE Transactions on electron devices} \textbf{1972},
  \emph{19}, 9 1056.

\bibitem{Mcintyre1973avalanche}
R.~J. McIntyre,
\newblock \emph{IEEE Transactions on Electron Devices} \textbf{1973},
  \emph{20}, 7 637.

\bibitem{Gulinatti2011}
A.~Gulinatti, I.~Rech, M.~Assanelli, M.~Ghioni, S.~Cova,
\newblock \emph{Journal of Modern Optics} \textbf{2011}, \emph{58}, 3-4 210.

\bibitem{Pancheri2014}
L.~Pancheri, D.~Stoppa, G.-F. Dalla~Betta,
\newblock \emph{IEEE Journal of Selected Topics in Quantum Electronics}
  \textbf{2014}, \emph{20}, 6 328.

\bibitem{Okuto1974}
Y.~Okuto, C.~R. Crowell,
\newblock \emph{Physical Review B} \textbf{1974}, \emph{10}, 10 4284.

\bibitem{Ghioni1996}
M.~Ghioni, S.~Cova, F.~Zappa, C.~Samori,
\newblock \emph{Review of scientific instruments} \textbf{1996}, \emph{67}, 10
  3440.

\bibitem{Takesue2006}
H.~Takesue, E.~Diamanti, C.~Langrock, M.~Fejer, Y.~Yamamoto,
\newblock \emph{Optics Express} \textbf{2006}, \emph{14}, 20 9522.

\bibitem{Ripamonti1985}
G.~Ripamonti, S.~Cova,
\newblock \emph{Solid-State Electronics} \textbf{1985}, \emph{28}, 9 925.

\bibitem{Lacaita1993}
A.~Lacaita, S.~Cova, M.~Ghioni, F.~Zappa,
\newblock \emph{IEEE Electron Device Letters} \textbf{1993}, \emph{14}, 7 360.

\bibitem{Spinelli1997}
A.~Spinelli, A.~L. Lacaita,
\newblock \emph{IEEE Transactions on Electron Devices} \textbf{1997},
  \emph{44}, 11 1931.

\bibitem{Tan2007}
C.~Tan, J.~Ng, G.~Rees, J.~David,
\newblock \emph{IEEE Journal of Selected Topics in Quantum Electronics}
  \textbf{2007}, \emph{13}, 4 906.

\bibitem{Ingargiola2009}
A.~Ingargiola, M.~Assanelli, A.~Gallivanoni, I.~Rech, M.~Ghioni, S.~Cova,
\newblock In \emph{Advanced Photon Counting Techniques III}, volume 7320.
  International Society for Optics and Photonics, \textbf{2009} 73200K.

\bibitem{Lacaita1990}
A.~Lacaita, M.~Mastrapasqua, M.~Ghioni, S.~Vanoli,
\newblock \emph{Applied Physics Letters} \textbf{1990}, \emph{57}, 5 489.

\bibitem{Assanelli2011}
M.~Assanelli, A.~Ingargiola, I.~Rech, A.~Gulinatti, M.~Ghioni,
\newblock \emph{IEEE Journal of Quantum Electronics} \textbf{2011}, \emph{47},
  2 151.

\bibitem{Lacaita1993photon}
A.~Lacaita, S.~Cova, A.~Spinelli, F.~Zappa,
\newblock \emph{Applied Physics Letters} \textbf{1993}, \emph{62}, 6 606.

\bibitem{Gulinatti2005}
A.~Gulinatti, P.~Maccagnani, I.~Rech, M.~Ghioni, S.~Cova,
\newblock \emph{Electronics Letters} \textbf{2005}, \emph{41}, 5 272.

\bibitem{Haitz1965}
R.~H. Haitz,
\newblock \emph{Journal of Applied Physics} \textbf{1965}, \emph{36}, 10 3123.

\bibitem{Cova1991}
S.~Cova, A.~Lacaita, G.~Ripamonti,
\newblock \emph{IEEE Electron Device Letters} \textbf{1991}, \emph{12}, 12 685.

\bibitem{Giudice2003}
A.~Giudice, M.~Ghioni, S.~Cova, F.~Zappa,
\newblock In \emph{33rd Conference on European Solid-State Device Research
  (ESSDERC'03)}. IEEE, \textbf{2003} 347--350.

\bibitem{Lacaita1993bremsstrahlung}
A.~L. Lacaita, F.~Zappa, S.~Bigliardi, M.~Manfredi,
\newblock \emph{IEEE Transactions on Electron Devices} \textbf{1993},
  \emph{40}, 3 577.

\bibitem{Gulinatti2012}
A.~Gulinatti, F.~Panzeri, I.~Rech, P.~Maccagnani, M.~Ghioni, S.~Cova,
\newblock In \emph{Quantum Sensing and Nanophotonic Devices IX}, volume 8268.
  International Society for Optics and Photonics, \textbf{2012} 82681D.

\bibitem{Rech2008}
I.~Rech, A.~Ingargiola, R.~Spinelli, I.~Labanca, S.~Marangoni, M.~Ghioni,
  S.~Cova,
\newblock \emph{Optics Express} \textbf{2008}, \emph{16}, 12 8381.

\bibitem{Ghioni2008}
M.~Ghioni, A.~Gulinatti, I.~Rech, P.~Maccagnani, S.~Cova,
\newblock In \emph{Quantum Sensing and Nanophotonic Devices V}, volume 6900.
  International Society for Optics and Photonics, \textbf{2008} 69001D.

\bibitem{Acconcia2017}
G.~Acconcia, I.~Rech, I.~Labanca, M.~Ghioni,
\newblock \emph{Electronics Letters} \textbf{2017}, \emph{53}, 5 328.

\bibitem{Labanca2018}
I.~Labanca, F.~Ceccarelli, A.~Gulinatti, M.~Ghioni, I.~Rech,
\newblock \emph{Electronics Letters} \textbf{2018}, \emph{54}, 10 644.

\bibitem{Mandai2012}
S.~Mandai, M.~W. Fishburn, Y.~Maruyama, E.~Charbon,
\newblock \emph{Optics Express} \textbf{2012}, \emph{20}, 6 5849.

\bibitem{Acerbi2019}
F.~Acerbi, S.~Gundacker,
\newblock \emph{Nuclear Instruments and Methods in Physics Research Section A:
  Accelerators, Spectrometers, Detectors and Associated Equipment}
  \textbf{2019}, \emph{926} 16.

\bibitem{Wang2017}
H.~Wang, Y.~He, Y.-H. Li, Z.-E. Su, B.~Li, H.-L. Huang, X.~Ding, M.-C. Chen,
  C.~Liu, J.~Qin, J.-P. Li, Y.-M. He, C.~Schneider, M.~Kamp, C.-Z. Peng,
  S.~H{\"{o}}fling, C.-Y. Lu, J.-W. Pan,
\newblock \emph{Nature Photonics} \textbf{2017}, \emph{11}, 6 361.

\bibitem{Restelli2010}
A.~Restelli, J.~C. Bienfang, C.~W. Clark, I.~Rech, I.~Labanca, M.~Ghioni,
  S.~Cova,
\newblock \emph{IEEE Journal of Selected Topics in Quantum Electronics}
  \textbf{2010}, \emph{16}, 5 1084.

\bibitem{Howland2013}
G.~A. Howland, J.~C. Howell,
\newblock \emph{Physical Review X} \textbf{2013}, \emph{3}, 1 1.

\bibitem{Lvovsky2009}
A.~I. Lvovsky, B.~C. Sanders, W.~Tittel,
\newblock \emph{Nature Photonics} \textbf{2009}, \emph{3}, 12 706.

\bibitem{Wang2016}
X.-L. Wang, L.-K. Chen, W.~Li, H.-L. Huang, C.~Liu, C.~Chen, Y.-H. Luo, Z.-E.
  Su, D.~Wu, Z.-D. Li, H.~Lu, Y.~Hu, X.~Jiang, C.-Z. Peng, L.~Li, N.-L. Liu,
  Y.-A. Chen, C.-Y. Lu, J.-W. Pan,
\newblock \emph{Physical Review Letters} \textbf{2016}, \emph{117}, 21 210502.

\bibitem{Kaneda2015}
F.~Kaneda, B.~G. Christensen, J.~J. Wong, H.~S. Park, K.~T. McCusker, P.~G.
  Kwiat,
\newblock \emph{Optica} \textbf{2015}, \emph{2}, 12 1010.

\bibitem{Senellart2017}
P.~Senellart, G.~Solomon, A.~White,
\newblock \emph{Nature Nanotechnology} \textbf{2017}, \emph{12}, 11 1026.

\bibitem{Politi2009}
A.~Politi, J.~Matthews, M.~Thompson, J.~O'Brien,
\newblock \emph{IEEE Journal of Selected Topics in Quantum Electronics}
  \textbf{2009}, \emph{15}, 6 1673.

\bibitem{Meany2015}
T.~Meany, M.~Gr{\"{a}}fe, R.~Heilmann, A.~Perez-Leija, S.~Gross, M.~J. Steel,
  M.~J. Withford, A.~Szameit,
\newblock \emph{Laser {\&} Photonics Reviews} \textbf{2015}, \emph{9}, 4 363.

\bibitem{Munoz2017}
P.~Mu{\~{n}}oz, G.~Mic{\'{o}}, L.~Bru, D.~Pastor, D.~P{\'{e}}rez,
  J.~Dom{\'{e}}nech, J.~Fern{\'{a}}ndez, R.~Ba{\~{n}}os, B.~Gargallo,
  R.~Alemany, A.~S{\'{a}}nchez, J.~Cirera, R.~Mas, C.~Dom{\'{i}}nguez,
\newblock \emph{Sensors} \textbf{2017}, \emph{17}, 9 2088.

\bibitem{McIntyre1985}
R.~J. McIntyre,
\newblock \emph{Measurement} \textbf{1985}, \emph{3}, 4 146.

\bibitem{Dautet1993}
H.~Dautet, P.~Deschamps, B.~Dion, A.~D. MacGregor, D.~MacSween, R.~J. McIntyre,
  C.~Trottier, P.~P. Webb,
\newblock \emph{Applied Optics} \textbf{1993}, \emph{32}, 21 3894.

\bibitem{Brida2012}
G.~Brida, I.~P. Degiovanni, M.~Genovese, F.~Piacentini, P.~Traina, A.~{Della
  Frera}, A.~Tosi, A.~{Bahgat Shehata}, C.~Scarcella, A.~Gulinatti, M.~Ghioni,
  S.~V. Polyakov, A.~Migdall, A.~Giudice,
\newblock \emph{Applied Physics Letters} \textbf{2012}, \emph{101}, 22 221112.

\bibitem{Ates2012}
S.~Ates, I.~Agha, A.~Gulinatti, I.~Rech, M.~T. Rakher, A.~Badolato,
  K.~Srinivasan,
\newblock \emph{Physical Review Letters} \textbf{2012}, \emph{109}, 14 147405.

\bibitem{Ates2013}
S.~Ates, I.~Agha, A.~Gulinatti, I.~Rech, A.~Badolato, K.~Srinivasan,
\newblock \emph{Scientific Reports} \textbf{2013}, \emph{3}, 1 1397.

\bibitem{Versteegh2014}
M.~A.~M. Versteegh, M.~E. Reimer, K.~D. J{\"{o}}ns, D.~Dalacu, P.~J. Poole,
  A.~Gulinatti, A.~Giudice, V.~Zwiller,
\newblock \emph{Nature Communications} \textbf{2014}, \emph{5}, 1 5298.

\bibitem{Jons2017}
K.~D. J{\"{o}}ns, L.~Schweickert, M.~A.~M. Versteegh, D.~Dalacu, P.~J. Poole,
  A.~Gulinatti, A.~Giudice, V.~Zwiller, M.~E. Reimer,
\newblock \emph{Scientific Reports} \textbf{2017}, \emph{7}, 1 1700.

\bibitem{Rogers2007}
D.~J. Rogers, J.~C. Bienfang, A.~Nakassis, H.~Xu, C.~W. Clark,
\newblock \emph{New Journal of Physics} \textbf{2007}, \emph{9} 319.

\bibitem{Zappa2007}
F.~Zappa, S.~Tisa, A.~Tosi, S.~Cova,
\newblock \emph{Sensors and Actuators A: Physical} \textbf{2007}, \emph{140}, 1
  103.

\bibitem{Stipcevic2013}
M.~Stipcevic, D.~Wang, R.~Ursin,
\newblock \emph{Journal of Lightwave Technology} \textbf{2013}, \emph{31}, 23
  3591.

\bibitem{SPCM2019}
{Excelitas Technologies Corp.},
\newblock {Single Photon Counting Modules Family Brochure}, \textbf{2019},
\newblock
  \urlprefix\url{{https://www.excelitas.com/file-download/download/public/67136?filename=PD\_SPCM
  Family\_brochure\_June2019.pdf}}.

\bibitem{COUNT2019}
{Laser Components GmbH},
\newblock {Single Photon Counting Modules COUNT-Series Datasheet},
  \textbf{2017},
\newblock
  \urlprefix\url{{https://www.lasercomponents.com/fileadmin/user\_upload/home/Datasheets/lcp/count-series.pdf}}.

\bibitem{SAP5002020}
{Laser Components GmbH},
\newblock {Silicon Geiger Mode Avalanche Photodiode SAP500 Datasheet},
  \textbf{2020},
\newblock
  \urlprefix\url{{https://www.lasercomponents.com/fileadmin/user\_upload/home/Datasheets/lc-apd/sap-series.pdf}}.

\bibitem{ID1202019}
{ID Quantique SA},
\newblock {ID120 Visible Single-Photon Detector Datasheet}, \textbf{2019},
\newblock
  \urlprefix\url{{https://marketing.idquantique.com/acton/attachment/11868/f-0238/1/-/-/-/-/ID120\_Brochure.pdf}}.

\bibitem{SPCM-TR2020}
{Excelitas Technologies Corp.},
\newblock {SPCM-AQRH-TR Family Datasheet}, \textbf{2020},
\newblock
  \urlprefix\url{{https://www.excelitas.com/file-download/download/public/60596?filename=PD\_Excelitas
  SPCM AQRH TR datasheet.pdf}}.

\bibitem{Rech2006}
I.~Rech, I.~Labanca, M.~Ghioni, S.~Cova,
\newblock \emph{Review of Scientific Instruments} \textbf{2006}, \emph{77}, 3
  033104.

\bibitem{SPCM-AQRH2020}
{Excelitas Technologies Corp.},
\newblock {SPCM-AQRH Family Datasheet}, \textbf{2020},
\newblock
  \urlprefix\url{{https://www.excelitas.com/file-download/download/public/60241?filename=PD\_Excelitas
  SPCM AQRH datasheet.pdf}}.

\bibitem{Ceccarelli2016}
F.~Ceccarelli, A.~Gulinatti, I.~Labanca, I.~Rech, M.~Ghioni,
\newblock \emph{IEEE Photonics Technology Letters} \textbf{2016}, \emph{28}, 9
  1002 .

\bibitem{Aull2018}
B.~F. Aull, E.~K. Duerr, J.~P. Frechette, K.~A. McIntosh, D.~R. Schuette, R.~D.
  Younger,
\newblock \emph{IEEE Journal of Selected Topics in Quantum Electronics}
  \textbf{2018}, \emph{24}, 2 1.

\bibitem{Ceccarelli2019}
F.~Ceccarelli, G.~Acconcia, A.~Gulinatti, M.~Ghioni, I.~Rech,
\newblock \emph{IEEE Photonics Technology Letters} \textbf{2019}, \emph{31}, 1
  102.

\bibitem{Ceccarelli2018}
F.~Ceccarelli, G.~Acconcia, I.~Labanca, A.~Gulinatti, M.~Ghioni, I.~Rech,
\newblock \emph{IEEE Photonics Technology Letters} \textbf{2018}, \emph{30}, 4
  391.

\bibitem{PDM2019}
{Micro Photon Devices S.r.l.},
\newblock {PDM Datasheet}, \textbf{2019},
\newblock
  \urlprefix\url{{http://www.micro-photon-devices.com/MPD/media/Datasheet/PDM.pdf}}.

\bibitem{Gulinatti2012red}
A.~Gulinatti, I.~Rech, F.~Panzeri, C.~Cammi, P.~Maccagnani, M.~Ghioni, S.~Cova,
\newblock \emph{Journal of Modern Optics} \textbf{2012}, \emph{59}, 17 1489.

\bibitem{Gulinatti2020}
A.~Gulinatti, F.~Ceccarelli, M.~Ghioni, I.~Rech,
\newblock {Custom silicon technology for SPAD-arrays with red-enhanced
  sensitivity and low timing jitter}, \textbf{2020}.

\bibitem{Ceccarelli2018redarray}
F.~Ceccarelli, A.~Gulinatti, I.~Labanca, M.~Ghioni, I.~Rech,
\newblock \emph{IEEE Photonics Technology Letters} \textbf{2018}, \emph{30}, 6
  557.

\bibitem{Ceccarelli2018timing}
F.~Ceccarelli, G.~Acconcia, A.~Gulinatti, M.~Ghioni, I.~Rech,
\newblock \emph{IEEE Photonics Technology Letters} \textbf{2018}, \emph{30}, 19
  1727.

\bibitem{Ghioni2008rce}
M.~Ghioni, G.~Armellini, P.~Maccagnani, I.~Rech, M.~K. Emsley, M.~S. Unlu,
\newblock \emph{IEEE Photonics Technology Letters} \textbf{2008}, \emph{20}, 6
  413.

\bibitem{Ghioni2009}
M.~Ghioni, G.~Armellini, P.~Maccagnani, I.~Rech, M.~K. Emsley, M.~S.
  {\"{U}}nl{\"{u}},
\newblock \emph{Journal of Modern Optics} \textbf{2009}, \emph{56}, 2-3 309.

\bibitem{Rochas2003}
A.~Rochas, M.~Gani, B.~Furrer, P.~A. Besse, R.~S. Popovic, G.~Ribordy,
  N.~Gisin,
\newblock \emph{Review of Scientific Instruments} \textbf{2003}, \emph{74}, 7
  3263.

\bibitem{Finkelstein2006}
H.~Finkelstein, M.~J. Hsu, S.~C. Esener,
\newblock \emph{IEEE Electron Device Letters} \textbf{2006}, \emph{27}, 11 887.

\bibitem{Niclass2007}
C.~Niclass, M.~Gersbach, R.~Henderson, L.~Grant, E.~Charbon,
\newblock \emph{IEEE Journal of Selected Topics in Quantum Electronics}
  \textbf{2007}, \emph{13}, 4 863.

\bibitem{Xiao2007}
Z.~Xiao, D.~Pantic, R.~Popovic,
\newblock In \emph{2007 International Solid-State Sensors, Actuators and
  Microsystems Conference}. IEEE, \textbf{2007} 1365--1368.

\bibitem{Faramarzpour2008}
N.~Faramarzpour, M.~J. Deen, S.~Shirani, Q.~Fang,
\newblock \emph{IEEE Transactions on Electron Devices} \textbf{2008},
  \emph{55}, 3 760.

\bibitem{Gersbach2009}
M.~Gersbach, J.~Richardson, E.~Mazaleyrat, S.~Hardillier, C.~Niclass, R.~K.
  Henderson, L.~Grant, E.~Charbon,
\newblock \emph{Solid-State Electronics} \textbf{2009}, \emph{53}, 7 803.

\bibitem{Field2010}
R.~M. Field, J.~Lary, J.~Cohn, L.~Paninski, K.~L. Shepard,
\newblock \emph{Applied Physics Letters} \textbf{2010}, \emph{97}, 21 211111.

\bibitem{Karami2010}
M.~A. Karami, M.~Gersbach, H.-J. Yoon, E.~Charbon,
\newblock \emph{Optics Express} \textbf{2010}, \emph{18}, 21 22158.

\bibitem{Pancheri2011}
L.~Pancheri, D.~Stoppa,
\newblock In \emph{2011 Proceedings of the European Solid-State Device Research
  Conference (ESSDERC)}. IEEE, \textbf{2011} 179--182.

\bibitem{Gersbach2012}
M.~Gersbach, Y.~Maruyama, R.~Trimananda, M.~W. Fishburn, D.~Stoppa, J.~A.
  Richardson, R.~Walker, R.~Henderson, E.~Charbon,
\newblock \emph{IEEE Journal of Solid-State Circuits} \textbf{2012}, \emph{47},
  6 1394.

\bibitem{Villa2014}
F.~Villa, D.~Bronzi, Y.~Zou, C.~Scarcella, G.~Boso, S.~Tisa, A.~Tosi, F.~Zappa,
  D.~Durini, S.~Weyers, U.~Paschen, W.~Brockherde,
\newblock \emph{Journal of Modern Optics} \textbf{2014}, \emph{61}, 2 102.

\bibitem{Dutton2016}
N.~A.~W. Dutton, I.~Gyongy, L.~Parmesan, S.~Gnecchi, N.~Calder, B.~R. Rae,
  S.~Pellegrini, L.~Grant, R.~K. Henderson,
\newblock \emph{IEEE Transactions on Electron Devices} \textbf{2016},
  \emph{63}, 1 189.

\bibitem{RonchiniXimenes2019}
A.~{Ronchini Ximenes}, P.~Padmanabhan, M.-J. Lee, Y.~Yamashita, D.-N. Yaung,
  E.~Charbon,
\newblock \emph{IEEE Journal of Solid-State Circuits} \textbf{2019}, \emph{54},
  11 3203.

\bibitem{Ximenes2018}
A.~R. Ximenes, P.~Padmanabhan, M.-J. Lee, Y.~Yamashita, D.-N. Yaung,
  E.~Charbon,
\newblock In \emph{2018 IEEE International Solid-State Circuits
  Conference-(ISSCC)}. IEEE, \textbf{2018} 96--98.

\bibitem{Gyongy2017}
I.~Gyongy, N.~Calder, A.~Davies, N.~A. Dutton, R.~R. Duncan, C.~Rickman,
  P.~Dalgarno, R.~K. Henderson,
\newblock \emph{IEEE Transactions on Electron Devices} \textbf{2017},
  \emph{65}, 2 547.

\bibitem{Ulku2018}
A.~C. Ulku, C.~Bruschini, I.~M. Antolovi{\'c}, Y.~Kuo, R.~Ankri, S.~Weiss,
  X.~Michalet, E.~Charbon,
\newblock \emph{IEEE Journal of Selected Topics in Quantum Electronics}
  \textbf{2018}, \emph{25}, 1 1.

\bibitem{Perenzoni2015}
M.~Perenzoni, N.~Massari, D.~Perenzoni, L.~Gasparini, D.~Stoppa,
\newblock \emph{IEEE Journal of Solid-State Circuits} \textbf{2015}, \emph{51},
  1 155.

\bibitem{Villa2014bis}
F.~Villa, R.~Lussana, D.~Bronzi, S.~Tisa, A.~Tosi, F.~Zappa, A.~{Dalla Mora},
  D.~Contini, D.~Durini, S.~Weyers, W.~Brockherde,
\newblock \emph{IEEE Journal of Selected Topics in Quantum Electronics}
  \textbf{2014}, \emph{20}, 6 364.

\bibitem{Field2014}
R.~M. Field, S.~Realov, K.~L. Shepard,
\newblock \emph{IEEE Journal of Solid-State Circuits} \textbf{2014}, \emph{49},
  4 867.

\bibitem{Braga2013}
L.~H.~C. Braga, L.~Gasparini, L.~Grant, R.~K. Henderson, N.~Massari,
  M.~Perenzoni, D.~Stoppa, R.~Walker,
\newblock In \emph{2013 IEEE International Solid-State Circuits Conference
  Digest of Technical Papers}. IEEE, \textbf{2013} 486--487.

\bibitem{Maruyama2013}
Y.~Maruyama, J.~Blacksberg, E.~Charbon,
\newblock \emph{IEEE Journal of Solid-State Circuits} \textbf{2013}, \emph{49},
  1 179.

\bibitem{Niclass2012}
C.~Niclass, M.~Soga, H.~Matsubara, S.~Kato, M.~Kagami,
\newblock \emph{IEEE Journal of Solid-State Circuits} \textbf{2012}, \emph{48},
  2 559.

\bibitem{Veerappan2011}
C.~Veerappan, J.~Richardson, R.~Walker, D.-U. Li, M.~W. Fishburn, Y.~Maruyama,
  D.~Stoppa, F.~Borghetti, M.~Gersbach, R.~K. Henderson, et~al.,
\newblock In \emph{2011 IEEE International Solid-State Circuits Conference}.
  IEEE, \textbf{2011} 312--314.

\bibitem{Zhang2018}
C.~Zhang, S.~Lindner, I.~M. Antolovi{\'c}, J.~M. Pavia, M.~Wolf, E.~Charbon,
\newblock \emph{IEEE Journal of Solid-State Circuits} \textbf{2018}, \emph{54},
  4 1137.

\bibitem{Hutchings2019}
S.~W. Hutchings, N.~Johnston, I.~Gyongy, T.~Al~Abbas, N.~A. Dutton, M.~Tyler,
  S.~Chan, J.~Leach, R.~K. Henderson,
\newblock \emph{IEEE Journal of Solid-State Circuits} \textbf{2019}, \emph{54},
  11 2947.

\bibitem{Stoppa2009}
D.~Stoppa, D.~Mosconi, L.~Pancheri, L.~Gonzo,
\newblock \emph{IEEE Sensors Journal} \textbf{2009}, \emph{9}, 9 1084.

\bibitem{Schwartz2008}
D.~E. Schwartz, E.~Charbon, K.~L. Shepard,
\newblock \emph{IEEE Journal of Solid-State Circuits} \textbf{2008}, \emph{43},
  11 2546.

\bibitem{Lubin2019}
G.~Lubin, R.~Tenne, I.~M. Antolovic, E.~Charbon, C.~Bruschini, D.~Oron,
\newblock \emph{Optics Express} \textbf{2019}, \emph{27}, 23 32863.

\bibitem{Piacentini2017}
F.~Piacentini, A.~Avella, E.~Rebufello, R.~Lussana, F.~Villa, A.~Tosi,
  M.~Gramegna, G.~Brida, E.~Cohen, L.~Vaidman, I.~P. Degiovanni, M.~Genovese,
\newblock \emph{Nature Physics} \textbf{2017}, \emph{13}, 12 1191.

\bibitem{Piacentini2016}
F.~Piacentini, A.~Avella, M.~P. Levi, R.~Lussana, F.~Villa, A.~Tosi, F.~Zappa,
  M.~Gramegna, G.~Brida, I.~P. Degiovanni, M.~Genovese,
\newblock \emph{Physical Review Letters} \textbf{2016}, \emph{116}, 18 180401.

\bibitem{Piacentini2016bis}
F.~Piacentini, A.~Avella, M.~P. Levi, M.~Gramegna, G.~Brida, I.~P. Degiovanni,
  E.~Cohen, R.~Lussana, F.~Villa, A.~Tosi, F.~Zappa, M.~Genovese,
\newblock \emph{Physical Review Letters} \textbf{2016}, \emph{117}, 17 170402.

\bibitem{Burri2013}
S.~Burri, D.~Stucki, Y.~Maruyama, C.~Bruschini, E.~Charbon, F.~Regazzoni,
\newblock In \emph{International Image Sensor Workshop (IISW)}. \textbf{2013}
  5--8.

\bibitem{Tisa2015}
S.~Tisa, F.~Villa, A.~Giudice, G.~Simmerle, F.~Zappa,
\newblock \emph{IEEE Journal of Selected Topics in Quantum Electronics}
  \textbf{2015}, \emph{21}, 3 23.

\bibitem{Eisele2011}
A.~Eisele, R.~Henderson, B.~Schmidtke, T.~Funk, L.~Grant, J.~Richardson,
  W.~Freude,
\newblock In \emph{Proceedings of International Image Sensor Workshop}.
  \textbf{2011} 278--280.

\bibitem{Niclass2010}
C.~Niclass, M.~Soga,
\newblock In \emph{2010 International Electron Devices Meeting}. IEEE,
  \textbf{2010} 14--3.

\bibitem{Steindl2017}
B.~Steindl, M.~Hofbauer, K.~Schneider-Hornstein, P.~Brandl, H.~Zimmermann,
\newblock \emph{IEEE Journal of Selected Topics in Quantum Electronics}
  \textbf{2017}, \emph{24}, 2 1.

\bibitem{Webster2012}
E.~A.~G. Webster, J.~Richardson, L.~Grant, D.~Renshaw, R.~K. Henderson,
\newblock \emph{IEEE Electron Device Letters} \textbf{2012}, \emph{33}, 5 694.

\bibitem{Takai2016}
I.~Takai, H.~Matsubara, M.~Soga, M.~Ohta, M.~Ogawa, T.~Yamashita,
\newblock \emph{Sensors} \textbf{2016}, \emph{16}, 4 459.

\bibitem{Webster2012bis}
E.~A.~G. Webster, L.~Grant, R.~K. Henderson,
\newblock \emph{IEEE Electron Device Letters} \textbf{2012}, \emph{33}, 11
  1589.

\bibitem{Veerappan2014}
C.~Veerappan, E.~Charbon,
\newblock \emph{IEEE Journal of Selected Topics in Quantum Electronics}
  \textbf{2014}, \emph{20}, 6 299.

\bibitem{Veerappan2016}
C.~Veerappan, E.~Charbon,
\newblock \emph{IEEE Transactions on Electron Devices} \textbf{2016},
  \emph{63}, 1 65.

\bibitem{Guerrieri2010}
F.~Guerrieri, S.~Tisa, A.~Tosi, F.~Zappa,
\newblock \emph{IEEE Photonics Journal} \textbf{2010}, \emph{2}, 5 759.

\bibitem{Sanzaro2018}
M.~Sanzaro, P.~Gattari, F.~Villa, A.~Tosi, G.~Croce, F.~Zappa,
\newblock \emph{IEEE Journal of Selected Topics in Quantum Electronics}
  \textbf{2018}, \emph{24}, 2 1.

\bibitem{Silverstone2016}
J.~W. Silverstone, D.~Bonneau, J.~L. O'Brien, M.~G. Thompson,
\newblock \emph{IEEE Journal of Selected Topics in Quantum Electronics}
  \textbf{2016}, \emph{22}, 6 390.

\bibitem{Qiang2018}
X.~Qiang, X.~Zhou, J.~Wang, C.~M. Wilkes, T.~Loke, S.~O'Gara, L.~Kling, G.~D.
  Marshall, R.~Santagati, T.~C. Ralph, J.~B. Wang, J.~L. O'Brien, M.~G.
  Thompson, J.~C.~F. Matthews,
\newblock \emph{Nature Photonics} \textbf{2018}, \emph{12}, 9 534.

\bibitem{Korzh2015}
B.~Korzh, C.~C.~W. Lim, R.~Houlmann, N.~Gisin, M.~J. Li, D.~Nolan,
  B.~Sanguinetti, R.~Thew, H.~Zbinden,
\newblock \emph{Nature Photonics} \textbf{2015}, \emph{9}, 3 163.

\bibitem{Liao2017}
S.-K. Liao, H.-L. Yong, C.~Liu, G.-L. Shentu, D.-D. Li, J.~Lin, H.~Dai, S.-Q.
  Zhao, B.~Li, J.-Y. Guan, W.~Chen, Y.-H. Gong, Y.~Li, Z.-H. Lin, G.-S. Pan,
  J.~S. Pelc, M.~M. Fejer, W.-Z. Zhang, W.-Y. Liu, J.~Yin, J.-G. Ren, X.-B.
  Wang, Q.~Zhang, C.-Z. Peng, J.-W. Pan,
\newblock \emph{Nature Photonics} \textbf{2017}, \emph{11}, 8 509.

\bibitem{Yuan2016}
Z.~L. Yuan, B.~Fr{\"{o}}hlich, M.~Lucamarini, G.~L. Roberts, J.~F. Dynes, A.~J.
  Shields,
\newblock \emph{Physical Review X} \textbf{2016}, \emph{6}, 3 031044.

\bibitem{Kaneda2019}
F.~Kaneda, P.~G. Kwiat,
\newblock \emph{Science Advances} \textbf{2019}, \emph{5}, 10 eaaw8586.

\bibitem{Portalupi2019}
S.~L. Portalupi, M.~Jetter, P.~Michler,
\newblock \emph{Semiconductor Science and Technology} \textbf{2019}, \emph{34},
  5 053001.

\bibitem{Kim2016}
J.-H. Kim, T.~Cai, C.~J.~K. Richardson, R.~P. Leavitt, E.~Waks,
\newblock \emph{Optica} \textbf{2016}, \emph{3}, 6 577.

\bibitem{Tanzilli2005}
S.~Tanzilli, W.~Tittel, M.~Halder, O.~Alibart, P.~Baldi, N.~Gisin, H.~Zbinden,
\newblock \emph{Nature} \textbf{2005}, \emph{437}, 7055 116.

\bibitem{Rakher2010}
M.~T. Rakher, L.~Ma, O.~Slattery, X.~Tang, K.~Srinivasan,
\newblock \emph{Nature Photonics} \textbf{2010}, \emph{4}, 11 786.

\bibitem{Lacaita1996}
A.~Lacaita, F.~Zappa, S.~Cova, P.~Lovati,
\newblock \emph{Applied Optics} \textbf{1996}, \emph{35}, 16 2986.

\bibitem{Hiskett2000}
P.~A. Hiskett, G.~S. Buller, A.~Y. Loudon, J.~M. Smith, I.~Gontijo, A.~C.
  Walker, P.~D. Townsend, M.~J. Robertson,
\newblock \emph{Applied Optics} \textbf{2000}, \emph{39}, 36 6818.

\bibitem{Acerbi2013}
F.~Acerbi, M.~Anti, A.~Tosi, F.~Zappa,
\newblock \emph{IEEE Photonics Journal} \textbf{2013}, \emph{5}, 2 6800209.

\bibitem{Ma2016}
J.~Ma, B.~Bai, L.-J. Wang, C.-Z. Tong, G.~Jin, J.~Zhang, J.-W. Pan,
\newblock \emph{Applied Optics} \textbf{2016}, \emph{55}, 27 7497.

\bibitem{Pellegrini2006}
S.~Pellegrini, R.~E. Warburton, L.~J.~J. Tan, J.~S. Ng, A.~B. Krysa, K.~Groom,
  J.~P.~R. David, S.~Cova, M.~J. Robertson, G.~S. Buller,
\newblock \emph{IEEE Journal of Quantum Electronics} \textbf{2006}, \emph{42},
  4 397.

\bibitem{Jiang2007}
X.~Jiang, M.~A. Itzler, R.~Ben-Michael, K.~Slomkowski,
\newblock \emph{IEEE Journal of Selected Topics in Quantum Electronics}
  \textbf{2007}, \emph{13}, 4 895.

\bibitem{Kang2004}
Y.~Kang, Y.-H. Lo, M.~Bitter, S.~Kristjansson, Z.~Pan, A.~Pauchard,
\newblock \emph{Applied Physics Letters} \textbf{2004}, \emph{85}, 10 1668.

\bibitem{Meng2016}
X.~Meng, S.~Xie, X.~Zhou, N.~Calandri, M.~Sanzaro, A.~Tosi, C.~H. Tan, J.~S.
  Ng,
\newblock \emph{Royal Society Open Science} \textbf{2016}, \emph{3}, 3 150584.

\bibitem{Zhang2015}
J.~Zhang, M.~A. Itzler, H.~Zbinden, J.-W. Pan,
\newblock \emph{Light: Science {\&} Applications} \textbf{2015}, \emph{4}, 5
  e286.

\bibitem{Ribordy1998}
G.~Ribordy, J.-D. Gautier, N.~Gisin, O.~Guinnard, H.~Zbinden,
\newblock \emph{Electronics Letters} \textbf{1998}, \emph{34}, 22 2116.

\bibitem{Bourennane1999}
M.~Bourennane, F.~Gibson, A.~Karlsson, A.~Hening, P.~Jonsson, T.~Tsegaye,
  D.~Ljunggren, E.~Sundberg,
\newblock \emph{Optics Express} \textbf{1999}, \emph{4}, 10 383.

\bibitem{Peng2007}
C.-Z. Peng, J.~Zhang, D.~Yang, W.-B. Gao, H.-X. Ma, H.~Yin, H.-P. Zeng,
  T.~Yang, X.-B. Wang, J.-W. Pan,
\newblock \emph{Physical Review Letters} \textbf{2007}, \emph{98}, 1 010505.

\bibitem{Walenta2012}
N.~Walenta, T.~Lunghi, O.~Guinnard, R.~Houlmann, H.~Zbinden, N.~Gisin,
\newblock \emph{Journal of Applied Physics} \textbf{2012}, \emph{112}, 6
  063106.

\bibitem{Frohlich2017}
B.~Fr{\"{o}}hlich, M.~Lucamarini, J.~F. Dynes, L.~C. Comandar, W.-S. Tam,
  A.~Plews, A.~W. Sharpe, Z.~Yuan, A.~J. Shields,
\newblock \emph{Optica} \textbf{2017}, \emph{4}, 1 163.

\bibitem{Korzh2014}
B.~Korzh, N.~Walenta, T.~Lunghi, N.~Gisin, H.~Zbinden,
\newblock \emph{Applied Physics Letters} \textbf{2014}, \emph{104}, 8 081108.

\bibitem{Liu2012}
Y.~Liu, L.~Ju, X.-L. Liang, S.-B. Tang, G.-L.~S. Tu, L.~Zhou, C.-Z. Peng,
  K.~Chen, T.-Y. Chen, Z.-B. Chen, J.-W. Pan,
\newblock \emph{Physical Review Letters} \textbf{2012}, \emph{109}, 3 030501.

\bibitem{Williams2019}
B.~P. Williams, J.~M. Lukens, N.~A. Peters, B.~Qi, W.~P. Grice,
\newblock \emph{Physical Review A} \textbf{2019}, \emph{99}, 6 062311.

\bibitem{Davanco2012}
M.~Davan{\c{c}}o, J.~R. Ong, A.~B. Shehata, A.~Tosi, I.~Agha, S.~Assefa,
  F.~Xia, W.~M.~J. Green, S.~Mookherjea, K.~Srinivasan,
\newblock \emph{Applied Physics Letters} \textbf{2012}, \emph{100}, 26 261104.

\bibitem{Dynes2008}
J.~F. Dynes, Z.~L. Yuan, A.~W. Sharpe, A.~J. Shields,
\newblock \emph{Applied Physics Letters} \textbf{2008}, \emph{93}, 3 031109.

\bibitem{Jiang2015}
X.~Jiang, M.~Itzler, K.~ODonnell, M.~Entwistle, M.~Owens, K.~Slomkowski,
  S.~Rangwala,
\newblock \emph{IEEE Journal of Selected Topics in Quantum Electronics}
  \textbf{2015}, \emph{21}, 3 5.

\bibitem{Kardyna2008}
B.~E. Kardyna{\l}, Z.~L. Yuan, A.~J. Shields,
\newblock \emph{Nature Photonics} \textbf{2008}, \emph{2}, 7 425.

\bibitem{Itzler2007}
M.~A. Itzler, R.~Ben-Michael, C.~F. Hsu, K.~Slomkowski, A.~Tosi, S.~Cova,
  F.~Zappa, R.~Ispasoiu,
\newblock \emph{Journal of Modern Optics} \textbf{2007}, \emph{54}, 2-3 283.

\bibitem{Tosi2009}
A.~Tosi, A.~{Dalla Mora}, F.~Zappa, S.~Cova,
\newblock \emph{Journal of Modern Optics} \textbf{2009}, \emph{56}, 2-3 299.

\bibitem{Tosi2012}
A.~Tosi, F.~Acerbi, M.~Anti, F.~Zappa,
\newblock \emph{IEEE Journal of Quantum Electronics} \textbf{2012}, \emph{48},
  9 1227.

\bibitem{Tosi2014}
A.~Tosi, N.~Calandri, M.~Sanzaro, F.~Acerbi,
\newblock \emph{IEEE Journal of Selected Topics in Quantum Electronics}
  \textbf{2014}, \emph{20}, 6 192.

\bibitem{Itzler2011}
M.~A. Itzler, X.~Jiang, M.~Entwistle, K.~Slomkowski, A.~Tosi, F.~Acerbi,
  F.~Zappa, S.~Cova,
\newblock \emph{Journal of Modern Optics} \textbf{2011}, \emph{58}, 3-4 174.

\bibitem{Yuan2007}
Z.~L. Yuan, B.~E. Kardynal, A.~W. Sharpe, A.~J. Shields,
\newblock \emph{Applied Physics Letters} \textbf{2007}, \emph{91}, 4 041114.

\bibitem{Yuan2010}
Z.~L. Yuan, A.~W. Sharpe, J.~F. Dynes, A.~R. Dixon, A.~J. Shields,
\newblock \emph{Applied Physics Letters} \textbf{2010}, \emph{96}, 7 071101.

\bibitem{Namekata2006}
N.~Namekata, S.~Sasamori, S.~Inoue,
\newblock \emph{Optics Express} \textbf{2006}, \emph{14}, 21 10043.

\bibitem{Zhang2010}
J.~Zhang, P.~Eraerds, N.~Walenta, C.~Barreiro, R.~Thew, H.~Zbinden,
\newblock In M.~A. Itzler, J.~C. Campbell, editors, \emph{SPIE Defense,
  Security + Sensing}, volume 7681. \textbf{2010} 76810Z--76810Z--8,
\newblock
  \urlprefix\url{http://proceedings.spiedigitallibrary.org/proceeding.aspx?doi=10.1117/12.862118}.

\bibitem{Zhang2009}
J.~Zhang, R.~Thew, C.~Barreiro, H.~Zbinden,
\newblock \emph{Applied Physics Letters} \textbf{2009}, \emph{95}, 9 091103.

\bibitem{Restelli2013}
A.~Restelli, J.~C. Bienfang, A.~Migdall,
\newblock \emph{Applied Physics Letters} \textbf{2013}, \emph{102}, 14 141104.

\bibitem{Comandar2015}
L.~C. Comandar, B.~Fr{\"{o}}hlich, J.~F. Dynes, A.~W. Sharpe, M.~Lucamarini,
  Z.~L. Yuan, R.~V. Penty, A.~J. Shields,
\newblock \emph{Journal of Applied Physics} \textbf{2015}, \emph{117}, 8
  083109.

\bibitem{Scarcella2015}
C.~Scarcella, G.~Boso, A.~Ruggeri, A.~Tosi,
\newblock \emph{IEEE Journal of Selected Topics in Quantum Electronics}
  \textbf{2015}, \emph{21}, 3 17.

\bibitem{Warburton2009}
R.~E. Warburton, M.~Itzler, G.~S. Buller,
\newblock \emph{Applied Physics Letters} \textbf{2009}, \emph{94}, 7 071116.

\bibitem{Thew2007}
R.~T. Thew, D.~Stucki, J.-D. Gautier, H.~Zbinden, A.~Rochas,
\newblock \emph{Applied Physics Letters} \textbf{2007}, \emph{91}, 20 201114.

\bibitem{Liu2008}
M.~Liu, C.~Hu, J.~C. Campbell, Z.~Pan, M.~M. Tashima,
\newblock \emph{IEEE Journal of Quantum Electronics} \textbf{2008}, \emph{44},
  5 430.

\bibitem{IDQube2020}
{ID Quantique SA},
\newblock {ID Qube NIR Datasheet}, \textbf{2020},
\newblock
  \urlprefix\url{{https://marketing.idquantique.com/acton/attachment/11868/f-926db6fe-7c84-4bed-92fa-6d90a2612e03/1/-/-/-/-/ID
  Qube NIR Gated Brochure.pdf}}.

\bibitem{PDM-IR2019}
{Micro Photon Devices S.r.l.},
\newblock {PDM-IR Datasheet}, \textbf{2019},
\newblock
  \urlprefix\url{{http://www.micro-photon-devices.com/MPD/media/Datasheet/PDM-IR
  Datasheet window.pdf}}.

\bibitem{Itzler2009}
M.~A. Itzler, X.~Jiang, B.~Nyman, K.~Slomkowski,
\newblock In M.~Razeghi, R.~Sudharsanan, G.~J. Brown, editors, \emph{SPIE
  Photonics West - OPTO}, volume 7222. \textbf{2009} 72221K--72221K--12,
\newblock
  \urlprefix\url{http://proceedings.spiedigitallibrary.org/proceeding.aspx?doi=10.1117/12.814669}.

\bibitem{Itzler2010}
M.~A. Itzler, X.~Jiang, B.~M. Onat, K.~Slomkowski,
\newblock In M.~Razeghi, R.~Sudharsanan, G.~J. Brown, editors, \emph{SPIE
  Photonics West - OPTO}, volume 7608. \textbf{2010} 760829--760829--13,
\newblock
  \urlprefix\url{http://proceedings.spiedigitallibrary.org/proceeding.aspx?doi=10.1117/12.843588}.

\bibitem{Lunghi2012}
T.~Lunghi, C.~Barreiro, O.~Guinnard, R.~Houlmann, X.~Jiang, M.~A. Itzler,
  H.~Zbinden,
\newblock \emph{Journal of Modern Optics} \textbf{2012}, \emph{59}, 17 1481.

\bibitem{Amri2016}
E.~Amri, G.~Boso, B.~Korzh, H.~Zbinden,
\newblock \emph{Optics Letters} \textbf{2016}, \emph{41}, 24 5728.

\bibitem{Sanzaro2016}
M.~Sanzaro, N.~Calandri, A.~Ruggeri, A.~Tosi,
\newblock \emph{IEEE Journal of Quantum Electronics} \textbf{2016}, \emph{52},
  7 1.

\bibitem{Cheng2011}
J.~Cheng, S.~You, S.~Rahman, Y.-H. Lo,
\newblock \emph{Optics Express} \textbf{2011}, \emph{19}, 16 15149.

\bibitem{Calandri2016}
N.~Calandri, M.~Sanzaro, L.~Motta, C.~Savoia, A.~Tosi,
\newblock \emph{IEEE Photonics Technology Letters} \textbf{2016}, \emph{28}, 16
  1767.

\bibitem{Haecker1971}
W.~Haecker, O.~Groezinger, M.~H. Pilkuhn,
\newblock \emph{Applied Physics Letters} \textbf{1971}, \emph{19}, 4 113.

\bibitem{Lacaita1994}
A.~Lacaita, P.~A. Francese, F.~Zappa, S.~Cova,
\newblock \emph{Applied Optics} \textbf{1994}, \emph{33}, 30 6902.

\bibitem{Buller1996}
G.~S. Buller, S.~J. Fancey, J.~S. Massa, A.~C. Walker, S.~Cova, A.~Lacaita,
\newblock \emph{Applied Optics} \textbf{1996}, \emph{35}, 6 916.

\bibitem{Vines2019}
P.~Vines, K.~Kuzmenko, J.~Kirdoda, D.~C.~S. Dumas, M.~M. Mirza, R.~W. Millar,
  D.~J. Paul, G.~S. Buller,
\newblock \emph{Nature Communications} \textbf{2019}, \emph{10}, 1.

\bibitem{Martinez2017}
N.~J.~D. Martinez, M.~Gehl, C.~T. Derose, A.~L. Starbuck, A.~T. Pomerene, A.~L.
  Lentine, D.~C. Trotter, P.~S. Davids,
\newblock \emph{Optics Express} \textbf{2017}, \emph{25}, 14 16130.

\bibitem{Sammak2016}
A.~Sammak, M.~Aminian, L.~K. Nanver, E.~Charbon,
\newblock \emph{IEEE Transactions on Electron Devices} \textbf{2016},
  \emph{63}, 1 92.

\bibitem{Loudon2002}
A.~Y. Loudon, P.~A. Hiskett, G.~S. Buller, R.~T. Carline, D.~C. Herbert, W.~Y.
  Leong, J.~G. Rarity,
\newblock \emph{Optics Letters} \textbf{2002}, \emph{27}, 4 219.

\bibitem{Kang2009}
Y.~Kang, H.~D. Liu, M.~Morse, M.~J. Paniccia, M.~Zadka, S.~Litski, G.~Sarid,
  A.~Pauchard, Y.~H. Kuo, H.~W. Chen, W.~S. Zaoui, J.~E. Bowers, A.~Beling,
  D.~C. McIntosh, X.~Zheng, J.~C. Campbell,
\newblock \emph{Nature Photonics} \textbf{2009}, \emph{3}, 1 59.

\bibitem{Tan2012}
Y.~H. Tan, C.~S. Tan,
\newblock \emph{Thin Solid Films} \textbf{2012}, \emph{520}, 7 2711.

\bibitem{Lu2011}
Z.~Lu, Y.~Kang, C.~Hu, Q.~Zhou, H.-D. Liu, J.~C. Campbell,
\newblock \emph{IEEE Journal of Quantum Electronics} \textbf{2011}, \emph{47},
  5 731.

\bibitem{Warburton2013}
R.~E. Warburton, G.~Intermite, M.~Myronov, P.~Allred, D.~R. Leadley,
  K.~Gallacher, D.~J. Paul, N.~J. Pilgrim, L.~J.~M. Lever, Z.~Ikonic, R.~W.
  Kelsall, E.~Huante-Ceron, A.~P. Knights, G.~S. Buller,
\newblock \emph{IEEE Transactions on Electron Devices} \textbf{2013},
  \emph{60}, 11 3807.

\bibitem{Tosi2007}
A.~Tosi, A.~{Dalla Mora}, F.~Zappa, S.~Cova,
\newblock In W.~Becker, editor, \emph{SPIE Optics East}, volume 6771.
  \textbf{2007} 67710P--67710P--12,
\newblock
  \urlprefix\url{http://proceedings.spiedigitallibrary.org/proceeding.aspx?doi=10.1117/12.734961}.

\bibitem{Prevedel2007}
R.~Prevedel, P.~Walther, F.~Tiefenbacher, P.~B{\"{o}}hi, R.~Kaltenbaek,
  T.~Jennewein, A.~Zeilinger,
\newblock \emph{Nature} \textbf{2007}, \emph{445}, 7123 65.

\bibitem{Walther2005}
P.~Walther, K.~J. Resch, T.~Rudolph, E.~Schenck, H.~Weinfurter, V.~Vedral,
  M.~Aspelmeyer, A.~Zeilinger,
\newblock \emph{Nature} \textbf{2005}, \emph{434}, 7030 169.

\bibitem{Bowers1986}
J.~E. Bowers, C.~A. Burrus,
\newblock \emph{Electronics Letters} \textbf{1986}, \emph{22}, 17 905.

\bibitem{Koester2006}
S.~J. Koester, J.~D. Schaub, G.~Dehlinger, J.~O. Chu,
\newblock \emph{IEEE Journal of Selected Topics in Quantum Electronics}
  \textbf{2006}, \emph{12}, 6 1489.

\bibitem{Zhu2009}
S.~Zhu, K.-W. Ang, S.~C. Rustagi, J.~Wang, Y.~Z. Xiong, G.~Q. Lo, D.~L. Kwong,
\newblock \emph{IEEE Electron Device Letters} \textbf{2009}, \emph{30}, 9 934.

\bibitem{Lu2019}
X.~Lu, Q.~Li, D.~A. Westly, G.~Moille, A.~Singh, V.~Anant, K.~Srinivasan,
\newblock \emph{Nature Physics} \textbf{2019}, \emph{15}, 4 373.

\bibitem{Sacher2015}
W.~D. Sacher, Y.~Huang, G.-Q. Lo, J.~K.~S. Poon,
\newblock \emph{Journal of Lightwave Technology} \textbf{2015}, \emph{33}, 4
  901.

\bibitem{Yanikgonul2018}
S.~Yanikgonul, V.~Leong, J.~R. Ong, C.~E. Png, L.~Krivitsky,
\newblock \emph{Optics Express} \textbf{2018}, \emph{26}, 12 15232.

\bibitem{Yanikgonul2020}
S.~Yanikgonul, V.~Leong, J.~R. Ong, C.~E. Png, L.~Krivitsky,
\newblock \emph{IEEE Journal of Selected Topics in Quantum Electronics}
  \textbf{2020}, \emph{26}, 2 1.

\bibitem{Malik2014}
A.~Malik, S.~Dwivedi, L.~{Van Landschoot}, M.~Muneeb, Y.~Shimura, G.~Lepage,
  J.~{Van Campenhout}, W.~Vanherle, T.~{Van Opstal}, R.~Loo, G.~Roelkens,
\newblock \emph{Optics Express} \textbf{2014}, \emph{22}, 23 28479.

\bibitem{Younis2016}
U.~Younis, S.~K. Vanga, A.~E.-J. Lim, P.~G.-Q. Lo, A.~A. Bettiol, K.-W. Ang,
\newblock \emph{Optics Express} \textbf{2016}, \emph{24}, 11 11987.

\bibitem{Soref2019}
R.~A. Soref, F.~{De Leonardis}, V.~M.~N. Passaro,
\newblock \emph{ACS Applied Nano Materials} \textbf{2019}, \emph{2}, 12 7503.

\bibitem{Elshaari2020}
A.~W. Elshaari, W.~H.~P. Pernice, K.~Srinivasan, O.~Benson, V.~Zwiller,
\newblock \emph{Nature Photonics} \textbf{2020}, \emph{14}, 5 285.

\bibitem{Kim2020}
J.-H. Kim, S.~Aghaeimeibodi, J.~Carolan, D.~Englund, E.~Waks,
\newblock \emph{Optica} \textbf{2020}, \emph{7}, 4 291.

\bibitem{Ke2017}
S.~Ke, S.~Lin, D.~Mao, Y.~Ye, X.~Ji, W.~Huang, C.~Li, S.~Chen,
\newblock \emph{Applied Optics} \textbf{2017}, \emph{56}, 16 4646.

\bibitem{Roelkens2005}
G.~Roelkens, J.~Brouckaert, D.~Taillaert, P.~Dumon, W.~Bogaerts, D.~{Van
  Thourhout}, R.~Baets, R.~N{\"{o}}tzel, M.~Smit,
\newblock \emph{Optics Express} \textbf{2005}, \emph{13}, 25 10102.

\end{thebibliography}


\begin{figure}
    \includegraphics[width=40mm,height=50mm]{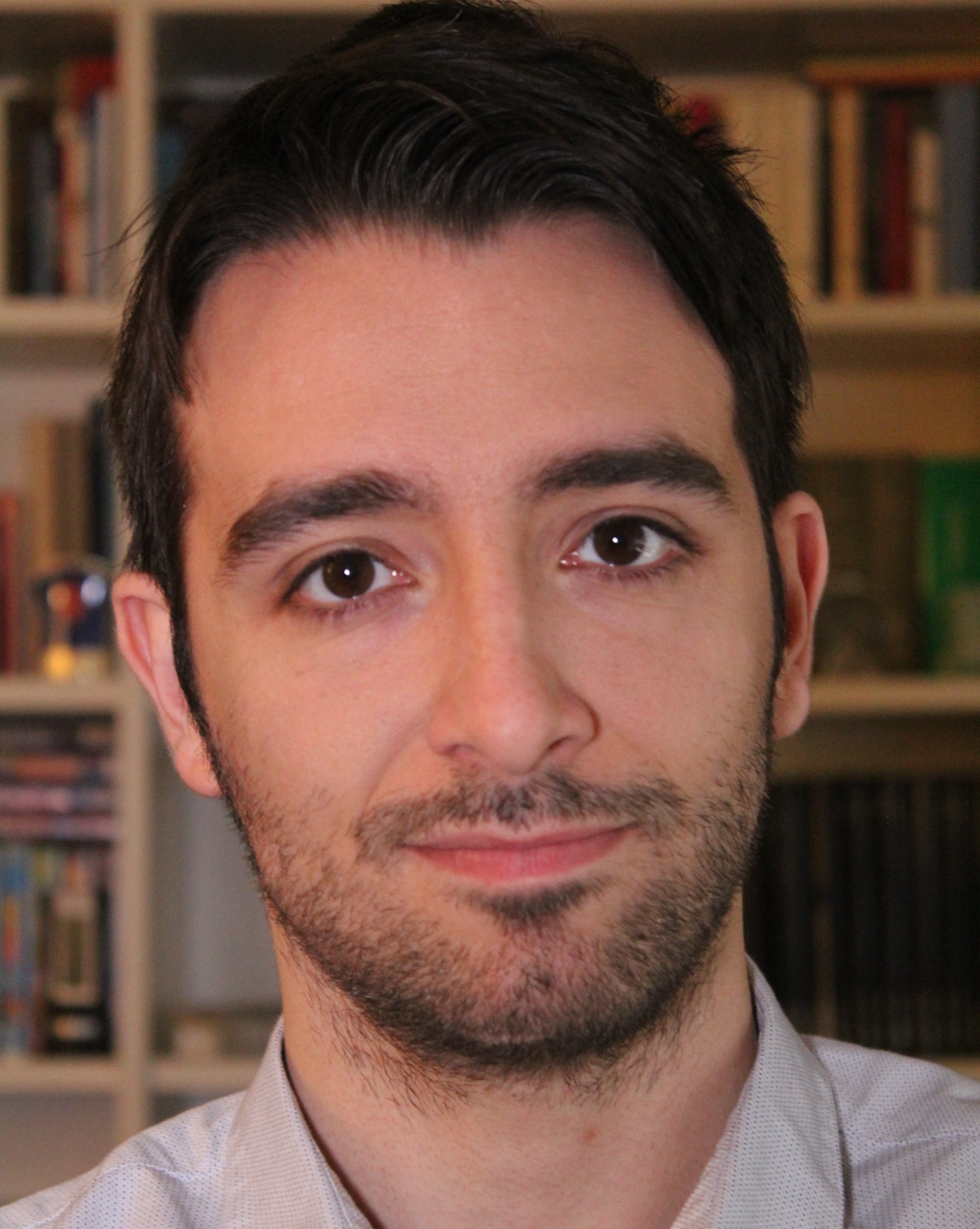}
    \caption*{Francesco Ceccarelli received the Master’s Degree (summa cum laude) in Electronics Engineering from the Politecnico di Milano (Italy) in 2014 and the Ph.D. (with honors) in Information Technology from the same university in 2018, with a dissertation on custom-technology single-photon avalanche diode arrays. Since 2020 he is permanent researcher at the Institute for Photonics and Nanotechnologies (IFN) of the Italian National Research Council (CNR), working on the development of reconfigurable integrated optical circuits for photonic quantum information processing.}
\end{figure}

\begin{figure}
    \includegraphics[width=40mm,height=50mm]{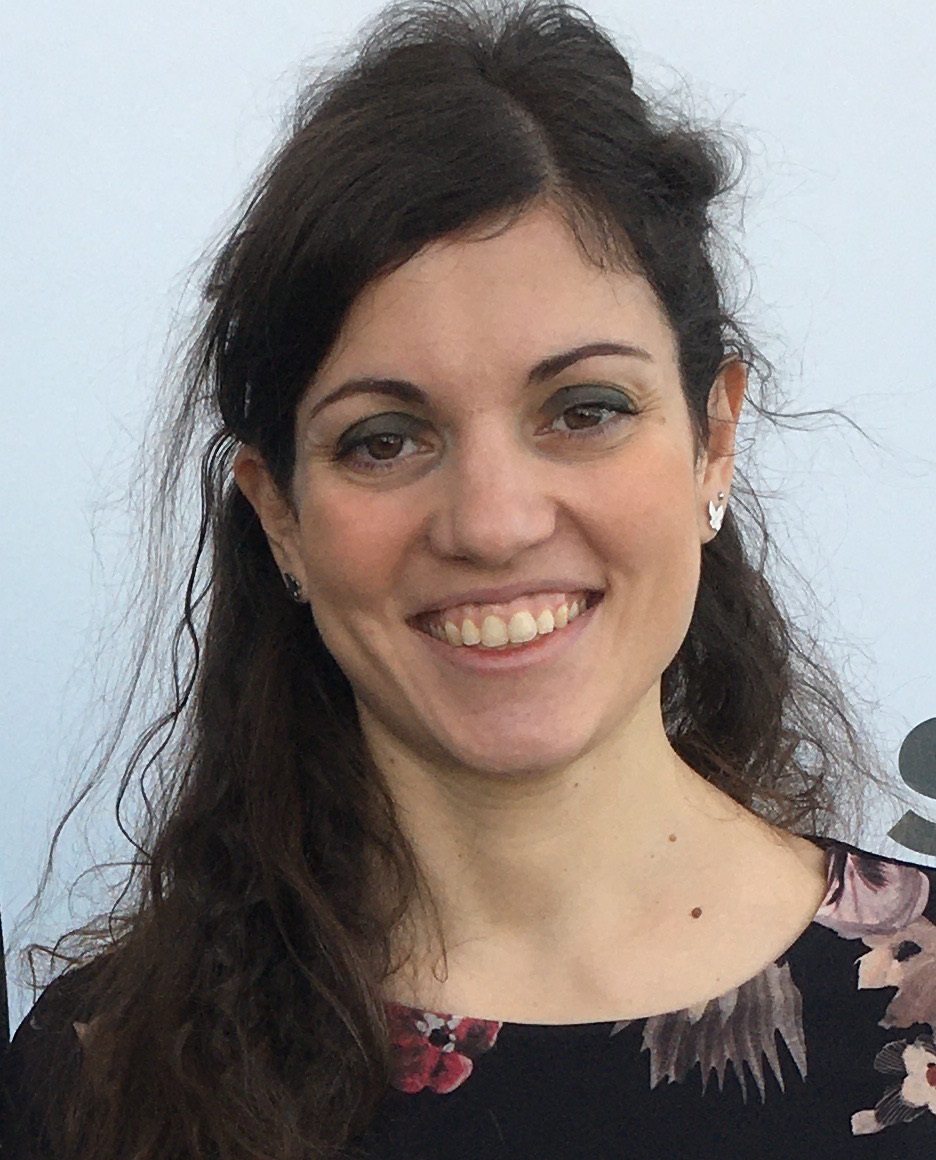}
    \caption*{Giulia Acconcia received her Master’s Degree (summa cum laude) in Electronics Engineering and her Ph.D. (with honors) in Information Technology from the Politecnico di Milano (Italy), in 2013 and 2017, respectively. Since 2020 she is a researcher at the Politecnico di Milano. Her current research interests include the development of fully-integrated electronics and systems to extract the best performance from custom single-photon avalanche diodes. She has been awarded SPIE Rising Researcher in 2019 and Young Investigator Award in 2020.}
\end{figure}

\begin{figure}
    \includegraphics[width=40mm,height=50mm]{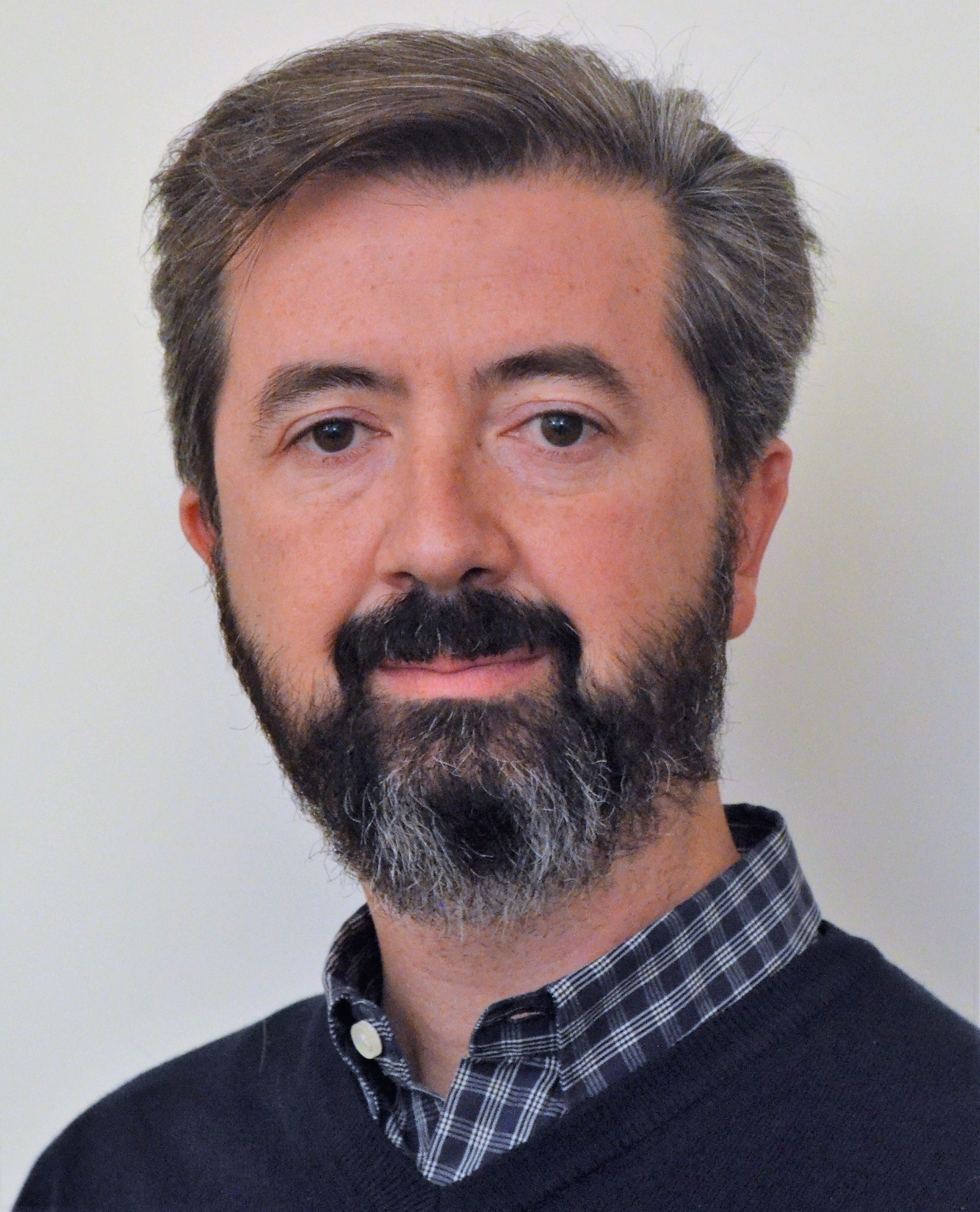}
    \caption*{Roberto Osellame is a Director of Research at the Institute for Photonics and Nanotechnologies (IFN) of the Italian National Research Council (CNR), and a Contract Professor at the Politecnico di Milano. He received his Ph.D. in Physics from the Politecnico di Torino (Italy) in 2000. His research interests focus on microfabrication of integrated photonic devices for such diverse applications as quantum technologies, lab-on-a-chip, and optical communications. He is a Fellow of the Optical Society of America.}
\end{figure}


\begin{figure}
    \textbf{Table of Contents}\\
    \medskip
    \includegraphics[width=55mm,height=50mm]{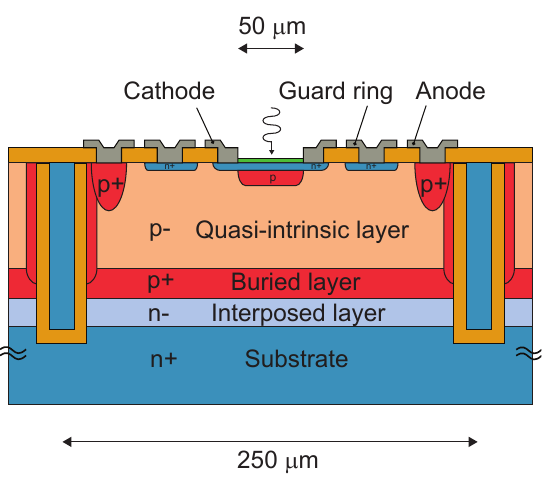}
    \medskip
    \caption*{Photonic quantum technologies promise a revolution of the world of information processing thanks to the many advantages of exploiting single photons as quantum information carriers. In this scenario, single-photon avalanche diodes (SPADs) play a key role. In this paper, we provide a thorough discussion of the recent progress made on these detectors, concluding with our vision of the future.}
\end{figure}

\end{document}